\documentclass[%
 reprint,
 superscriptaddress,
nofootinbib,
 amsmath,amssymb,
 aps,
 pra,
]{revtex4-2}

\usepackage{graphicx}
\usepackage{dcolumn}
\usepackage{bm}
\usepackage{url}
\usepackage[colorlinks,citecolor=blue,linkcolor=blue,urlcolor=blue]{hyperref}
\usepackage{enumitem}

\usepackage{changes}
\definecolor{darkviolet}{rgb}{0.58, 0.0, 0.83}
\definecolor{pakistangreen}{rgb}{0.0, 0.4, 0.0}
\definechangesauthor[name={Hector}, color=blue]{H}
\definechangesauthor[name={Francois}, color=red]{F}
\definechangesauthor[name={Lock}, color=darkviolet]{L}
\usepackage{framed}

\usepackage{mathtools}

\def\f{\frac}
\def\k{\kappa}

\def\ve{\varepsilon}
\def\be{\beta}
\def\D{\Delta}

\def\t{\tau}
\def\a{\alpha}
\def\mP{\mathcal{P}}
\def\mI{\mathcal{I}}
\def\mS{\mathcal{S}}
\def\mJ{\mathcal{J}}
\def\s{\sigma}
\def\kb{k_B}
\def\d{\delta}
\def\p{\partial}
\def\l{\lambda}
\def\L{\Lambda}
\def\G{\Gamma}
\def\o{\Omega}

\def\g{\gamma}

\newcommand{\avg}[1]{\langle #1 \rangle}

\newcommand{\llangle}{\langle\!\langle}
\newcommand{\rrangle}{\rangle\!\rangle}
\newcommand{\avgg}[1]{\llangle #1 \rrangle}

\newcommand{\Bigllangle}{\Big\langle\!\!\Big\langle}
\newcommand{\Bigrrangle}{\Big\rangle\!\!\Big\rangle}
\newcommand{\Bavgg}[1]{\Bigllangle #1 \Bigrrangle}
\newcommand{\Bavg}[1]{\Big\langle #1 \Big\rangle}

\usepackage{amsthm}
\theoremstyle{remark}
\newtheorem{remark}{Technical remark}

\begin{document}

\title{A variational formulation of stochastic thermodynamics: Spatially extended systems}

\author{H\'ector Vaquero del Pino}
\email{hector001@e.ntu.edu.sg}
\affiliation{Division of Physics and Applied Physics, School of Physical and Mathematical Sciences,
Nanyang Technological University, 21 Nanyang Link, Singapore 637371}

\author{Fran\c{c}ois Gay-Balmaz}
\email{francois.gb@ntu.edu.sg}
\affiliation{
 Division of Mathematical Sciences, School of Physical and Mathematical Sciences,
Nanyang Technological University, 21 Nanyang Link, Singapore 637371
}

\author{Hiroaki Yoshimura}
\affiliation{School of Science and Engineering, Waseda University. 3–4–1, Okubo, Shinjuku, Tokyo 169-8555, Japan}

\author{Lock Yue Chew}
\email{lockyue@ntu.edu.sg}
\affiliation{Division of Physics and Applied Physics, School of Physical and Mathematical Sciences,
Nanyang Technological University, 21 Nanyang Link, Singapore 637371}

\date{\today}

\begin{abstract}
Stochastic field theories are often constructed phenomenologically, without a systematic assessment of thermodynamic consistency or local detailed balance. This may hinder a physical description of irreversibility at the field-theoretic level beyond the standard statistical formulation of stochastic thermodynamics. Here, we develop a variational formulation for thermodynamically consistent stochastic field theories by extending Hamilton’s principle of classical field theory. Introducing the second law as an axiom yields a consistent local thermodynamic structure in which novel fluctuation-dissipation relations emerge naturally, ensuring local detailed balance. Remarkably, the resulting entropy production takes the same form as in standard stochastic thermodynamics, up to a reformulation in an extended phase space incorporating both configurational and thermal variables. This correspondence extends key results, including individual trajectory-level thermodynamics and fluctuation theorems. The construction is formulated within a unified geometric framework based on a generalized Lagrange-d’Alembert principle, providing a top-down connection between phenomenological modeling and thermodynamic consistency. Potential applications include thermodynamically consistent modeling of complex fluids, Lagrangian reduction by symmetry in continuum systems, and structure-preserving numerical schemes for stochastic partial differential equations.
\end{abstract}

\maketitle


\section{\label{sec:intro} Introduction}

Variational mechanics, deduced from Hamilton’s Principle, has served as the cornerstone for a large class of physical theories, from classical and quantum mechanics to field theory and continuum systems  \cite{landau_mechanics,landau_fields,lanczos1986variational,Feynman_PI,Weinberg1995,marsden_hughes_1983}. Moreover, it provides a direct connection with geometric structures, enabling a geometric axiomatization of theoretical physics \cite{Jost2009,abraham_marsden_1987,Arnold1978,goldstein2002classical,Synge1963PrinciplesOM}.  Recently, variational mechanics has been extended to thermodynamics, incorporating irreversible processes and entropy evolution \cite{GBYo2019}. The generality and systematic character of this approach highlight its power as a modeling paradigm.

On the other hand, stochastic thermodynamics (ST) has emerged as the underlying foundation to classical thermodynamics, describing stochastic dynamics ranging from nanoscale to living systems out of equilibrium \cite{Seifert_2012,VANDENBROECK20156}, where the conventional tools of statistical mechanics do not apply. This framework builds on the extensions of the first and second laws to the stochastic regime. Remarkably, it has achieved an individual-trajectory level description of thermodynamics \cite{Seifert_2005,Sekimoto2010}, a (generalized) Clausius equality relation between heat and entropy production at the stochastic level \cite{Horowitz_2014,Horowitz_esposito}, and fluctuation theorems valid arbitrarily far from equilibrium \cite{Seifert_2012}, which have been verified in several experimental platforms \cite{PhysRevLett.89.050601,PhysRevLett.96.070603,PhysRevLett.104.218103,PhysRevLett.133.090402}. Under the assumption of timescale separation, the observable (slow) and thermal (fast) degrees of freedom (DoFs) are typically modeled by Markovian dynamics, such as master equations (for discrete-time systems) or Langevin equations (for continuous-time systems) \cite{Seifert_2012}.

Despite its success, ST does not prescribe how models should be constructed, in contrast to other thermodynamic frameworks such as projection methods~\cite{zwanzig,Grabert1982}. While ST provides a consistent information-theoretic characterization of irreversibility \cite{pelitti,VANDENBROECK20156,Seifert_2005,Seifert_2012,ActiveFields,Fodor_2022} and a complementary energetic formulation \cite{Sekimoto2010}, the connection between stochastic energetics and entropy production is not systematic at the level of model construction. The reconciliation between the energetics and entropy production is mediated by the principle of local detailed balance (LDB), which encodes the time reversal symmetry of the microscopic laws of Physics, or microreversibility \cite{Maes_LDB,Maes2003}. As such, LDB is thus essential for constructing physically meaningful and thermodynamically interpretable models \cite{ActiveFields,PhysRevX.15.021050,Maes_LDB}. 

This issue becomes more challenging for spatially extended systems whose dynamics are described by stochastic partial differential equations (SPDEs). For collective phenomena, stochastic field theories have proven powerful, yet they are typically formulated in a top-down manner, guided by symmetry arguments~ \cite{Kardar2007,Mazenko2006,ojalvo,Ma2018,Tauber2014,hohenberg}, due to the well-known difficulties associated with coarse-graining microscopic dynamics \cite{PhysRevX.15.021050}. This practice can lead to thermodynamically inconsistent models \cite{Esposito2012,esposito}, in which entropy production is quantified solely through information-theoretic measures \cite{Fodor_2022,ActiveFields}, rather than through the system energetics. 

For continuous-variable systems considered here, LDB manifests itself in the Langevin equation through the fluctuation–dissipation relation (FDR). Throughout this work, we refer to the FDR of the second kind, which links random and dissipative forces \cite{Broeck_1,Broeck_2,Polettini, VANDENBROECK20156}, and should not be identified with the FDR of the first kind, which concerns the system’s response to an external perturbation~\cite{Maes_FDR}. Crucially, imposing LDB in the set-up of Markov dynamics implies an FDR of the second kind, ensuring the thermodynamic consistency of the associated Langevin dynamics; this, in turn, implies an FDR of the first kind \textit{near} equilibrium \cite{Maes_FDR,Jung2021}. 

In this work, we develop a novel modeling paradigm for the ST of spatially extended systems, providing a geometric framework that retains the generality of Hamilton’s Principle while systematically enforcing thermodynamic consistency. As a top-down approach, it recovers Hamilton’s Principle of classical field theory as the reversible limiting case---thereby differentiating it from other variational methods---while endowing stochastic field theories with a rigorous geometric structure. This is achieved through a generalized Lagrange–d’Alembert principle, building on the variational structure introduced in Ref. \cite{HVP_partI}, in analogy with its macroscopic (deterministic) counterpart \cite{GBpart2}, by incorporating entropy production as a nonlinear nonholonomic constraint.

The key insight of this framework is the introduction of thermodynamic displacements, dual (conjugate) to the thermal variables, which provide a natural definition of the variational constraint and allow for a variational derivation of the associated dissipative dynamics. The resulting variational structure yields a complete set of evolution equations for both configurational and thermal variables. Introducing the second law as an axiom then systematically yields a consistent thermodynamic structure, ensuring LDB. Remarkably, the resulting entropy production assumes the same form as in standard ST, up to a reformulation in an extended phase space. Finally, the framework provides a systematic treatment of spatial boundary conditions, both at the level of the equations of motion and of the associated functional Fokker–Planck equation.

Beyond its role as a modeling tool, the variational principle further allows for a geometric treatment of Lie group symmetries, as well as the associated Noether conserved quantities and symmetry reduction processes \cite{Marsden1999,GAYBALMAZ2009}, providing a rigorous top-down route to stochastic field theories. We emphasize that we are concerned with Lie groups acting on the configuration manifold, such as symmetries, as opposed to the Markov semigroups that govern the Fokker–Planck equation. For clarity, the present paper addresses continuum non-fluid systems, while symmetry reduction for complex fluids will be illustrated in subsequent work. Moreover, the examples used to illustrate the method comprise the infinite-dimensional extensions of those presented in Ref. \cite{HVP_partI}, allowing for a clear comparison of results and for highlighting the new features inherent to the additional spatial degrees of freedom (DoFs).

From an information-theoretic perspective, the formulation of ST is independent of system dimensionality; accordingly, arguments already developed in Ref. \cite{HVP_partI} will be referred to without repetition, and only those involving technical differences will be explicitly discussed. For notational simplicity, explicit spatial and/or functional dependencies are omitted when unambiguous.

The paper is organized as follows. Section \ref{sec:ST} provides an overview of the framework of ST for fields, which can be skipped by readers already familiar with the topic, except perhaps for subsection \ref{sec:ST_caveats}. Section \ref{sec:VP} introduces the variational principle for ST and illustrates it with examples of (infinite-dimensional) closed and open systems. Section \ref{sec:discussion} offers a discussion of our findings, and Section \ref{sec:conclusion} summarizes the main conclusions of this study. Additional analytical derivations are organized into five appendices at the end. 

\section{\label{sec:ST} Stochastic thermodynamics of fields}
In this section, we extend the framework of ST to spatially extended (infinite-dimensional) systems, following the structure established in Ref. \cite{HVP_partI}: we first summarize the representations of SPDEs, then derive the corresponding stochastic entropy production and energetics, and finally introduce the notion of thermodynamic consistency.

\subsection{\label{sec:representations} Representations of SPDEs}
The stochastic dynamics of spatially extended continuous-time systems admit three equivalent representations: the functional Langevin equation, the functional Fokker–Planck equation, and the path-integral formulation. The equations of motion are expressed as SPDEs, formally understood as the continuum limit of an underlying (finite-dimensional) theory defined on the lattice into which the space is discretized \cite{ojalvo}. This limit, however, is not always well defined~\cite{Hairer2012,Ryser12}, in which case the discrete description must be retained. For ST, where fields represent coarse-grained mesoscale variables, i.e. effective degrees of freedom (DoFs), the discrete formulation is consistent with the short-scale cutoff introduced by coarse-graining and therefore does not entail any conceptual issue \cite{Cates_2022}. Throughout this work, the continuum notation is thus employed as a shorthand for the discrete theory, while we note that a mathematically rigorous meaning of SPDEs can be established in certain cases \cite{Hairer2014}.

Let us define the coarse-grained $d$-dimensional vector field~\cite{WuWang}
\begin{equation}
    \phi^a_\l = \f{1}{\D V}\int _{V_\l} d\mathbf{x}\, \phi^a (\mathbf{x}),
\end{equation}
where $\D V $ gives the cell volume, $a=(1,...,d)$ denotes component index and $\l = \{(i,j,k)\in(1,...,K)\}$ denotes the spatial cell index located at $\mathbf{x}=(i\D x,j \D y, k\D z)$, with its corresponding spatial domain $V_\l$. The lattice thus represents the discretization of the spatial domain $\mathcal{D}\subset\mathbb{R}^3$. The continuum limit then corresponds to \cite{ojalvo,WuWang}
\begin{equation}
    \phi^a (\mathbf{x}) = \lim_{\D V\rightarrow 0} \phi^a_\l.
\end{equation}

The phase space is a Hilbert space $\Omega := L^2(\mathcal{D};\mathbb{R}^d)$, such that $\bm{\phi}\in \Omega$ is a square-integrable vector field 
$\bm{\phi}:\mathcal{D}\rightarrow\mathbb{R}^d$, 
equipped with the real inner product $(\cdot|\cdot):\o\times\o \rightarrow\mathbb{R}$
\begin{equation}\label{eq:inner_prod}
    (\bm{\varphi}|\bm{\phi})\equiv \int_\mathcal{D} d\mathbf{x}\, \varphi^a (\mathbf{x})\phi^a (\mathbf{x}).
\end{equation}
Repeated indices are summed over according to Einstein’s convention. The functional Langevin equation in the Hilbert space $\Omega $ takes the general form \cite{ojalvo,WuWang}:
\begin{equation}\label{eq:Langevin}
    d\phi^a (\mathbf{x},t) = F^a (\mathbf{x},t)[\bm{\phi}]dt+g^a_\k(\mathbf{x},t)[\bm{\phi}]\circ dW^\k (\mathbf{x},t),
\end{equation}
where $[\cdot]$ denotes functional dependence, $ F^a (\mathbf{x},t)[\bm{\phi}]$ is a generalized force, $g^a_\k(\mathbf{x},t)[\bm{\phi}]$ modulates the noise amplitude and the gaussian white noise verifies 
\begin{subequations}\label{eq:gauss_noise}
\begin{align}
    &\avg{dW^\k (\mathbf{x},t)}=0,\\
    &\avg{dW^\k (\mathbf{x},t)dW^\s (\mathbf{x'},t') }=2C^{\k\s}(\mathbf{x},\mathbf{x}';t)\d(t-t')dt.
\end{align}
\end{subequations}
The product $\circ$ denotes the Stratonovich convention (SC), which we adopt throughout this work since it corresponds to the white-noise limit of real noise with vanishing correlation time \cite{ojalvo}. The components of the $Q$-Wiener process in the Hilbert space $\mathcal{U}:= L^2(\mathcal{D}; \mathbb{R}^m)$ are indexed by $\k,\s$ while $a,b$ denote field components in $\mathbb{R}^d$ \cite{DaPrato_Zabczyk_1992}. While noise is delta-correlated in time, we do not impose uncorrelated noise in space for the sake of generality. 

We clarify the notation through an example. A functional is a map $\mathcal{F}:\o\rightarrow \mathbb{R}$, e.g. a polynomial $\mathcal{F}[\bm{\phi}]=\int d\mathbf{x}\, \sum_n a_n|\bm{\phi}(\mathbf{x})|^n$, while an operator, such as the force field $ \bm{F} (\mathbf{x},t)[\bm{\phi}]=  \sum_n a_n\bm{\nabla}^n\bm{\phi}(\mathbf{x},t)$, is a map $\bm{F}:\o\rightarrow \o$. The generalized force in Eq. \eqref{eq:Langevin} may be any (nonlinear) partial differential or integro-differential operator \cite{WuWang}, as well as a space-time driving $\bm{f}(\mathbf{x},t)$.

\begin{remark}[Noise setting] 
To avoid inconsistencies in the formulation of SPDEs, let us recall the precise setting of the noise. 
The covariance kernel $C^{\kappa\sigma}(\mathbf{x},\mathbf{x}';t)$ defines a nonnegative, self-adjoint operator on the Hilbert space $\mathcal{U}$, and specifies the spatial covariance of the $\mathcal{U}$-valued process, with spatially white noise recovered in the limiting case $C^{\k\s}(\mathbf{x},\mathbf{x}')\rightarrow \delta(\mathbf{x}-\mathbf{x}')$. 
A $Q$-Wiener process is a well-defined $\mathcal{U}$-valued Gaussian process whenever $\operatorname{Tr} \bm{C}<+\infty$; otherwise one obtains only a cylindrical Wiener process, which cannot be realized in $\mathcal{U}$ but still admits a consistent stochastic integral formulation \cite{DaPrato_Zabczyk_1992}. 
In the trace-class case, $W^\k(\mathbf{x},t)$ can be expanded as an infinite series of mutually independent Brownian motions $W_B (t)$ \cite{DaPrato_Zabczyk_1992}, 
\begin{equation}\label{eq:Qwiener}
W^\kappa(\mathbf{x},t)=\sum_{l=0}^\infty b^\kappa_l(\mathbf{x})\, W^l_B(t),
\end{equation}
so that arbitrary correlations are reproduced as 
\begin{equation}
C^{\kappa\sigma}(\mathbf{x},\mathbf{x}')=\sum_{l=0}^\infty b^\kappa_l(\mathbf{x})\,b^\sigma_l(\mathbf{x}').
\end{equation}
The operators $g[\bm{\phi}]:\mathcal{U}\to\Omega$ appearing in Eq.~\eqref{eq:Langevin} thus determine how the noise enters the system, and their specification is required to uniquely define the SPDE in the SC. In this particular case we assume $g[\bm{\phi}]$ involves a single-point coupling, i.e. for $\bm{u}\in \mathcal{U}$:
\begin{equation}
\left(g[\bm{\phi}] \cdot \bm{u}\right)(\mathbf{x})^a= \sum_{\kappa=1}^m g^a_\kappa [\bm{\phi}](\mathbf{x}) u (\mathbf{x})^\kappa, 
\end{equation}
although the more general multiple-point coupling (and discretized) version is given in Appendix \ref{app:discrete_theory}. Finally, we stress that the Gaussian driving process is independent of the system state  \cite{DaPrato_Zabczyk_1992,Oksendal2003}; consequently, the covariance $C^{\kappa\sigma}(\mathbf{x},\mathbf{x}';t)$ is fixed externally and cannot depend functionally on the evolving field $\bm{\phi}$.
\end{remark}

The associated covariance operator in infinite dimensions, which gives the covariance of the random forces, is
\begin{equation}\label{eq:cov_matrix}
    D^{ab}(\mathbf{x},\mathbf{x}',t)[\bm{\phi}]=g^{a}_{\k}(\mathbf{x},t)[\bm{\phi}]g^{b}_{\s}(\mathbf{x}',t)[\bm{\phi}]C^{\k\s}(\mathbf{x},\mathbf{x}';t).
\end{equation}
and is positive-definite iff 
\begin{equation}\label{eq:PD}
    \int d\mathbf{x}d\mathbf{x}'\, \varphi^a(\mathbf{x})D^{ab}(\mathbf{x},\mathbf{x}',t)[\bm{\phi}] \varphi^b(\mathbf{x}') \geq 0 ,
\end{equation}
with equality restricted to $\varphi \equiv 0$. The inverse, assuming it exists, is denoted by $M_{ab}(\mathbf{x},\mathbf{x}',t)[\bm{\phi}]\equiv [D^{ab}(\mathbf{x},\mathbf{x}',t)]^{-1}[\bm{\phi}]$, being defined as \cite{WuWang,HANGGI} 
\begin{equation}\label{eq:inverse_def}
    \int d\mathbf{x}''\,M_{ab}(\mathbf{x},\mathbf{x}'',t)[\bm{\phi}] D^{bc}(\mathbf{x}'',\mathbf{x}',t)[\bm{\phi}]=\d_a^{c}\d(\mathbf{x}-\mathbf{x}').
\end{equation}

The corresponding functional Fokker-Planck equation (FPE) is \cite{ojalvo,gardiner1985handbook}: 
\begin{widetext}
\begin{equation}\label{eq:FPE}
    \begin{aligned}
        &\p_t \mP_t [\bm{\phi}] = -\int_\mathcal{D} d\mathbf{x}\, \f{\d}{\d \phi^a (\mathbf{x})}\bigg\{ \left(F^a(\mathbf{x},t)[\bm{\phi}] -g^a_\k(\mathbf{x},t)[\bm{\phi}]\int_\mathcal{D} d\mathbf{x}' \f{\d }{\d \phi^b (\mathbf{x}')}g^{b}_{\s}(\mathbf{x}',t)[\bm{\phi}]C^{\k\s}(\mathbf{x},\mathbf{x}';t)\right)\mP_t [\bm{\phi}] \\
        &\phantom{\p_t \mP_t [\bm{\phi}] =} \qquad -\int_\mathcal{D} d\mathbf{x}' D^{ab}(\mathbf{x},\mathbf{x}',t)[\bm{\phi}]\f{\d \mP_t [\bm{\phi}]}{\d \phi^b (\mathbf{x}')}\bigg\}\\
        & \phantom{\p_t \mP_t [\bm{\phi}] } = -\int_\mathcal{D} d\mathbf{x}\, \f{\d }{\d \phi^a (\mathbf{x})}j^a (\mathbf{x},t)[\bm{\phi}],
    \end{aligned}
\end{equation}
\end{widetext}
where $\mP_t [\bm{\phi}]$ denotes the probability density functional (PDF) of a certain configuration $\bm{\phi}$ at time $t$, and $\bm{j} (\mathbf{x},t)[\bm{\phi}]$ the probability density current. While the Langevin equation \eqref{eq:Langevin} provides dynamical information at the individual-trajectory level, the FPE \eqref{eq:FPE} provides the ensemble level information. In addition to an initial condition, the FPE requires boundary conditions (BCs) in both the functional phase space $\Omega$ and the spatial domain $\mathcal{D}$; a summary of common choices is provided in Appendix~\ref{app:FPE_BC}.

An additional representation which is applied in ST is the path integral (PI) formulation. The PI gives the transition probability from the initial configuration $\bm{\phi}(t_0)\equiv\bm{\phi}_0$ to the final configuration $\bm{\phi}(t)\equiv \bm{\phi}_t$ by integrating over all intermediate states \cite{Cugliandolo_2019}:
\begin{subequations}
\begin{align}
     &\mP_t [\bm{\phi}]=\int _{\Omega}d\bm{\phi}_0\, K(\bm{\phi},t;\bm{\phi}_0 ,t_0) \mP_0[\bm{\phi}_0],\label{eq:propagator}\\
    &K(\bm{\phi}_t,t;\bm{\phi}_0 ,t_0) = \int _{\bm{\phi}_0}^{\bm{\phi}_t}\mathcal{D}\bm{\phi}\,P[\bm{\phi}|\bm{\phi}_0]. 
\end{align}
\end{subequations}
The path integral measure, defined for a discretized temporal grid $\{t_n = \epsilon n\}_{n=0}^N$, is given by $\mathcal{D}\bm{\phi}\equiv \prod_{n=1}^{N-1}d\bm{\phi}_n$.

The path-probability, which gives the probability weight associated with a particular trajectory starting at state $\bm{\phi}_0$, can be defined as
\begin{equation}
  P[\bm{\phi}|\bm{\phi}_0]=\mathcal{N}\{\bm{\phi}_n\}\,  \exp (-\mathcal{A}[\bm{\phi}]), 
\end{equation}
where $\mathcal{A}[\bm{\phi}]$ is the action functional and $\mathcal{N}\{\bm{\phi}_n\}$ the normalization factor. For a given interpretation (Stratonovich/Ito), each Langevin equation corresponds to a unique Fokker--Planck equation, so the PDF $\mathcal{P}_t[\bm{\phi}]$ is uniquely defined. In the continuous limit, this only uniquely fixes $K(\bm{\phi},t;\bm{\phi}_0,t_0)$, leaving freedom to choose different normalizations $\mathcal{N}\{\bm{\phi}_n\}$ or actions $\mathcal{A}[\bm{\phi}]$ as long as Eq.~\eqref{eq:propagator} holds \cite{Wissel1979}. Consequently, multiple path integral expressions for the same noise interpretation exist in the literature \cite{Onsager_Machlup,Lau_2007,Cates_2022,Cugliandolo_2017,Cugliandolo_2019,Graham1977,Graham2,Deininghaus1979}, though this ambiguity vanishes in ST once the normalization $\mathcal{N}\{\bm{\phi}_n\}$ is properly accounted for \cite{HVP_partI}.

A convenient expression for the PI in the context of ST is the Onsager-Machlup (OM) action \cite{Onsager_Machlup,Lau_2007,Cates_2022,Itami2017}. The propagator and the action\footnote{Note the action is written in the continuous time limit.} are given, respectively, by: 
\begin{widetext}
\begin{subequations}\label{eq:PI_multid}
\begin{align}
& K(\bm{\phi}_t,t;\bm{\phi}_0 ,t_0) = \int _{\bm{\phi}_0}^{\bm{\phi}_t}\mathcal{D}\bm{\phi} \, \prod_{n=0}^{N-1}\f{1}{\det \left(D^{ab}_{n+1/2}\right)^{1/2}(4\pi\epsilon/\D V^2)^{Kd/2}}\exp (-\mathcal{A}[\bm{\phi}]) \\
    &\mathcal{A}[\bm{\phi}]=\int_{t_0}^{t} \int_\mathcal{D} d\t d\mathbf{x}\bigg[ \f{1}{4}\left(\p_\t \phi^a (\mathbf{x},\t) - F'^a (\mathbf{x},\t)[\bm{\phi}]\right)\int_\mathcal{D} d\mathbf{x}' M_{ab}(\mathbf{x},\mathbf{x}',\t)[\bm{\phi}]\left(\p_\t \phi^b (\mathbf{x}',\t) -  F'^b (\mathbf{x}',\t)[\bm{\phi}]\right)  \nonumber\\ 
    &\phantom{\mathcal{A}[\bm{\phi}]=} +\f{1}{2}\f{\d F^a (\mathbf{x},\t)[\bm{\phi}]}{\d \phi^a (\mathbf{x})} 
     +\f{1}{4}\int_\mathcal{D} d\mathbf{x}'\left(\f{\d g^{a}_{\k}(\mathbf{x},\t)[\bm{\phi}]}{\d \phi^b (\mathbf{x}')}\f{\d g^{b}_{\sigma} (\mathbf{x}',\t)[\bm{\phi}]}{\d \phi^a (\mathbf{x})}-\f{\d g^{a}_{\k} (\mathbf{x},\t)[\bm{\phi}]}{\d \phi^a (\mathbf{x})}\f{\d g^{b}_{\sigma} (\mathbf{x}',\t)[\bm{\phi}]}{\d \phi^b (\mathbf{x}')}\right)C^{\k\s}(\mathbf{x},\mathbf{x}';\t)\bigg].\label{eq:OM}
\end{align}
\end{subequations}
\end{widetext}
Here, $F'^a (\mathbf{x},t)[\bm{\phi}]$ denotes the drift term in Eq.~\eqref{eq:FPE}
\begin{align}
    F'^a &(\mathbf{x},t)[\bm{\phi}] := F^a(\mathbf{x},t)[\bm{\phi}] \nonumber\\
    & -g^a_\k(\mathbf{x},t)[\bm{\phi}]\int_\mathcal{D} d\mathbf{x}' \f{\d }{\d \phi^b (\mathbf{x}')}g^{b}_{\s}(\mathbf{x}',t)[\bm{\phi}]C^{\k\s}(\mathbf{x},\mathbf{x}';t),
\end{align}
including both the deterministic and noise-induced drifts. The advantage of the OM form lies in the fact that its measure is symmetric under time reversal: for the midpoint convention ($a=1/2$), the time-reversed configurations $\hat{\bm{\phi}}_{n+a} = \bm{\phi}_{n+(1-a)}$ coincide with the forward trajectory, so only the action terms contribute to the entropy production \cite{Cates_2022}.

We emphasize once more that the continuum equations may be ill-defined. In fact, the discrete formulation provides the fundamental description, from which the continuum expressions used in the main text are derived. For completeness, the derivations and the discrete form of the equations presented here are given in Appendix~\ref{app:discrete_theory}.

\subsection{\label{sec:stoch_entropy_prod} Stochastic entropy production}
In ST, entropy production (EP) is defined as time reversal symmetry breaking (TRSB), and is quantitatively measured by the Kullback-Leibler (KL) divergence---or statistical distance---between the probability distributions of a forward process and its time-reversed counterpart \cite{Parrondo_2009, pelitti, VANDENBROECK20156}:
\begin{equation}\label{eq:KL}
    \langle \Delta s_{\rm tot} \rangle :=  D_{KL}[P[\bm{\phi}]||\hat{P}[\bm{\phi}]]= \int \mathcal{D}\phi\, P[\bm{\phi}] \ln \f{P[\bm{\phi}]}{\hat{P}[\bm{\phi}]},
\end{equation}
where---setting $t_0 = 0$ without loss of generality---$P[\bm{\phi}]$  denotes the probability of the forward trajectory, and $\hat{P}[\bm{\phi}]$ the probability of the backward or time-reversed trajectory, defined by reversing phase space variables in time. Fields that are odd are those that change sign under time reversal, i.e. $\hat{\bm{\phi}}_-(\mathbf{x},\t) = -\bm{\phi}(\mathbf{x},u)$, such as momentum or velocity field, while fields that are even (like density field) remain unchanged i.e. $\hat{\bm{\phi}}_+(\mathbf{x},\t) = +\bm{\phi}(\mathbf{x},u)$, with $u=t-\tau$. 

The KL divergence is an invariant nonnegative flat operator \cite{KL}, whose definition \eqref{eq:KL} already implies both the second law of ST $ \langle \Delta s_{\rm tot} \rangle \geq 0$, and the integral fluctuation theorem (IFT) \cite{Seifert_2005}:
\begin{equation}\label{eq:IFT}
    \langle e^{-\Delta s_{\rm tot}}\rangle = \int\mathcal{D}\bm{\phi}\, P[\bm{\phi}] \f{\hat{P}[\bm{\phi}]}{P[\bm{\phi}]}= \int  \mathcal{D}\bm{\phi}\, \hat{P}[\bm{\phi}] = 1
\end{equation}
where the last equality is drawn from the normalization condition and the fact that summing over all paths do not distinguish between forward and backward trajectories \cite{Seifert_2012}. 

The EP for an individual trajectory is obtained by undoing the path average:
\begin{subequations}
\begin{align}
    &\Delta s_{\rm tot} = \text{ln}\f{P[\bm{\phi}]}{\hat{P}[\bm{\phi}]} = \Delta s_m + \Delta s_{sys}, \label{eq:Delta_s_tot}\\
    &\Delta s_m  := \text{ln}\f{P[\bm{\phi}|\bm{\phi}_0]}{P[\hat{\bm{\phi}}|\bm{\phi}_t]},\quad  \Delta s_{sys}:= \text{ln}\frac{\mathcal{P}_{t_0}[\bm{\phi}_0]}{\mathcal{P}_{t}[\bm{\phi}_t]} . \label{eq:EP}
\end{align}
\end{subequations}
The first term in Eq. \eqref{eq:EP} represents the stochastic EP in the medium $(\Delta s_m )$, while the second term accounts for the stochastic entropy change of the system $(\Delta s_{sys} )$. The probability distributions $\mP_t$ are governed by the corresponding FPE. Upon taking the ensemble average of the system contribution, one recovers the Shannon entropy $\mS$ \cite{Seifert_2012}:
\begin{subequations}\label{eq:stoch_shannon}
    \begin{align}
      &  s_{sys}(t)[\bm{\phi}]= - \ln \mP_t[\bm{\phi}],\\
     & \mS(t) := \avg{s_{sys}(t)[\bm{\phi}]} = -\int d\bm{\phi}\, \mP_t[\bm{\phi}]\ln \mP_t[\bm{\phi}] .
    \end{align}
\end{subequations}
Taking the time derivative of the total EP yields the stochastic entropy production rate (EPR):
\begin{equation} \label{eq:dot_s_tot}
    \dot s_{\rm tot}(t) [\bm{\phi}]= \dot s_{sys} (t) [\bm{\phi}]+ \dot s_m (t)[\bm{\phi}].
\end{equation}

We illustrate the EPR by applying the framework to a model of underdamped dynamics with Hamiltonian
\begin{equation}\label{eq:Hamiltonian}
    H[\bm{\phi},\bm{\pi}]=\f{1}{2}\|\bm{\pi}\|^2 + \varepsilon[\bm{\phi},\l(t)] ,
\end{equation} 
where $\varepsilon[\bm{\phi},\lambda]$ is the potential energy functional, $\lambda(t)$ is a prescribed time-dependent protocol, and the conjugate momentum is defined as $\pi^a(\mathbf{x},t) \equiv \p_t {\phi}^a(\mathbf{x},t)$. The equations of motion are
\begin{subequations}\label{eq:ST_model}
    \begin{align}
    &\pi^a (\mathbf{x},t) = \p_t\phi^a (\mathbf{x},t),\\
    &\p_{t}\pi^a (\mathbf{x},t) =  F^a (\mathbf{x},t)[\bm{\phi},\bm{\pi}] +g^{a}_{\k}(\mathbf{x},t)[\bm{\phi},\bm{\pi}] \circ\zeta^\k (\mathbf{x},t)  ,
\end{align}
\end{subequations}
where we assume independent noise, i.e. $C^{\k\s}(\mathbf{x},\mathbf{x}';t)=\d^{\k\s}C(\mathbf{x},\mathbf{x}')$. Even and odd components of the force are defined as $F^a_\pm=(F^a\pm \hat F^a)/2$, with
\begin{equation}
    \hat{F}^a (\mathbf{x})[\bm{\phi},\bm{\pi}, \l(t)] = F^a (\mathbf{x})[\bm{\phi},-\bm{\pi}, \l(u)] .
\end{equation}
The even force includes both conservative and non-conservative contributions,   $F^a_+ =   - \d_{\phi^a (\mathbf{x})} \varepsilon[\bm{\phi},\l(t)] + \mathrm{f}^a_+ (\mathbf{x},t)[\bm{\phi},\bm{\pi}]$, where $\mathbf{f}_+ $ denotes the non-conservative component, as could be an external manipulation by an agent \cite{Kwon_2016} or feedback cooling \cite{Munakata_2012}.

Applying the tools of Sec. \ref{sec:representations} we define the effective odd drift and the associated FPE as
\begin{subequations}
    \begin{align}
        &  F'^a_- (\mathbf{x},t)[\bm{\phi},\bm{\pi}]= F^a_-(\mathbf{x},t)[\bm{\phi},\bm{\pi}] \nonumber\\
        & \phantom{F'^a_- (\mathbf{x},t) =} -g^{a}_{\k}(\mathbf{x},t)[\bm{\phi},\bm{\pi}]\int_{\mathcal{D}} d\mathbf{x}' \f{\d g^{b}_{\k}(\mathbf{x}',t)[\bm{\phi},\bm{\pi}]}{\d \pi^b (\mathbf{x}')}C(\mathbf{x},\mathbf{x}'), \\
        & \p_t \mP_t [\bm{\phi},\bm{\pi}] = -\int_{\mathcal{D}} d\mathbf{x}\, \bigg(\f{\d j_{\bm{\phi}}^a (\mathbf{x},t)[\bm{\phi},\bm{\pi}]}{\d \phi^a (\mathbf{x})}  + \f{\d j_{\bm{\pi}}^a (\mathbf{x},t)[\bm{\phi},\bm{\pi}]}{\d \pi^a (\mathbf{x})}\bigg).
    \end{align}
\end{subequations}
The probability density currents are
\begin{subequations}
    \begin{align}
    & j_{\bm{\phi}}^a (\mathbf{x},t)[\bm{\phi},\bm{\pi}] = \pi^a (\mathbf{x},t)\mP_t [\bm{\phi},\bm{\pi}],\\
    & j_{\bm{\pi}}^a = j_{r}^{a} + j_{d}^{a}  ,\\
    & j_{r}^{a}(\mathbf{x},t)[\bm{\phi},\bm{\pi}] =  F^a_+(\mathbf{x},t)[\bm{\phi},\bm{\pi}] \mP_t [\bm{\phi},\bm{\pi}],\\
    & j_{d}^{a}(\mathbf{x},t)[\bm{\phi},\bm{\pi}] = F'^a_- (\mathbf{x},t)[\bm{\phi},\bm{\pi}]\mP_t [\bm{\phi},\bm{\pi}] \nonumber\\
    &\phantom{j_{d}^{a}(\mathbf{x},t)[\bm{\phi},\bm{\pi}] =} -\int_{\mathcal{D}} d\mathbf{x}' D^{ab}(\mathbf{x},\mathbf{x}',t)[\bm{\phi},\bm{\pi}]\f{\d \mP_t [\bm{\phi},\bm{\pi}]}{\d \pi^b (\mathbf{x}')} \label{eq:odd_drift},
\end{align}
\end{subequations}
with $ \bm{j}_{r} , \bm{j}_{d}$ being the reversible and dissipative currents, respectively, and $ D^{ab}(\mathbf{x},\mathbf{x}',t)[\bm{\phi},\bm{\pi}]=g^{a}_{\k}(\mathbf{x},t)[\bm{\phi},\bm{\pi}]g^{b}_{\k}(\mathbf{x}',t)[\bm{\phi},\bm{\pi}]C(\mathbf{x},\mathbf{x}')$. From this point on we omit the explicit functional dependence for compact notation and denote $\Psi = (\bm{\phi}, \bm{\pi})$. The system EPR yields:
\begin{widetext}  
\begin{equation}  
\begin{aligned}
    \dot s_{sys}(t)  &= -\f{d}{dt} \ln \mP_t = -\f{\p_t \mP}{ \mP_t}-\f{1}{\mP_t}\int_{\mathcal{D}} d\mathbf{x}\, \f{\d\mP_t}{\d \Psi^a (\mathbf{x})}\p_t {\Psi}^a \\
    &= -\f{1}{\mP_t}\bigg\{\p_t \mP + \int_{\mathcal{D}} d\mathbf{x}\, \f{\d\mP_t}{\d \phi^a (\mathbf{x})}\pi^a(\mathbf{x},t)   + \int_{\mathcal{D}} d\mathbf{x}d\mathbf{x}'\, \p_t\pi^a (\mathbf{x},t) M_{ab}(\mathbf{x},\mathbf{x}',t)\left(F'^b_- (\mathbf{x}',t)\mP_t - j_d^{b} (\mathbf{x}',t)\right)\bigg\},
\end{aligned}
\end{equation}
\end{widetext}
where Eq. \eqref{eq:odd_drift} was inverted applying Eq. \eqref{eq:inverse_def}:
\begin{equation}
    \f{\d\mP_t}{\d \pi^a (\mathbf{x})} = \int_{\mathcal{D}} d\mathbf{x}'\, M_{ab}(\mathbf{x},\mathbf{x}',t)\left(F'^b_- (\mathbf{x}',t)\mP_t - j_d^{b} (\mathbf{x}',t)\right).
\end{equation}

The medium EP is derived by applying the PI \eqref{eq:PI_multid} in Eq. \eqref{eq:EP}, recalling that in the SC $\mathcal{N}\{\Psi_n\}=\mathcal{N}\{\hat {\Psi}_n\}$:
\begin{equation}
\begin{aligned}
     &\Delta s_m = \text{ln}\frac{P[\Psi|\Psi_0 ]}{P[\hat{\Psi}|\Psi_t ]}=\text{ln}\frac{\mathcal{N}\{\Psi_n\}\, e^{-\mathcal{A}[\Psi]}}{\mathcal{N}\{\hat {\Psi}_n\}\, e^{-\hat{\mathcal{A}}[\Psi]}}=\hat{\mathcal{A}}[\Psi]-\mathcal{A}[\Psi] \\
     &\phantom{\Delta s_m} = \int_{t_0 , {\mathcal{D}}}^t d\t d\mathbf{x}\bigg[ -\f{\d \mathrm{f}^a_+ (\mathbf{x},\t)}{\d \pi^a (\mathbf{x})}\\
    &+ \int_{\mathcal{D}} d\mathbf{x}'\left(\p_\t \pi^a (\mathbf{x},\t) - F^a_+ (\mathbf{x},\t)\right) M_{ab}(\mathbf{x},\mathbf{x}',\t) F'^b_- (\mathbf{x}',\t) \bigg],
\end{aligned}
\end{equation}
where the backward action is given by $\hat{\mathcal{A}}[\Psi] = \mathcal{A}[\bm{\phi}(\mathbf{x},u),-\bm{\pi}(\mathbf{x},u)]$, with $u=t-\t$. Since the integration limits coincide under this change of variables, we can relabel the dummy index $du = d\t$, allowing direct subtraction of the integrands \cite{Vaquero2025GSI}. Taking the time derivative gives the medium EPR:
\begin{equation}\label{eq:stoch_s_m}
\begin{aligned}
      &\dot s_m(t)[\Psi] =  \int_{\mathcal{D}} d\mathbf{x}\bigg[ -\f{\d \mathrm{f}^a_+ (\mathbf{x},t)}{\d \pi^a (\mathbf{x})}\\
      &+ \int_{\mathcal{D}} d\mathbf{x}'\left(\p_t \pi^a (\mathbf{x},t) - F^a_+ (\mathbf{x},t)\right) M_{ab}(\mathbf{x},\mathbf{x}',t) F'^b_- (\mathbf{x}',t) \bigg].  
\end{aligned}
\end{equation}
Finally, adding both contributions gives the total EPR:
\begin{equation}\label{eq:total_EPR}
\begin{aligned}
    \dot s_{\rm tot}&(t)[\Psi] = \dot s_{sys} + \dot s_m \\
    =& -\f{\p_t \mP_t }{\mP_t} -\f{1}{\mP_t}\int_{\mathcal{D}} d\mathbf{x}\bigg[ \f{\d\mP_t}{\d \phi^a (\mathbf{x})}\pi^a(\mathbf{x},t)\\
    &- \int_{\mathcal{D}} d\mathbf{x}'\left(\p_t \pi^a (\mathbf{x},t)  - F^a_+ (\mathbf{x},t)\right) M_{ab}(\mathbf{x},\mathbf{x}',t) j^{b}_d (\mathbf{x}',t) \bigg] \\
    & -\int_{\mathcal{D}} d\mathbf{x}\bigg[ \f{\d \mathrm{f}^a_+ (\mathbf{x},t)}{\d \pi^a (\mathbf{x})} +  \f{F^a_+ (\mathbf{x},t)}{\mP_t}\f{\d \mP_t}{\d \pi^a (\mathbf{x})}\bigg].
\end{aligned}
\end{equation}
It can be seen how the stochastic total EPR is not constrained to be nonnegative, since the second law applies on the ensemble level rather than the individual trajectory level \cite{Seifert_2012}.

Computing the average EPR requires a two-step process, involving a conditional average over all trajectories which are at time $t$ in a configuration $\Psi(\mathbf{x})$, followed by an ensemble average \cite{Seifert_2005,Seifert_2012}. The first step involves the integral over paths which gives $\avg{\p_t{\bm{\pi}}(\mathbf{x},t) | \Psi (\mathbf{x}),t}=\bm{j}_{\bm{\pi}}(\mathbf{x},t)[\Psi]/\mP_t [\Psi]$, required to project the time derivatives defined along a stochastic trajectory into the ensemble space where the PDF $\mP_t [\Psi]$ is defined. The second step is taken by averaging over the phase space $\Omega$.  As a result, the double average yields, for some general functional $h$: 
\begin{equation}\label{eq:two_step_avg}
    \Bavgg{h[\Psi] \dot{\bm{\pi}}}(\mathbf{x},t)=\int_{\o}d\Psi \, h(\mathbf{x},t)[\Psi] \bm{j}_{\bm{\pi}}(\mathbf{x},t)[\Psi], 
\end{equation}
where $d\Psi=d\bm{\phi}d\bm{\pi}$. See Ref. \cite[App. A]{HVP_partI} for a proof of this relation. While that derivation is carried out for finite-dimensional systems, the argument extends directly to the present case: it relies only on general properties of Markov processes, the time discretization underlying the SC, and functional Gaussian integrals, which can be understood as the continuum limit of their discrete counterparts \cite{WuWang,peskinQFT,bellac_quantum_1992,parisi_statistical_1988}. Even if the functional determinant diverges, such divergences cancel in the final result.

In the following derivations, we take into account the natural boundary conditions in $\o$ (see Appendix \ref{app:FPE_BC}). The average total EPR is then: 
\begin{equation}\label{eq:dot_S_tot}
    \dot S_{\rm tot} (t)= \dot \mS (t) + \dot S_m (t), 
\end{equation}
where we define $\dot S_{\rm tot}(t) :=\avgg{\dot s_{\rm tot}} $ and $ \dot S_m (t) := \avgg{\dot s_m}$ for
\begin{align}
    &\avgg{\dot s_{\rm tot}} = \int_{\o,\mathcal{D}}d\Psi d\mathbf{x} d\mathbf{x}'\f{j^{a}_d (\mathbf{x},t)}{\mP_t} M_{ab}(\mathbf{x},\mathbf{x}',t) j^{b}_d (\mathbf{x}',t) , \label{eq:avg_total_EPR} \\
     &\avgg{\dot s_m} = \int_{\o,\mathcal{D}}d\Psi d\mathbf{x} \bigg[ d\mathbf{x}'\,j_{d}^{a} (\mathbf{x},t) M_{ab}(\mathbf{x},\mathbf{x}',t)F'^b_- (\mathbf{x}',t) \bigg]    \nonumber \\
     &  \phantom{\avgg{\dot s_m} =} - \Bavg{\int_{\mathcal{D}} d\mathbf{x}\, \f{\d \mathrm{f}^a_+ (\mathbf{x},t)[\Psi]}{\d \pi^a (\mathbf{x})}}       .\label{eq:s_m_definition}
\end{align} 
Notably, only the third term in Eq. \eqref{eq:total_EPR} survives the averaging process. Equation~\eqref{eq:s_m_definition} makes explicit that the double average and the spatial integral over $\mathbf{x}$ commute. Provided the covariance matrix $D^{ab}$ is positive-definite, the average total EPR yields a quadratic form and hence verifies $\dot S_{\rm tot} \geq 0$, with equality being restricted to equilibrium where dissipative currents vanish $(\bm{j}_d = \bm{0})$ \cite{HVP_partI}.

Moreover, $\dot \mS = \avgg{\dot s_{sys}}$, which can be shown by directly differentiating Shannon entropy:
\begin{equation}
\begin{aligned}
     \dot \mS  &= - \int_\o d\Psi\, \ln \mP_t\, \f{\p \mP_t}{\p t}\\
&= -\int_{\o,\mathcal{D}}d\Psi d\mathbf{x}\left[\f{j_{\bm{\phi}}^a (\mathbf{x},t)}{ \mP_t}\f{\d \mP_t}{\d \phi^a (\mathbf{x})}  + \f{j_{\bm{\pi}}^a (\mathbf{x},t)}{ \mP_t}\f{\d \mP_t}{\d \pi^a (\mathbf{x})}\right].
\end{aligned}
\end{equation}
Applying functional integration by parts, the first term vanishes due to independence of phase space variables. The reversible current contribution gives
\begin{equation}
\begin{aligned}
\int_{\o,\mathcal{D}}d\Psi d\mathbf{x}\, \f{j_{r}^{a} (\mathbf{x},t)}{ \mP_t}\f{\d \mP_t}{\d \pi^a (\mathbf{x})} = -\Bavg{\int_{\mathcal{D}} d\mathbf{x}\, \f{\d \mathrm{f}^a_+ (\mathbf{x},t)}{\d \pi^a (\mathbf{x})}}.
\end{aligned}
\end{equation}
For the dissipative current contribution:
\begin{equation}
    \begin{aligned}
     &\int_{\o,\mathcal{D}}d\Psi d\mathbf{x}\, \f{j_{d}^{a} (\mathbf{x},t)}{ \mP_t }\f{\d \mP_t }{\d \pi^a (\mathbf{x})} \\
    &=\int_{\o,\mathcal{D}}d\Psi d\mathbf{x}d\mathbf{x}'\, \f{j_{d}^{a} (\mathbf{x},t)}{ \mP_t}\\
    &\phantom{==}\times M_{ab}(\mathbf{x},\mathbf{x}',t)\left( F'^b_- (\mathbf{x}',t)\mP_t - j_d^{b} (\mathbf{x}',t)\right).
    \end{aligned}
\end{equation}
Gathering the results, the average system EPR gives: 
\begin{equation}\label{eq:dS_multid}
\begin{aligned}
    \dot \mS =\avgg{\dot s_{\rm tot}} - \avgg{\dot s_m} = \avgg{\dot s_{sys}}.
\end{aligned}
\end{equation}
Thus, whereas the Shannon entropy is a state function, the medium entropy—--and hence the total entropy—--is intrinsically path-dependent, so no state function $S_m$ exists whose time derivative gives the average medium EP for all times \cite{HVP_partI}.

All results presented in this section are directly analogous to their finite-dimensional counterparts derived in Ref. \cite{HVP_partI}. The crucial difference is that, owing to the presence of spatial degrees of freedom, one can recover the corresponding entropy densities and correlation functions by lifting the spatial averages over $\mathbf{x}$ and $\mathbf{x}'$ \cite{WuWang}.

\subsection{Stochastic energetics}\label{sec:ST_energetics}
Stochastic energetics deals with the stochastic extension of the first law, i.e. the energetics along individual trajectories. We illustrate it with the system Eq. \eqref{eq:ST_model}. Following Sekimoto's approach \cite{Sekimoto2010}, we can derive the first law from the Hamiltonian of the system given by Eq. \eqref{eq:Hamiltonian}:
\begin{equation}\label{eq:1law}
\begin{aligned}
    & \f{dH[\Psi] }{dt} =  \int_{\mathcal{D}} d\mathbf{x}\,\pi^a \left(\dot\pi^a + \d_{\phi^a } \varepsilon[\bm{\phi},\l(t)]\right) + \p_\l \varepsilon[\bm{\phi},\l(t)] \dot\l \\
    & \phantom{ \f{dH[\Psi] }{dt}} =  \int_{\mathcal{D}} d\mathbf{x}\,\pi^a  \left(  F^a_-[\Psi] +g^{a}_{\k}[\Psi]\circ\zeta^\k\right) \\
    &\phantom{ \f{dH[\Psi] }{dt}=} +  \int_{\mathcal{D}} d\mathbf{x}\,\pi^a \mathrm{f}^a_+ [\Psi] +\p_\l \varepsilon[\bm{\phi},\l(t)] \dot\l.
\end{aligned}
\end{equation}
Then, heat and work can be identified as (in units $\kb = 1$) \cite{Seifert_2012,Fodor_2022, Sekimoto2010}:
\begin{subequations}
\begin{align}
& \dot q(t)[\Psi] = \int_{\mathcal{D}} d\mathbf{x}\,\pi^a  \left(  F^a_-[\Psi] +g^{a}_{\k}[\Psi]\circ\zeta^\k\right) ,\label{eq:stoch_q}\\
    &\dot w (t)[\Psi] = \int_{\mathcal{D}} d\mathbf{x}\,\pi^a \mathrm{f}^a_+ [\Psi] +\p_\l \varepsilon[\bm{\phi},\l(t)] \dot\l . \label{eq:stoch_w} 
\end{align}
\end{subequations}
This identification stems from the fact that heat involves irreversible forces, that is, forces that contribute to EP \eqref{eq:avg_total_EPR}, meanwhile work involves reversible forces that do not explicitly contribute to the total EP \cite{Horowitz_2014}. Again, this is analogous to the finite-dimensional case \cite{HVP_partI}, except that the global quantities are obtained as spatial integrals of local densities. 

For a local damping coefficient $\gamma(\mathbf{x},t)$ with dissipative force
$F^a_- (\mathbf{x},t)= -\gamma^{ab}(\mathbf{x},t)\pi_b(\mathbf{x},t)$, and Gaussian noise that is white in space and time, $C(\mathbf{x},\mathbf{x'})=\delta(\mathbf{x}-\mathbf{x}')$, the medium EP is obtained from Eq.~\eqref{eq:stoch_s_m} using the extended Einstein relation  $T\gamma^{ab}(\mathbf{x},t) = D^{ab}(\mathbf{x},t)$ \cite{Cates_2022},
\begin{equation}
\begin{aligned}
      &\dot s_m(t)[\Psi] =  \int_{\mathcal{D}} d\mathbf{x}\bigg[ -\f{\d \mathrm{f}^a_+ (\mathbf{x},t)[\Psi]}{\d \pi^a (\mathbf{x})}\\
      &\phantom{\dot s_m(t)[\Psi] } -\f{\pi^a (\mathbf{x},t)}{T}  \left(  F^a_-(\mathbf{x},t)[\Psi] +g^{a}_{\k}(\mathbf{x},t)[\Psi]\circ\zeta^\k(\mathbf{x},t)\right) \bigg]\\
      &\phantom{\dot s_m(t)[\Psi] } = -\f{\dot q(t)[\Psi]}{T} - \dot \mI(t)[\Psi] ,
\end{aligned}
\end{equation}
recovering the Clausius relation in agreement with the first law Eq. \eqref{eq:1law}, as well as an additional contribution which measures the lack of phase space volume preservation by reversible/non-dissipative interactions:
\begin{equation}\label{eq:entropy_pump}
    \dot{\mI}=\frac{1}{2}\left[\d_{\Psi}\cdot \bm{\mathfrak{F}}[\Psi]-\d_{\hat{\Psi}}\cdot \hat{\bm{\mathfrak{F}}}[\Psi]\right]= \d_{\psi^-_k}\mathfrak{F}^+_k + \d_{\psi^+_k}\mathfrak{F}^-_k .
\end{equation}
Here  $\bm{\mathfrak{F}}$ collectively denotes all the deterministic forces or fluxes acting on the system, and $\psi^\pm$ distinguishes even $(\psi^+)$ from odd $(\psi^-)$ variables, with $\dot{\mI} = \d_{\bm{\pi}}\cdot\mathbf{f}_+[\Psi]$. The physical interpretation depends on the specific force; in the present example it corresponds to entropy pumping, arising from the velocity-dependent control force exerted by an external agent \cite{Kim&Qian}. The information-theoretic interpretation of entropy pumping was established as a minimal information requirement for the agent to perform the momentum-dependent force $\mathbf{f}_+$ \cite{Horowitz_2014}, an interpretation subsequently adopted \cite{Kwon_2016,Munakata_2012,Mandal_2017,Baiesi_2015, Chun_2018}. 

Hence, the EP in the medium includes both dissipative and information terms \cite{Kim&Qian,Parrondo2015, Dabelow_2019, Cafaro, Sagawa_2012, Esposito_2011, Horowitz_esposito}: 
\begin{equation}\label{eq:Gen_clausius}
    \D s_m = -\int\f{dq}{T}-\D \mI ,
\end{equation}
and we refer to Eq. \eqref{eq:Gen_clausius} as the generalized Clausius relation \cite{Horowitz_2014,Kwon_2016}. We define the medium as \textit{all} unobserved DoFs, including both the system’s heat bath and external agents; thus, the heat bath is a subset of the medium. As specified below, relation~\eqref{eq:Gen_clausius} follows from TRSB and is not constrained to underdamped dynamics.

\subsection{\label{sec:ST_caveats} Thermodynamic consistency}

The definition of entropy in ST through its information-theoretic formulation inherently satisfies the second law and the fluctuation theorems, due to the properties of the KL-divergence. This holds for any Markovian dynamics, not necessarily corresponding to a physical process \cite{Seifert_2012}, and hence does not constitute a true thermodynamic law \cite{HVP_partI}. This point has been emphasized by identifying the total EP as an information measure, the informatic entropy production rate (IEPR), which provides a measure of irreversibility (TRSB) but cannot be linked to the system energetics \cite{Fodor_2022,Cates_2022}.

As raised in Ref.~\cite{HVP_partI}, this issue stems from the definition of medium entropy itself in ST  [Eq. \eqref{eq:EP}]. The requirement of LDB is expressed as the condition that, for all trajectories $\Psi(t)$ \cite{Maes2003},
\begin{equation}\label{eq:LDB}
   \ln\frac{P[\Psi|\Psi_0 ]}{P[\hat{\Psi}|\Psi_t ]}=-\f{\D q}{T} -\D \mI , 
\end{equation}
where the heat bath entropy change is defined as $\D s = -\D q/T$ (Clausius relation). The entropy $\D s$ has a clear energetic interpretation and coincides with the  bath’s Boltzmann entropy change \cite{xing2025}. Condition \eqref{eq:LDB} holds for any model satisfying LDB \cite{Maes_LDB}, for both continuous- and discrete-time dynamics (see e.g. \cite{Crooks1999Excursions}), from which the standard properties of thermodynamic consistency follow. Under LDB, the stochastic medium entropy defined in Eq. \eqref{eq:EP} satisfies the generalized Clausius relation [Eq. \eqref{eq:Gen_clausius}] and therefore acquires a physical interpretation. It is worth reiterating that the quantity defined in Eq. \eqref{eq:EP} is fundamentally an information-theoretic measure of TRSB, and only under the additional constraint of LDB can it be identified with the physical entropy change of the medium.

Thus, LDB provides the principle that reconciles the statistical and physical descriptions of entropy \cite{Broeck_1,Broeck_2,Polettini}. In contrast, phenomenological models lack a systematic top-down assessment of LDB \cite{Seifert_2012}, requiring heat and entropy to be inferred indirectly from the system dynamics. Moreover, if not all \textit{thermodynamically} relevant DoFs are taken into account, LDB is violated, the FDRs break down, and EP becomes ill-defined due to inconsistent coarse-graining \cite{PhysRevX.15.021050,PhysRevResearch.6.013190,Cates_2022}.

Hence, under the requirement of LDB [Eq. \eqref{eq:LDB}], we refer to a thermodynamically consistent model as that which verifies \cite{HVP_partI}: 
\begin{enumerate}[label=(\alph*)]
    \item The second law, i.e. non-decreasing average total EPR $(\dot S_{\rm tot} \geq 0)$ \cite{Eckart,GBYo2019,Esposito_2011,Broeck_2}.
    \item The generalized Clausius relation [Eq. \eqref{eq:Gen_clausius}] \cite{hiddenS,Kwon_2016, Cates_2022, ActiveFields, Horowitz_2014}.
    \item If the system is isolated, it must satisfy detailed balance or an equilibrium steady state of the associated FPE or master equation \cite{zwanzig,Kardar2007,Mazenko2006,Lau_2007,Broeck_2,Pep_generic}.
\end{enumerate}

\section{\label{sec:VP} Variational principle of ST of fields}
In this section, we develop the variational principle for ST of infinite-dimensional systems, building on the geometric structure presented in \cite{HVP_partI} together with \cite{GBpart2}. We begin by reviewing the structure of the thermodynamic phase space, the nonholonomic constraints and the emergence of LDB. Then, we illustrate the method with subsequent examples.

\subsection{Thermodynamic phase space}
In order to derive the SPDEs for configurational and thermal variables, the phase space $\Omega$ is extended to incorporate all (thermodynamically) relevant DoFs \cite{GBpart2,GBYo2019,HVP_partI}, determining the complete system state at all times. The thermal variable central to this formulation is the thermodynamic entropy field $s$, an independent variable that captures the (internal) hidden DoFs, i.e. the heat bath macrostate, as well as the path-dependence of the medium entropy, see Sec.~\ref{sec:stoch_entropy_prod}. The evolution of $s$, governed by heat exchange with the system, follows the Clausius relation $T\,ds=-dq$.

More generally, the thermal fields are denoted by $\varphi_\alpha$, which encompass both observable macroscopic quantities---such as the mass density $\rho$ of a chemical species---as well as the thermodynamic entropy $s$. The collection of configurational and thermal fields span the thermodynamic phase space, which is defined as $\Omega := \{[\bm{\phi},\bm{\pi}, \varphi_\alpha]\} = \{[\Psi]\}$, where now $\Psi$ denotes the complete state of the system in $\o$. Associated with each phase space variable is a conjugate variable, referred to as the (generalized) thermodynamic displacement $\Lambda^\alpha$, which span the dual space $\o^*$. These displacements play a central role in our formulation, as they enable the derivation of the general form of the dynamical equations through a variational principle \cite{GBYo2019}. 

Unlike in the standard framework of ST, the PDF is now defined over the full thermodynamic phase space $\Omega$, taking the form $\mP_t[\Psi]$. The associated probability measure becomes $\mP_t [\Psi]d\Psi$ and the 
PI measure $\mathcal{D} \Psi =\mathcal{D}\bm{\phi}\mathcal{D}\bm{\pi}\mathcal{D}\varphi_\a\d[\bm{\pi}-\dot{\bm{\phi}}]$. Then, the stochastic system entropy and the Shannon entropy are defined as follows:
\begin{subequations}\label{eq:shannon_VP}
\begin{align}
      &  s_{sys}(t)[\Psi]= - \ln \mP_t[\Psi],\\
     & \mS(t) := \avg{s_{sys}(t)[\Psi]} = -\int_\o d\Psi\, \mP_t[\Psi]\ln \mP_t[\Psi] .
    \end{align}
\end{subequations}
The FPE is denoted by 
\begin{equation}\label{eq:FPE_VM}
    \p_t\mP_t [\Psi] = -\int_\mathcal{D} d\mathbf{x}\, \f{\d }{\d \Psi^a (\mathbf{x})} j^a (\mathbf{x},t)[\Psi]
\end{equation}
where $\d_\Psi$ denotes the functional derivative with respect to the thermodynamic phase space variables. Note that the distribution $\mP_t$ encodes the statistics of the bath macrostate as well, due to its dependence on the thermodynamic entropy $s$  \cite{HVP_partI}. 

An additional crucial variable in this formulation is the medium entropy, denoted by $\Sigma$, which is analogous to the conventional ST medium entropy $s_m$. It comprises both the thermodynamic entropy $s$ and the entropy flow $s_f = \Sigma - s$ arising from interactions with the external environment. The introduction of $\Sigma$ enables an unambiguous distinction between these different contributions. As will be shown below, within the variational formulation the evolution of the medium entropy is systematically determined by a nonholonomic constraint imposed through the Lagrange-d’Alembert principle.

\subsection{\label{sec:variational_structure} Lagrange-d'Alembert principle}
In Ref. \cite{GBYo2019}, nonequilibrium thermodynamics was formulated via a generalized Lagrange–d’Alembert principle of nonholonomic mechanics \cite{Bloch2015}, where EP, expressed as the sum of irreversible contributions, enters as a nonlinear nonholonomic constraint. This framework was extended to the stochastic regime in Ref. \cite{HVP_partI}, and we now further generalize it to infinite-dimensional systems.

Recall that Hamilton's Principle defines the dynamics of a physical system by the variational principle:
\begin{subequations}  
\begin{align}
    &\delta \mathcal{A}[\bm{\phi}] = \delta \int_{t_1}^{t_2}dt\, L [\bm{\phi}, \p_t\bm{\phi}]=0,\\
     &\d \bm{\phi}(t_1) = \d \bm{\phi}(t_2)=0, \\
    & L [\bm{\phi}, \p_t \bm{\phi}] = \int_\mathcal{D} d\mathbf{x}\, \ell(\bm{\phi}, \p_t\bm{\phi},  \bm{\nabla} \bm{\phi}),
\end{align}
\end{subequations}
with $\ell$ denoting the Lagrangian density. The Lagrangian functional $L [\bm{\phi}, \p_t \bm{\phi}]:T\o \cong\o\times\o\rightarrow \mathbb{R} $ encodes the physical information in terms of the kinetic and potential energies. While $\ell$ is assumed local here, the framework, along with most results presented below, extends straightforwardly to nonlocal $\ell$.

The first step to formulate a Lagrange-d'Alembert principle for ST is extending the Lagrangian to the complete thermodynamic phase space by including the internal energy $\varepsilon[\bm{\phi},\varphi_\a]$ to account for thermal contributions, which may be formally derived from the Boltzmann entropy in the microcanonical ensemble~\cite{tuckerman2010}:
\begin{equation}\label{eq:Lagrangian}
\begin{aligned}
L[\Psi,\p_t \Psi]&=K[\bm{\phi}, \p_t\bm{\phi}]-\varepsilon[\bm{\phi},\varphi_\a]\\
&= \int_{\mathcal{D}}d\mathbf{x}\,\ell(\Psi, \partial_t\Psi, \bm{\nabla}\Psi),
\end{aligned}
\end{equation}
with $\ell$ denoting the Lagrangian density. For example, if the bath is modeled as an ideal or real gas, $\ve[\rho,V,s]$ would be given by the conventional tools of statistical mechanics as in e.g. Ref.~\cite{tuckerman2010} $(\ve\sim e^{s/c})$, which is nonlinear in $s$ and thus requires the additional dependence of the Lagrangian. In particular, the functional dependence of $L$ on the fields arises through their values and their first-order derivatives. This internal energy $\varepsilon$ encodes the properties of the heat bath, such as its temperature or chemical potential, respectively,
\begin{equation}\label{eq:T_mu}
    T(\mathbf{x},t)[\Psi]:= -\f{\d L[\Psi]}{\d s(\mathbf{x},t)},\quad \mu(\mathbf{x},t)[\Psi]:= -\f{\d L[\Psi]}{\d \rho(\mathbf{x},t)}.
\end{equation}
Importantly, this definition of temperature emerges directly from the internal energy and enters consistently in the FDR, unlike previous treatments where 
$T$ was introduced as a positive parameter satisfying $T\gamma^{ab}(\mathbf{x},t) = D^{ab}(\mathbf{x},t)$ and only interpreted \textit{a posteriori} as the bath temperature \cite{HVP_partI}.

\begin{remark}[Functional derivatives]
In this work, the functional derivatives $\delta L[\Psi]/\delta \Psi$ are defined from the condition
\[
\frac{d}{d\varepsilon}\bigg|_{\varepsilon=0} L[\Psi+\varepsilon\delta\Psi,\partial_t\Psi ]= \int_\mathcal{D} d\mathbf{x}\frac{\delta L}{\delta \Psi} \delta \Psi,\;\;\forall \delta\Psi|_{\partial\mathcal{D}}=0,
\]
where we restrict the variations to vanish on the boundary and here $\varepsilon$ denotes a parameter.
For the class of Lagrangians defined from a density $\ell$ as in \eqref{eq:T_mu}, one has
\begin{equation}
\begin{aligned}
    \left.\frac{d}{d\varepsilon} \right|_{\varepsilon=0}L[\Psi+\varepsilon\delta\Psi,\partial_t\Psi ]&= \int_{\mathcal{D}} d\mathbf{x}\,\bigg[ \frac{\partial \ell}{\partial\Psi } \delta \Psi +\frac{\partial \ell}{\partial\bm{\nabla}\Psi } \delta \bm{\nabla}\Psi\bigg] \\
    &=\int_{\mathcal{D}}d\mathbf{x}\,\bigg( \frac{\partial \ell}{\partial\Psi }  -\bm{\nabla}\cdot\frac{\partial \ell}{\partial\bm{\nabla}\Psi }\bigg) \delta \Psi  \\
    &\quad + \int_{\p \mathcal{D}}  d\bm{\s}\, \frac{\partial \ell}{\partial\bm{\nabla}\Psi }\cdot \bm{n}\d  \Psi,
\end{aligned}
\end{equation}
for arbitrary variations $\delta\Psi$. Restricting to $\delta\Psi|_{\partial\mathcal{D}}=0$, one gets the general expression
\begin{equation}\label{delta_derivative}
\frac{\delta L}{\delta\Psi}=\frac{\partial \ell}{\partial\Psi }  -\bm{\nabla}\cdot\frac{\partial \ell}{\partial\bm{\nabla}\Psi }
\end{equation}
for the functional derivative, appearing for instance in \eqref{eq:T_mu}.
\end{remark}

\begin{remark}[Natural boundary conditions]
Related to the above, when applying the variational principle, the first variation of the action reads
\begin{equation}
\begin{aligned}
    \frac{d}{d\varepsilon}&\bigg|_{\varepsilon=0}\int_{t_1}^{t_2} dt L[\Psi+\varepsilon\delta\Psi,\partial_t\Psi+\varepsilon\delta\partial_t\Psi ]\\
    &=\int_{t_1 , \mathcal{D}}^{t_2} dtd\mathbf{x}\,\bigg( \frac{\partial \ell}{\partial\Psi }  -\bm{\nabla}\cdot\frac{\partial \ell}{\partial\bm{\nabla}\Psi }-\partial_t\frac{\partial \ell}{\partial\partial_t\Psi }\bigg) \delta \Psi  \\
    &\quad + \int_{t_1 ,\p \mathcal{D}}^{t_2} dt d\bm{\s}\, \frac{\partial \ell}{\partial\bm{\nabla}\Psi }\cdot \bm{n}\d  \Psi ,
\end{aligned}
\end{equation}
where we assumed that $\delta\Psi$ vanishes at $t=t_1,t_2$.
If no further constraints are imposed on $\d \Psi|_{\p\mathcal{D}}$ (e.g. Dirichlet BCs), the boundary term must vanish independently, leading to the natural BCs
\begin{equation}\label{eq:induced_grad_BC}
    \frac{\partial \ell}{\partial\bm{\nabla}\Psi }\cdot \bm{n}\bigg|_{\p \mathcal{D}} =0.
\end{equation}
This may induce BCs on configuration and thermal variables, such as
\begin{equation}\label{eq:T_mu_BC}
    \dfrac{\partial \ell}{\partial\bm{\nabla}\bm{\phi}}\cdot \bm{n} \bigg|_{\p \mathcal{D}}=0,\quad\dfrac{\partial \ell}{\partial\bm{\nabla}\rho}\cdot \bm{n} \bigg|_{\p \mathcal{D}}=0,\quad \dfrac{\partial \ell}{\partial\bm{\nabla} s }\cdot \bm{n} \bigg|_{\p \mathcal{D}}=0 .
\end{equation}
This set of \textit{natural BCs}, among others, arise from the variational principle and are automatically compatible with isolated systems. By contrast, \textit{essential BCs}, such as periodic BCs on $\bm{\phi}$, are imposed directly as modeling choices rather than derived from the variational principle.
\end{remark}

\begin{remark}[Hamilton–Pontryagin principle]   
To guarantee that time derivatives are well-defined stochastic differentials under the chosen integration convention, we employ the Hamilton–Pontryagin principle~\cite{HamiltonPontryagin} to derive the implicit Euler–Lagrange equations for $\dot{\bm{\phi}}, \dot{\bm{\pi}}, \dot \varphi_\alpha$. For the Lagrangian \eqref{eq:Lagrangian}, the only second-order kinematic constraint is $\dot{\bm{\phi}}= \bm{u}$, enforced by promoting $\pi$ to a Lagrange multiplier via the term $\int dtd\mathbf{x}\, \bm{\pi}\cdot (\dot{\bm{\phi}} - \bm{u})$, in line with the variational formulation of implicit Lagrangian systems in~\cite{HamiltonPontryagin}. 
\end{remark}

To consistently recover the reversible dynamics, the variational principle must enforce conservation of the thermal variables in the absence of irreversible processes, i.e., $\dot \varphi_\alpha = 0$. This is achieved by coupling each $\varphi_\alpha$ to a conjugate displacement $\Lambda^\alpha$ via Lagrange multipliers $\int dtd\mathbf{x}\, \varphi_\alpha \dot\Lambda^\alpha$  \cite{GBYo2019}. For entropy, only the multiplier $\int dtd\mathbf{x}\, (s-\Sigma) \dot\G$ is required, where $\G$ is conjugate to $s$; the minus sign guarantees $\dot \Sigma = \dot s$ for a closed system \cite{HVP_partI}.

Thus, the complete set of dynamical equations are obtained by taking variations of the (Hamilton-Pontryagin modified) action given by the extended Lagrangian, e.g. Eq. \eqref{eq:Lagrangian},
\begin{align}\label{eq:VM_action}
    \delta \int_{t_1}^{t_2} dt\int_\mathcal{D}d\mathbf{x} \bigg(\ell(\Psi, \p_t \Psi,  \bm{\nabla} \Psi)  + \bm{\pi}\cdot (\dot{\bm{\phi}} - \bm{u})  \nonumber\\
    + \varphi_\a\dot \L^\a+(s-\Sigma)\dot \Gamma \bigg) = 0 , 
\end{align}
subject to the nonholonomic constraints (to be defined). External manipulations can be accounted for by including an external non-conservative force $\mathbf{f}_+$ via the principle of virtual work, or a potential $U[\bm{\phi},\l(t)]$ for a time-dependent driving \cite{HVP_partI}.

By introducing a thermodynamic phase space $\Omega$ equipped with a set of thermal variables, and its dual $\Omega^*$ comprising their corresponding conjugate displacements, the EP can incorporate different irreversible processes within the same geometric structure, based on the contraction of a generalized friction force $J_\alpha$ and its associated generalized displacement $d\Lambda^\alpha$. This ansatz mirrors the geometric form of mechanical friction \cite{GBYo2019,HVP_partI}, while enabling the variational derivation of the associated SPDEs.

This structure is then embedded into a variational formulation using a Lagrange-d'Alembert type principle for nonequilibrium thermodynamics, as developed in \cite{GBYo2019}, where the critical curve condition is subject to both \textit{kinematic} and \textit{variational} constraints. It is worth noting that we employ the general variational framework for open systems, see \cite{GBYo2018,gawlik24}. These are defined as, in the weak formulation:
\begin{widetext}
\begin{equation}\label{eq:nonhol_constraints}
\begin{aligned}
     & \int_{\mathcal{D}}d\mathbf{x}\, w\f{\d L}{\d s}\dot \Sigma = \int_{\mathcal{D}}d\mathbf{x}\, w  \bigg[ (J_\a+\xi_\a)\cdot \mathbf{d}\dot \Lambda^\a  +  \mathcal{I}_\be\dot\Gamma^\be\bigg]  -  \int_{\p\mathcal{D}}d\bm{\s}\, w (\dot \Lambda^\a-X^\a_{ext})(J_\a+\xi_\a) \cdot\bm{n}\quad &&\forall w \in \L^0(\mathcal{D}), \\
    &\int_{\mathcal{D}}d\mathbf{x}\, w \f{\d L}{\d s}\d \Sigma=\int_{\mathcal{D}}d\mathbf{x}\, w \bigg[ (J_\a+\xi_\a)\cdot \mathbf{d}\d \Lambda^\a + \mathcal{I}_\be\delta\Gamma^\be \bigg] -  \int_{\p\mathcal{D}}d\bm{\s}\, w \d \Lambda^\a (J_\a +\xi_\a) \cdot\bm{n} &&\forall w \in \L^0(\mathcal{D}),
\end{aligned}  
\end{equation}
\end{widetext}
where $\a$ and $\be$ denote irreversible processes and entropy flows, respectively, with thermodynamic fluxes $J_\a$ and thermodynamic affinities $X^\a$, together with a thermodynamic affinity $X^\a_{ext}$ associated with the exterior. Here $w$ is a test function.

Then, the (generalized) thermodynamic displacements $\Lambda^\a$ are such that $\dot \Lambda^\a = X^\a$ \cite{GBYo2019}. Here $d\bm{\s}$ denotes the surface element, and recall $\Sigma$ refers to the medium entropy.

The operator $\mathbf{d}\equiv\bm{\nabla}$ acts on $\L^\alpha$ so that its contraction with $J_\alpha$ yields a scalar density. If $J_\alpha$ were instead a scalar flux, the resulting constraints would be chosen of the same form as in finite-dimensional (discrete) systems, as given in~\cite{HVP_partI}.

Unlike the deterministic formulation, thermodynamic consistency in ST requires each dissipative flux $J_\a$ to be complemented by its stochastic counterpart $\xi_\a$, as dictated by rigorous projection-operator methods~\cite{zwanzig,Grabert1982,Kardar2007,Mazenko2006}. Coarse-graining hides microscopic DoFs whose unresolved contributions must be accounted for by noise to correctly capture EP \cite{HVP_partI}.

In Sec.~\ref{sec:ST} we showed that, in Langevin dynamics, an additional information flux contributes to the medium entropy. Interpreted as an external entropy flux $\mI_\be$ [Eq.~\eqref{eq:Gen_clausius}], it enters the constraint structure \eqref{eq:nonhol_constraints} through coupling to the conjugate entropy displacement $\G$, adding the terms $\mI_\be \dot\G^\be$ and $\mI_\be \d\G^\be$ in the kinematic and variational constraints. The flux $\mI_\be$ is given by Eq.~\eqref{eq:entropy_pump} extended to the thermodynamic phase space. Since entropy pumping arises from non-dissipative mechanisms, it contributes only through deterministic terms ($\xi_\be\equiv 0$) \cite{HVP_partI}.

These two types of constraints are systematically related: variational constraints follow from the kinematic constraints by substituting $\dot \Lambda^\a \rightsquigarrow\d \Lambda^\a$ and $(\dot \Lambda^\a-X^\a_{ext}) \rightsquigarrow \d \Lambda^\a$, where the exterior contribution vanishes since $\d t = 0$ by definition---variations in Hamilton's principle are taken at fixed time \cite{GBYo2018,Bloch2015}. 

To summarize the role and interpretation of the variational structure: a thermodynamic phase space is defined on which the system Lagrangian is valued, enabling a variational treatment of thermal variables and a consistent definition of thermodynamic affinities. The critical curve condition [Eq.~\eqref{eq:VM_action}] is extended to recover conservation laws for macroscopic fields. Irreversibility is incorporated variationally through nonholonomic constraints, exhibiting a clear geometric structure given by the contraction between fluxes $J_\a\in \o$ and conjugate displacements $\L^\a\in \o^*$. This structure restricts admissible trajectories and variations to those admitting a well-defined energetic interpretation of the associated EP.

From a physical perspective, the constraints [Eq.~\eqref{eq:nonhol_constraints}] define $\Sigma$ through the generalized Clausius relation of ST [Eq.~\eqref{eq:Gen_clausius}], as discussed in Sec.~\ref{sec:ST_energetics}. This construction assumes timescale separation, large bath size and microreversibility---conditions underlying the stochastic description, the Clausius relation, and the energetic interpretation of EP; see Ref.~\cite{xing2025} for a detailed statistical-mechanical derivation of the foundations of ST. Importantly, the constraints determine $\Sigma$, not $s$, thereby enabling an unambiguous distinction between thermodynamic EP and entropy flow. The dynamics of $s$ are nevertheless fully specified: the critical curve condition [Eq.~\eqref{eq:VM_action}] yields the equations of motion for all variables in a self-contained manner \cite{HVP_partI}.

So far, the relation between the stochastic $\{\xi_\a\}$ and dissipative $\{J_\a\}$ fluxes has not been specified. These are determined by imposing the second law as an axiom on the dynamical equations derived from the variational formulation, thereby providing closure. The second law is defined as
\begin{equation}\label{eq:2nd_law_VP}
    \dot{\rm S}_{\rm tot}(t) :=  \frac{d}{dt} \mS(t) + \int d\mathbf{x}\,\avgg{ \dot \Sigma}(\mathbf{x},t)\geq 0.
\end{equation}
The definition \eqref{eq:2nd_law_VP} mirrors that of ST [Eq.~\eqref{eq:dot_S_tot}], except that the inequality is \textit{not} inherently satisfied by definition of $\dot{\rm S}_{\rm tot}$, contrary to the standard definition in ST as given in Eq. \eqref{eq:avg_total_EPR}. Moreover, we denote the stochastic counterpart of the total EPR by 
\begin{equation}
    \dot{s}_{\rm tot}(t) = - \frac{d}{dt}\ln \mP_t[\Psi] + \int d\mathbf{x}\, \dot \Sigma(\mathbf{x},t) ,
\end{equation}
satisfying $\dot{\rm S}_{\rm tot}(t) = \avgg{\dot{s}_{\rm tot}(t)}$. As a remark, although the second law is generally imposed locally in macroscopic thermodynamics~\cite{GBpart2,groot}, in the present framework it is imposed globally; the rationale for this choice will be discussed subsequently.

In Ref.~\cite[Sec.~IIIC]{HVP_partI} it was shown how LDB emerges from this axiom, which is here summarized. First, the information-theoretic entropies in the extended phase space are defined as
\begin{subequations}\label{eq:inf_entropy_VP}
\begin{align}
    &  \langle \Delta\mathfrak{s}_{\rm tot} \rangle := D_{KL}[P[\Psi]||P[\hat{\Psi}]], \\
    &      \Delta\mathfrak{s}_m =\ln\frac{P[\Psi|\Psi_0 ]}{P[\hat{\Psi}|\Psi_t ]},
\end{align}
\end{subequations}
for a trajectory  $\Psi:[0,t]\subset\mathbb{R}\rightarrow \o$. Then, by using $  \avgg{\dot{\mathfrak{s}}_{\rm tot}}=\dot \mS  + \avgg{\dot{\mathfrak{s}}_m}$ in~\eqref{eq:2nd_law_VP}, the total EPR functional yields
\begin{equation}\label{eq:s_tot_functional}
\begin{aligned}
       \dot {\rm S}_{\rm tot}(t)  &=\dot{\mathcal{S}}(t)+ \int_\mathcal{D} d\mathbf{x}\avgg{\dot{\Sigma}}(\mathbf{x},t)\\
       &=\avgg{\dot{\mathfrak{s}}_{\rm tot}}- \avgg{\dot{\mathfrak{s}}_m}+ \int_\mathcal{D} d\mathbf{x}\avgg{\dot{\Sigma}}(\mathbf{x},t)\\
        &=\avgg{\dot{\mathfrak{s}}_{\rm tot}} -\int_{\o,\mathcal{D}} d\Psi d\mathbf{x} \:\bm{\mathfrak{J}}\cdot\bm{\Xi},
\end{aligned}
\end{equation}
where $\bm{\mathfrak{J}}$ denotes the dissipative probability current densities and $\bm{\Xi}$ a state-dependent functional. This simplification can be verified on a case-by-case basis. Then, in the formal argument of \cite[App.~B]{HVP_partI} it was shown how the second law axiom requires $\bm{\Xi}\equiv 0$, or equivalently, 
\begin{equation}\label{eq:LDB_VP}
   \ln\frac{P[\Psi|\Psi_0 ]}{P[\hat{\Psi}|\Psi_t ]}=\Delta \Sigma , 
\end{equation}
which is the requirement of LDB in the extended phase space. Here we denoted $\D \Sigma=\int dt d\mathbf{x} \dot\Sigma(\mathbf{x},t)$. This result implies  $\langle \Delta s_{\rm tot} \rangle=\langle \Delta\mathfrak{s}_{\rm tot} \rangle$, reconciling the information-theoretic and physical descriptions of entropy. As a consequence, $s_{\rm tot}$ inherits the relations satisfied by $\mathfrak{s}_{\rm tot}$, which is further discussed in Sec. \ref{sec:under_open}. From condition \eqref{eq:LDB_VP} arises the FDRs, as it is demonstrated in subsequent examples.

\begin{remark}[Global vs local second law]
    The second law of ST, $\langle \Delta\mathfrak{s}_{\rm tot} \rangle\geq 0$, introduced in Sec.~\ref{sec:stoch_entropy_prod}, admits an information-theoretic formulation that does not distinguish between systems of different dimensionality. Consequently, the argument of Ref.~\cite{HVP_partI} extends directly to the field setting, where the second law naturally takes the form of a global inequality rather than a local one.  Moreover, this formulation enables a consistent treatment of systems with nonlocal spatial interactions, see Ref.~\cite{Giorgi2026} and others therein, which will be introduced through the variational approach. Nevertheless, the FDRs enforcing LDB~\eqref{eq:LDB_VP} will be shown to retain an effective local structure [Eqs.~\eqref{eq:FDR_closed}, \eqref{eq:FDR_inter}, and \eqref{eq:FDR_open}].
\end{remark}

The specific nature of noise correlations has not yet been specified. Throughout this work, $\{\xi_\a\}$ are assumed to be zero-mean Gaussian white noises to enable analytical tractability, and our discussion of the results is restricted to this setting. Nevertheless, the Clausius definition of entropy should remain valid for more general stochastic processes \cite{Maes_LDB,PhysRevLett.108.210601}, implying that the structure of the constraints [Eq.~\eqref{eq:nonhol_constraints}] is preserved, comprising a dissipative drift and its associated fluctuations, completely analogous to the finite-dimensional case \cite{HVP_partI}.

Since in Sec.~\ref{sec:representations} we emphasized the subtleties associated with the continuum limit, one may question the validity of the variational principle in this setting. Importantly, the principle admits a natural spatial discretization, in the sense that the underlying lattice formulation is fully compatible with the continuum description. Far from being a drawback, this discretized formulation provides a systematic route to structure-preserving numerical schemes \cite{Gawlik2020,Gawlik_2021,gawlik24,GAWLIK2025114336}, ensuring that the essential geometric and thermodynamic properties of the theory are retained at the computational level.

Subsequent sections analyze three different examples: the underdamped dynamics of closed systems (\ref{sec:under_closed}), multicomponent systems (\ref{sec:inter}), and the underdamped dynamics of open systems (\ref{sec:under_open}). 

\subsection{\label{sec:under_closed} Underdamped dynamics of closed system}
We first illustrate the variational approach with a generalization of the model of Sec.~\ref{sec:ST}. We work in the underdamped regime to retain the full dynamical structure since, in a general setting, the overdamped approximation does not always provide a reliable account of EP and may fail to capture key contributions \cite{entropy_anomaly}. An example of such underdamped dynamics is a stochastic wave equation with thermal noise, like the stochastic Klein-Gordon equation \cite{oppenheim2025}. As a remark, we term closed or isolated systems as those conserving the total energy derived from the Lagrangian, i.e., the observed system plus its own internal bath or hidden DoFs.

The Lagrangian of the vector field $\bm{\phi}$ immersed in a heat bath is given by
\begin{equation}\label{eq:closed_lagrangian}
\begin{aligned}
    L[\bm{\phi},\partial_t\bm{\phi},s]&= \int_\mathcal{D} d\mathbf{x} \,\ell( \bm{\phi}, \partial_t\bm{\phi}, \bm{\nabla}\bm{\phi}, s)\\
&= \int_\mathcal{D}d\mathbf{x}\frac{1}{2}|\partial_t\bm{\phi}|^2 - \int_\mathcal{D} d\mathbf{x} \,e(\bm{\phi},\bm{\nabla}\bm{\phi}, s)\\
&=: \f{1}{2}\|\p_t\bm{\phi}\|^2-\varepsilon[\bm{\phi},s],
\end{aligned}
\end{equation}
where $\mathcal{D}\subset \mathbb{R}^3$ denotes a bounded spatial domain. For instance, we could consider $e(\bm{\phi}, \bm{\nabla}\bm{\phi}, s)= \kappa\frac{1}{2}|\bm{\nabla}\bm{\phi}|^2 + \epsilon(\bm{\phi}, s)$. The variational formulation then involves taking variations of the (Hamilton-Pontryagin modified) action, subject to Lagrange-d’Alembert nonholonomic constraints for the entropy $\Sigma$:
\begin{equation}\label{eq:VP_closed}
\begin{aligned}
    &\delta \int_{t_1,\mathcal{D}}^{t_2} dtd\mathbf{x} \bigg(\ell( \bm{\phi}, \bm{u}, \bm{\nabla}\bm{\phi}, s)  + \bm{\pi}\cdot (\dot{\bm{\phi}} - \bm{u}) +(s-\Sigma)\dot \Gamma \bigg) = 0 , \\
     & \f{\d L}{\d s}\dot \Sigma = ({\rm f}^a_- [\Psi] + g^{a}_{\k}[\Psi] \circ\zeta^\k )\cdot \dot \phi_a , \\
    & \f{\d L}{\d s}\d \Sigma=({\rm f}^a_- [\Psi] + g^{a}_{\k}[\Psi] \circ\zeta^\k )\cdot \d \phi_a   ,
\end{aligned}  
\end{equation}
where $\Psi = (\bm{\phi},\bm{\pi}, s)$. Here $\zeta^\k$ is a Gaussian independent $Q$-Wiener process [Eq. \eqref{eq:gauss_noise}] with $C^{\k\s} (\mathbf{x},\mathbf{x}';t)= \d^{\k\s}C(\mathbf{x},\mathbf{x}')$. The dissipative force is ${\rm f}^a_- (\mathbf{x},t)[\Psi]=-\int  d\mathbf{x}'\gamma^{ab}(\mathbf{x},\mathbf{x}',t)[\Psi]\pi_b (\mathbf{x}',t)$ and $\g^{ab}$ is a nonlocal dissipation coefficient functional. 

The variational principle yields:
\begin{subequations}\label{eq:EoM_closed}
    \begin{align}
    &\pi^a   = \p_t\phi^a ,\\
    & T[\bm{\phi},s] := \f{\d \varepsilon [\bm{\phi},s]}{\d s }=\p_t \G ,\\
    & \p_t \Sigma = \p_t s, \\
    &\p_{t}\pi^a  = -\f{\d \varepsilon}{\d \phi^a}+ {\rm f}^a_- [\Psi] + g^{a}_{\k}[\Psi] \circ\zeta^\k   ,\\
    & \p_t s = -\f{\pi_a }{T}({\rm f}^a_-[\Psi] + g^{a}_{\k}[\Psi] \circ\zeta^\k) ,
\end{align}
\end{subequations}
together with the natural BC given by
\begin{equation}\label{eq:natural_BC_phi}
\frac{\partial e}{\partial\bm{\nabla}\bm{\phi}}\cdot \boldsymbol{n}\bigg|_{\p \mathcal{D}}=0
\end{equation}
associated to the variations $\delta\bm{\phi}$ on the boundary, see \eqref{eq:induced_grad_BC}. The derivation can be found in Appendix \ref{app:closed}.
We note that
\begin{equation}
    \f{\d \varepsilon}{\d \phi^a}= \f{\partial e}{\partial \phi^a } - \bm{\nabla} \cdot \f{\partial e}{\partial \bm{\nabla} \phi^a }
\end{equation}
in the momentum equation, see \eqref{delta_derivative}. Following the same approach as in Sec. \ref{sec:ST_energetics} we find the first law:
\begin{subequations}\label{eq:1law_closed}
\begin{align}
    & e_{\rm tot}=\frac{1}{2}|\bm{\pi}|^2 + e(\bm{\phi}, \bm{\nabla}\bm{\phi},s) \Rightarrow\\
    &  \p_t e_{\rm tot} (\mathbf{x},t) =  \bm{\nabla}\cdot \left(\frac{\partial e}{\partial \bm{\nabla} \bm{\phi}}\cdot\bm{\pi}\right), \\
    & \dot E (t)=\f{d}{dt}\int_{\mathcal{D}} d\mathbf{x} \, e_{\rm tot}=\int_{\partial\mathcal{D}} d\boldsymbol{\sigma}\,\left(\frac{\partial e}{\partial \bm{\nabla} \bm{\phi}}\cdot\bm{n}\right)\cdot\bm{\pi}=0,
\end{align}
\end{subequations}
where the boundary term vanishes due to the natural BC \eqref{eq:natural_BC_phi}. Hence the system, consisting of both configurational $(\bm{\phi})$ and thermal $(s)$ DoFs is closed.

To derive the FDRs we impose the second law. The probability density currents of the FPE [Eq. \eqref{eq:FPE_VM}] for the system \eqref{eq:EoM_closed} are:
\begin{subequations}\label{eq:closed_currents}
    \begin{align}
    & j_{\bm{\phi}}^a (\mathbf{x},t)[\Psi] = \pi^a (\mathbf{x},t)\mP_t [\Psi],\\
    & j_{\bm{\pi}}^a = j_{r}^{a} + j_{d}^{a}  ,\\
    & j_{r}^{a}(\mathbf{x},t)[\Psi] =  -\d_{\phi^a} \varepsilon(\mathbf{x},t)[\bm{\phi},s] \mP_t [\Psi],\\
    & j_{d}^{a}(\mathbf{x},t)[\Psi] = F'^a_- (\mathbf{x},t)[\Psi]\mP_t [\Psi] \nonumber\\
    & \qquad -\int_{\mathcal{D}} d\mathbf{x}' D^{ab}(\mathbf{x},\mathbf{x}',t)[\Psi] \left(\f{\d \mP_t [\Psi]}{\d \pi^b (\mathbf{x}')} - \f{\pi_b}{T}\f{\d \mP_t [\Psi]}{\d s (\mathbf{x}')}\right),\label{eq:j_d_closed}\\
    & j_{s} (\mathbf{x},t)[\Psi] =  -\f{\pi_a }{T} j_{d}^{a}(\mathbf{x},t)[\Psi],
\end{align}
\end{subequations}
with $   D^{ab}(\mathbf{x},\mathbf{x}',t)[\Psi]=g^{a}_{\k}(\mathbf{x},t)[\Psi]g^{b}_{\k}(\mathbf{x}',t)[\Psi]C(\mathbf{x},\mathbf{x'})$, and the effective odd drift
\begin{equation}\label{eq:odd_force}
\begin{aligned}
     &F'^a_-(\mathbf{x},t)  =  {\rm f}^a_- (\mathbf{x},t) -g^{a}_{\k}(\mathbf{x},t)[\Psi]\\
    &\times \int_{\mathcal{D}} d\mathbf{x}'\bigg[ \f{\d g^{b}_{\k}(\mathbf{x}',t)[\Psi]}{\d \pi^b (\mathbf{x}')} - \pi^b\f{\d [g^{b}_{\k}/T](\mathbf{x}',t)[\Psi]}{\d s (\mathbf{x}')} \bigg] C(\mathbf{x},\mathbf{x'}).
\end{aligned}
\end{equation}

The medium EPR then yields, applying the double average \eqref{eq:two_step_avg}:
\begin{equation}
    \avgg{\dot \Sigma}(\mathbf{x},t)=\avgg{\dot s}(\mathbf{x},t) = -\int_{\o}d\Psi \,  \f{\pi_a }{T} j_{d}^{a}(\mathbf{x},t)[\Psi],
\end{equation}
which can be shown to verify
\begin{equation}
\begin{aligned}    
    \f{d}{dt}\avg{s}(t) &= \int_\o d\Psi s  \p_t\mP_t [\Psi] = \int_{\o,\mathcal{D}}d\Psi d\mathbf{x}\, \bm{j}(\mathbf{x},t)[\Psi] \cdot\f{\d s}{\d\Psi (\mathbf{x})} \\
    & = \int_{\o,\mathcal{D}}d\Psi d\mathbf{x}\, j_s (\mathbf{x},t) = \int_{\mathcal{D}} d\mathbf{x}\avgg{\dot s}(\mathbf{x},t).
 \end{aligned}
\end{equation}
Remarkably, this implies that the total EPR is a total time derivative \cite{HVP_partI}
\begin{equation}\label{eq:real_time_rate}
  \dot {\rm S}_{\rm tot}(t)= \frac{d}{dt}(\mathcal{S}+  \langle s\rangle),
\end{equation}
in contrast with the standard ST setting, i.e., the results of Sec. \ref{sec:ST}. This implies that, for isolated systems, entropy behaves as a function of state, as it is expected near equilibrium.
The system EPR gives (see Appendix \ref{app:closed}):
\begin{align}
     \dot \mS(t) =& \int_{\o,\mathcal{D}}d\Psi d\mathbf{x}d\mathbf{x}'\bigg[\f{j_{d}^{a} (\mathbf{x},t)}{\mP_t} M_{ab}(\mathbf{x},\mathbf{x}',t)j_{d}^{b} (\mathbf{x}',t) \nonumber\\
     &- j_{d}^{a} (\mathbf{x},t) M_{ab}(\mathbf{x},\mathbf{x}',t)F'^b_- (\mathbf{x}',t)\bigg],
\end{align}
where $M_{ab}$ is the inverse of $D^{ab}$ following Eq. \eqref{eq:inverse_def}. Adding both contributions, the total EPR yields
\begin{equation}
    \begin{aligned}
      & \dot{\rm S}_{\rm tot}(t)  =   \int_{\o,\mathcal{D}}\frac{d\Psi}{\mP_t} d\mathbf{x}d\mathbf{x}'j_{d}^{a} (\mathbf{x},t) M_{ab}(\mathbf{x},\mathbf{x}',t)j_{d}^{b} (\mathbf{x}',t) \\
     & \phantom{\dot{\rm S}_{\rm tot}(t) =} -\int_{\o,\mathcal{D}}d\Psi d\mathbf{x}\,  j_{d}^{a}(\mathbf{x},t)\Xi_a (\mathbf{x},t),\\
     &\Xi_a (\mathbf{x},t)=  \f{\pi_a (\mathbf{x},t)}{T(\mathbf{x},t)}+\int_{\mathcal{D}}d\mathbf{x'} M_{ab}(\mathbf{x},\mathbf{x}',t)F'^b_- (\mathbf{x}',t).
    \end{aligned}
\end{equation}
Enforcing the second law \eqref{eq:2nd_law_VP} requires $\Xi_a\equiv 0$ as shown in Sec. \ref{sec:variational_structure}, giving the FDR and the associated total
EPR:
\begin{align}
   &  F'^a_- (\mathbf{x},t)[\Psi]=-\int_{\mathcal{D}}  d\mathbf{x}' D^{ab}(\mathbf{x},\mathbf{x}',t)[\Psi]\f{\pi_b}{T}(\mathbf{x}',t) \implies \\
   &  \dot{\rm S}_{\rm tot}(t) = \int_{\o,\mathcal{D}}\frac{d\Psi}{\mP_t} d\mathbf{x}d\mathbf{x}' j_{d}^{a} (\mathbf{x},t) M_{ab}(\mathbf{x},\mathbf{x}',t)j_{d}^{b} (\mathbf{x}',t) \geq 0.
\end{align}
Notably, the inequality $\dot S_{\rm tot} \geq 0$, which follows from the positive-definiteness of $D^{ab}$~\eqref{eq:PD}, does not in general hold locally (i.e., upon lifting the integral over $d\mathbf{x}$) unless the random forces are spatially uncorrelated, $C(\mathbf{x},\mathbf{x}') = \d(\mathbf{x}-\mathbf{x}')$. This illustrates how nonlocal effects can lead to apparent local violations of the second law. Additionally, the FDR can be regarded as effectively local, while still incorporating the net effect of nonlocal correlations.

For the particular dissipative force of this scenario the FDR takes the form
\begin{align}\label{eq:FDR_closed}
   & \int_{\mathcal{D}}  d\mathbf{x}'\gamma^{ab}(\mathbf{x},\mathbf{x}',t)\pi_b (\mathbf{x}',t) = \int_{\mathcal{D}}  d\mathbf{x}' D^{ab}(\mathbf{x},\mathbf{x}',t)\f{\pi_b}{T}(\mathbf{x}',t) \nonumber\\
   & - g^{a}_{\k}(\mathbf{x},t)\int_{\mathcal{D}} d\mathbf{x}'\bigg[ \f{\d g^{b}_{\k}(\mathbf{x}',t)}{\d \pi^b (\mathbf{x}')} - \pi^b\f{\d [g^{b}_{\k}/T](\mathbf{x}',t)}{\d s (\mathbf{x}')} \bigg]C(\mathbf{x},\mathbf{x}').
\end{align}
This is precisely the infinite-dimensional generalization of the FDR in \cite[Eq. (56)]{HVP_partI}, which was shown to be consistent with previous results in the literature \cite{Dubkov2009}. Note that for additive noise and $T$ independent of $s$ it simplifies to
\begin{equation}
     \int_{\mathcal{D}}  d\mathbf{x}'\gamma^{ab}(\mathbf{x},\mathbf{x}',t)\pi_b (\mathbf{x}',t) = \int_{\mathcal{D}}  d\mathbf{x}' D^{ab}(\mathbf{x},\mathbf{x}',t)\f{\pi_b}{T}(\mathbf{x}',t),
\end{equation}
which is the Einstein relation (extended to nonlocal correlations). The FDR \eqref{eq:FDR_closed} also has a strong form, which derives from lifting the integral over $\mathbf{x'}$, giving $T(\mathbf{x}',t)\gamma^{ab}(\mathbf{x},\mathbf{x}',t) = D^{ab}(\mathbf{x},\mathbf{x}',t)$.

Thermodynamic consistency requires the FPE to satisfy an equilibrium state when the system is closed. This is verified applying the FDR \eqref{eq:FDR_closed}, together with a general equilibrium distribution ansatz, as it was shown in \cite{HVP_partI}. Its generalization to spatially extended systems corresponds to the equilibrium functional 
\begin{equation}\label{eq:equilibrium_ansatz}
    \mP_{eq}[\Psi] = f(E[\Psi])e^{\int d\mathbf{x} s(\mathbf{x})},
\end{equation}
where $f(E)$ is a function of the total energy functional. Through an analogous computation it can be shown that \cite{HVP_partI}
\begin{equation}
     \f{\d }{\d \Psi (\mathbf{x})}\cdot\bm{j}_{eq} (\mathbf{x},t)[\Psi] = 0\Rightarrow \p_t \mP_{eq}[\Psi] = 0.
\end{equation}
This ansatz aligns with the maximum entropy principle in statistical mechanics as well as with the equilibrium distribution in fluctuation theory \cite{chandler1987,Ruppeiner,Landau,Einstein,kubo1967}, with an additional prefactor accounting for energy conservation. The distribution \eqref{eq:equilibrium_ansatz} is general and holds for coupled system-bath dynamics as well, and in fact, reproduces the Einstein's distribution derived from projection operator methods in the energy representation~\cite{Pep_generic}. The extended phase space, including the bath macrostate (specified via either $s$ or $\ve$), is key to achieving this result.

Such form can be understood from energy conservation as well as vanishing EPR at equilibrium, i.e. consistency with reversible processes. Indeed, since $f(E)$ is time-independent, taking logarithms on both sides of Eq. \eqref{eq:equilibrium_ansatz} and taking the difference at $t_1, t_2$ gives 
\begin{align}
    &\D {\int d\mathbf{x} s(\mathbf{x})} = \D \ln \mP= - \D s_{sys} \Rightarrow \nonumber\\
    &\D s_{\rm tot} = \D s_{sys}+\D {\int d\mathbf{x} s(\mathbf{x})} = 0,
\end{align}
such that the stochastic total EP vanishes along any individual trajectory at equilibrium.

Moreover, in the particular case of constant temperature, we can split the total energy as $E[\Psi]=H_0 [\bm{\phi},\bm{\pi}]+T\int d\mathbf{x} s(\mathbf{x})$. Then, using Eq. \eqref{eq:1law_closed} it follows $\mP(\mathcal{E})=\mP_0(\mathcal{E})$ and:
\begin{subequations}
    \begin{align}
    &f(\mathcal{E}) = \f{ \mP_0(\mathcal{E})}{Z(\mathcal{E})},\\
    &Z(\mathcal{E})=\int_\Omega d\Psi\, e^{\int d\mathbf{x} s(\mathbf{x})}\d (\mathcal{E}-E[\Psi]),
\end{align}
\end{subequations}
where $ \mP_0(\mathcal{E})$ is the initial energy distribution and $Z(\mathcal{E})$ a generalized partition function. The partition function can be simplified by using the property \cite{Arfken, Tarski_1984}:
\begin{equation}
    \d (H[\bm{\phi}]) =\sum_i\f{\d [\bm{\phi}-\bm{\phi}_i]}{|\d_{\bm{\phi}}H|}\quad \text{for}\quad H[\bm{\phi}_i]=0,
\end{equation}
which, assuming a uniform $ \mP_0(\mathcal{E})$, yields
\begin{align}
     & Z(\mathcal{E})= \f{1}{|T|}e^{\mathcal{E}/T}\int_\o d\Psi \,e^{-H_0[\bm{\phi},\bm{\pi}]/T} =\f{1}{|T|}e^{\mathcal{E}/T}\; Z_{MB},\\
     & \mP_{eq} [\Psi] \propto \f{e^{{\int d\mathbf{x} s(\mathbf{x})}-\mathcal{E}/T}}{Z_{MB}} = \f{e^{-H_0[\bm{\phi},\bm{\pi}]/T}}{Z_{MB}}.
\end{align}
This is the Maxwell-Boltzmann (MB) canonical distribution for fields \cite{chandler1987,Landau}. Note that a constant temperature $T$ corresponds to an infinite bath whose macroscopic state remains unchanged, hence the equilibrium distribution (MB) no longer depends on $s$. 

\subsection{\label{sec:inter} Multicomponent systems}
Let us consider a two-component continuum exchanging mass and heat in a bounded domain $\mathcal{D}\subset \mathbb{R}^3$.
The state of each species is given by the respective mass density $\rho_k$ and entropy $s_k$ $(k=1,2)$ and we denote $\Psi = (\{\rho_k\} , \{s_k\})$. The Lagrangian and corresponding densities are then 
\begin{equation}\label{eq:lagrangian_inter}
    \begin{aligned}
        L[\rho_k,s_k]& = \int_\mathcal{D} d\mathbf{x}\, \ell(\rho_k, s_k , \bm{\nabla} \rho_k)\\
        &= -\varepsilon [\rho_k,s_k]= -\int_\mathcal{D} d\mathbf{x}\, e(\rho_k, s_k , \bm{\nabla} \rho_k).
    \end{aligned}   
\end{equation}
The mass and entropy fluxes are denoted by $\bm{\mJ}_{12}(\mathbf{x},t)[\Psi]=-\bm{\mJ}_{21}(\mathbf{x},t)[\Psi]$ and $\bm{J}_{12}(\mathbf{x},t)[\Psi]=\bm{J}_{21}(\mathbf{x},t)[\Psi]$ respectively, with the stochastic counterparts $\bm{\xi}_{12}(\mathbf{x},t)[\Psi]$ and $\bm{\eta}_{12}(\mathbf{x},t)[\Psi]$. The symmetry of the mass fluxes follows directly from total mass conservation, $\sum_k \bm{\mJ}_k = 0$. By contrast, the symmetry imposed on the entropy fluxes is not required by conservation laws; it is introduced here only as a modeling choice. Note that the variational framework is independent of such assumption and can be applied to other settings (e.g. asymmetric entropy fluxes). The associated constraints are then given by
\begin{equation}\label{constraints_D}
    \frac{\d L}{\d s_k}\d \Sigma_k=(\bm{J}_{k}+\bm{\eta}_{k})\cdot\bm{\nabla}\d\Gamma^k +(\bm{\mJ}_{k}+\bm{\xi}_{k})\cdot\bm{\nabla}\d W^k ,
\end{equation}
where the variables $\Gamma^k,W^k$ are conjugate variables with $s_k$ (or $\Sigma_k$) and $\rho_k$, respectively. 

The variational principle defined by the action \eqref{eq:VM_action} and the associated constraints \eqref{eq:nonhol_constraints} in this scenario takes the form:
\begin{equation}\label{eq:VP_inter}
\begin{aligned}
    & \delta \int_{t_1 , \mathcal{D}}^{t_2} dtd\mathbf{x} \left(\ell(\rho_k,s_k,\bm{\nabla}\rho_k) +  \rho_k\dot{W}^k + (s_k-\Sigma_k) \dot\Gamma^k\right)=0,\\
    & \frac{\d L}{\d s_k}\d \Sigma_k=(\bm{J}_{k}+\bm{\eta}_{k})\cdot\bm{\nabla}\d\Gamma^k +(\bm{\mJ}_{k}+\bm{\xi}_{k})\cdot\bm{\nabla}\d W^k, \\
    & \frac{\d L}{\d s_k}\dot \Sigma_k=(\bm{J}_{k}+\bm{\eta}_{k})\cdot\bm{\nabla}\dot\Gamma^k +(\bm{\mJ}_{k}+\bm{\xi}_{k})\cdot\bm{\nabla}\dot W^k .
\end{aligned}
\end{equation}
Taking variations gives the following system:
\begin{equation}\label{eq:EoM_interc}
    \begin{aligned}
    & T^k(\mathbf{x},t)[\rho_k,s_k] := \f{\d \varepsilon [\rho_k,s_k]}{\d s_k (\mathbf{x},t)}=\p_t \G^k (\mathbf{x},t),\\
    & \mu^k(\mathbf{x},t)[\rho_k,s_k] := \f{\d \varepsilon [\rho_k,s_k]}{\d \rho_k (\mathbf{x},t)}=\p_t W^k (\mathbf{x},t),\\
    &  \p_t \Sigma_k = \p_t s_k + \bm{\nabla}\cdot (\bm{J}_{12}+\bm{\eta}_{12}), \\
    & \p_t \rho_1 =\bm{\nabla}\cdot(\bm{\mJ}_{12}+\bm{\xi}_{12}) ,\\
    & \p_t \rho_2 = -\bm{\nabla}\cdot (\bm{\mJ}_{12}+\bm{\xi}_{12}) ,\\
    & T^1\p_t \Sigma_1=- (\bm{J}_{12}+\bm{\eta}_{12})\cdot\bm{\nabla} T^1 +(\bm{\mJ}_{12}+\bm{\xi}_{12})\cdot\bm{\nabla}\mu^1 ,\\
    &  T^2 \p_t \Sigma_2=-(\bm{J}_{12}+\bm{\eta}_{12})\cdot\bm{\nabla} T^2 -(\bm{\mJ}_{12}+\bm{\xi}_{12})\cdot\bm{\nabla}\mu^2,
    \end{aligned}
\end{equation}
together with the Neumann (no-flux) BCs $(\bm{J}_{12}+\bm{\eta}_{12}) \cdot\bm{n}|_{\p\mathcal{D}} = (\bm{\mJ}_{12}+\bm{\xi}_{12})\cdot\bm{n}|_{\p\mathcal{D}} = 0$ which follows from the variations $\delta \Gamma^k$ and $\delta W^k$ on $\partial\mathcal{D}$ and the natural BC $\partial e/\partial\bm{\nabla}\rho_k\cdot \boldsymbol{n}|_{\p\mathcal{D}}=0$ which follow from the variation $\delta\rho_k$ on $\partial\mathcal{D}$. See Appendix \ref{app:inter} for the derivation.

Let us assume the stochastic fluxes are of the form
\begin{subequations}
    \begin{align}
        & \xi^{a}_{12}=g^{a}_{\k}(\mathbf{x},t)[\Psi]\circ\zeta^\k_M (\mathbf{x},t) ,\\
        & \eta^{a}_{12}=\eta^{a}_{\k}(\mathbf{x},t)[\Psi]\circ\zeta^\k_H (\mathbf{x},t),
    \end{align}
\end{subequations}
where $a$ is the spatial $\mathbb{R}^3$ index, and $\bm{\zeta} = (\bm{\zeta}_M , \bm{\zeta}_H)$ is a well-defined $Q$-Wiener process cross-correlated as
\begin{align}\label{cross_correlation}
   \avg{\bm{\zeta}(\mathbf{x},t)\otimes\bm{\zeta}(\mathbf{x}',t')}&=2\begin{bmatrix}
        \mathbb{I} & \bm{C}(\mathbf{x},\mathbf{x}';t)^T  \\
        \bm{C}(\mathbf{x},\mathbf{x}';t) & \mathbb{I} 
    \end{bmatrix}\d(t-t') \nonumber\\
    &=2\mathfrak{C}(\mathbf{x},\mathbf{x}';t)\d(t-t'). 
\end{align}
We further allow state-dependent correlations
\begin{equation}\label{J_noise}
\begin{aligned}
     & g^{a}_{\k}(\mathbf{x},t)[\Psi]g^{b}_{\k}(\mathbf{x}',t)[\Psi]=G^{ab}(\mathbf{x},\mathbf{x}',t)[\Psi],\\
      & \eta^{a}_{\k}(\mathbf{x},t)[\Psi]\eta^{b}_{\k}(\mathbf{x}',t)[\Psi]=K^{ab}(\mathbf{x},\mathbf{x}',t)[\Psi].
\end{aligned}
\end{equation}
Cross-correlations imply a common hidden physical mechanism behind the heat and mass fluxes, which will give rise to cross-effects \cite{HVP_partI}.

The first law for this system takes the form
\begin{subequations}
\begin{align}
    & \partial_t e  =  \bm{\nabla} \cdot \big((\bm{\mJ}_{12}+\bm{\xi}_{12})(\mu^1 - \mu^2)\big) \nonumber\\
     &\phantom{e=}  -\bm{\nabla}\cdot\big((\bm{J}_{12}+\bm{\eta}_{12})(T^1 + T^2)\big)  + \bm{\nabla}\cdot \left(\f{\p e}{\p \bm{\nabla}\rho_k}\p_t{\rho}_k\right), \\
    & \dot \varepsilon (t)=\f{d}{dt}\int_{\mathcal{D}}d\mathbf{x}\, e  = \int_{\partial\mathcal{D}} d\boldsymbol{\sigma}\,\left(\f{\p e}{\p \bm{\nabla}\rho_k}\cdot\bm{n}\right)\cdot\p_t{\rho}_k  \nonumber\\
    & \phantom{\dot E (t)=}+ \int_{\partial\mathcal{D}} d\boldsymbol{\sigma}\,(\mu^1 - \mu^2)(\bm{\mJ}_{12}+\bm{\xi}_{12})\cdot \bm{n} \nonumber\\
    &\phantom{\dot E (t)=} -\int_{\partial\mathcal{D}} d\boldsymbol{\sigma}\,(T^1 + T^2)(\bm{J}_{12}+\bm{\eta}_{12})\cdot \bm{n}  = 0. \label{eq:1law_inter}
\end{align}
\end{subequations}
The first boundary term vanishes since $\p e / \p \bm{\nabla}\rho_k\cdot\bm{n}|_{\p\mathcal{D}}=0$, while the last two terms vanish due to no-flux BCs. Thus, internal energy is conserved and the system renders closed (isolated). Notably, the thermodynamic EPR $(\p_t s_k)$ in Eq. \eqref{eq:EoM_interc} here represents the stochastic extension of its deterministic analog  (see Ref. \cite{GBpart2}).

The FPE [Eq. \eqref{eq:FPE_VM}] is defined by the probability density currents:
\begin{subequations}
\begin{align}
     & j_{\rho_1}[\Psi] = \bm{\nabla}\cdot \bm{\mathfrak{J}}_\rho ,\\
    & j_{\rho_2}[\Psi] = -\bm{\nabla}\cdot \bm{\mathfrak{J}}_\rho ,\\
    & j_{s_1}[\Psi] = -\f{\bm{\nabla}\cdot(T^1\bm{\mathfrak{J}}_s)}{T^1}  +\f{\bm{\nabla} \mu^1}{T^1}\cdot \bm{\mathfrak{J}}_\rho ,\\
     & j_{s_2}[\Psi] = -\f{\bm{\nabla}\cdot(T^2\bm{\mathfrak{J}}_s)}{T^2} -\f{\bm{\nabla} \mu^2}{T^2}\cdot \bm{\mathfrak{J}}_\rho  .
\end{align}
\end{subequations}
The specific form of the dissipative mass and entropy probability currents $\bm{\mathfrak{J}}_\rho,\bm{\mathfrak{J}}_s$ are given in Appendix \ref{app:inter}. Note that the no-flux condition $(\bm{J}_{12}+\bm{\eta}_{12})\cdot\bm{n}|_{\p\mathcal{D}}=0$ requires both contributions to vanish individually, i.e.,
$\bm{J}_{12}\cdot\bm{n}|_{\p\mathcal{D}}=\bm{\eta}_{12}\cdot\bm{n}|_{\p\mathcal{D}}=0$, since $\bm{J}_{12}$ is deterministic while $\bm{\eta}_{12}$ is stochastic (an analogous argument holds for the mass fluxes). These conditions then imply $\bm{\mathfrak{J}}_\rho \cdot\bm{n}|_{\p\mathcal{D}}=\bm{\mathfrak{J}}_s \cdot\bm{n}|_{\p\mathcal{D}}  =0$ by direct computation.

Applying the double average the medium EPR gives:
  \begin{align}
     &  \sum_k \avgg{\dot \Sigma_k} (\mathbf{x},t)= \int_{\o} d\Psi \bigg[\left(\f{\bm{\nabla} \mu^1}{T^1}-\f{\bm{\nabla} \mu^2}{T^2}\right)\cdot\bm{\mathfrak{J}}_\rho (\mathbf{x},t) \nonumber\\
       &\phantom{\sum_k \avgg{\dot \Sigma_k} (\mathbf{x},t)=}  -\left(\frac{\bm{\nabla} T^1}{T^1}+\frac{\bm{\nabla} T^2}{T^2}\right)\cdot\bm{\mathfrak{J}}_s(\mathbf{x},t) \bigg],
  \end{align}
  which again, remarkably, satisfies
  \begin{equation}\label{eq:real_time_rate_2}
  \begin{aligned}
       \sum_k\f{d}{dt}\avg{s_k}(t) &=  \sum_k \int_{\mathcal{D}} d\mathbf{x}\avgg{\dot s_k } (\mathbf{x},t)\\
       &=  \sum_k \int_{\mathcal{D}} d\mathbf{x} \avgg{\dot \Sigma_k }(\mathbf{x},t),
       \end{aligned}
  \end{equation}
  in a similar way with Eq.~\eqref{eq:real_time_rate}; as an isolated system that relaxes to a thermal state, its EP becomes path-independent. The no-flux BCs are essential for identity \eqref{eq:real_time_rate_2}; otherwise entropy fluxes appear and the equality no longer holds.
  The system EPR gives (see Appendix \ref{app:inter}):
\begin{align}
   \dot \mS(t) &= \int_{\o,\mathcal{D}} \f{d\Psi}{\mP_t}  d\mathbf{x} d\mathbf{x'}   \begin{bmatrix}
        \bm{\mathfrak{J}}_\rho (\mathbf{x},t) \\ \bm{\mathfrak{J}}_s  (\mathbf{x},t)
    \end{bmatrix}^T  \mathbb{L}^{-1}_\Psi (\mathbf{x},\mathbf{x'})\begin{bmatrix}
        \bm{\mathfrak{J}}_\rho(\mathbf{x'},t)  \\ \bm{\mathfrak{J}}_s(\mathbf{x'},t)  
    \end{bmatrix} \nonumber\\
    &- \int_{\o,\mathcal{D}} d\Psi  d\mathbf{x} d\mathbf{x'}   \begin{bmatrix}
        \bm{\mathfrak{J}}_\rho (\mathbf{x},t) \\ \bm{\mathfrak{J}}_s  (\mathbf{x},t)
    \end{bmatrix}^T  \mathbb{L}^{-1}_\Psi (\mathbf{x},\mathbf{x'})\begin{bmatrix}
         \bm{\mJ}'_{12} (\mathbf{x'},t) \\  \bm{J}'_{12} (\mathbf{x'},t) 
    \end{bmatrix} ,
\end{align}
where $\bm{\mJ}'_{12} =\bm{\mJ}_{12} - \bm{\nu}_\rho$ and $\bm{J}'_{12} =\bm{J}_{12} - \bm{\nu}_s$ in the second term above denote the effective fluxes and $\bm{\nu}_k$ the noise-induced drifts as given in Appendix \ref{app:inter}. The matrix $\mathbb{L}^{-1}_\Psi$ is the inverse, according to Eq. \eqref{eq:inverse_def}, of
\begin{align}\label{eq:L_inter}
    &\mathbb{L}_\Psi =\begin{bmatrix}
        \vspace{3mm} G^{ab}(\mathbf{x},\mathbf{x'}) & g^a_\k (\mathbf{x})C^{\k\s} (\mathbf{x},\mathbf{x'})\eta^b_\s(\mathbf{x'})\\ 
        \eta^a_\k (\mathbf{x})C^{\k\s} (\mathbf{x},\mathbf{x'})g^b_\s(\mathbf{x'}) & K^{ab}(\mathbf{x},\mathbf{x'})
    \end{bmatrix}.
\end{align}
The inverse is well defined since the square matrix $\mathbb{L}$ is symmetric and positive-definite, as we shall argue in subsequent discussion.
Adding both contributions yields the total EPR functional:
\begin{equation}
    \begin{aligned}
       &  \dot{\rm S}_{\rm tot}(t) = \int_{\o,\mathcal{D}} \f{d\Psi}{\mP_t}  d\mathbf{x}  d\mathbf{x'}   \begin{bmatrix}
        \bm{\mathfrak{J}}_\rho (\mathbf{x},t) \\ \bm{\mathfrak{J}}_s  (\mathbf{x},t)
    \end{bmatrix}^T  \mathbb{L}^{-1}_\Psi (\mathbf{x},\mathbf{x'})\begin{bmatrix}
        \bm{\mathfrak{J}}_\rho(\mathbf{x'},t)  \\ \bm{\mathfrak{J}}_s(\mathbf{x'},t)  
    \end{bmatrix} \\
    &\phantom{ \dot{\rm S}_{\rm tot}(t) =} - \int_{\o,\mathcal{D}} d\Psi d\mathbf{x}      \begin{bmatrix}
        \bm{\mathfrak{J}}_\rho (\mathbf{x},t) \\ \bm{\mathfrak{J}}_s  (\mathbf{x},t)
    \end{bmatrix}\cdot \bm{\Xi}(\mathbf{x},t),\\
   & \bm{\Xi}=\begin{bmatrix}
        \vspace{2mm}\left(\f{\bm{\nabla} \mu^2}{T^2}-\f{\bm{\nabla} \mu^1}{T^1}\right)(\mathbf{x},t) \\ \left(\frac{\bm{\nabla} T^1}{T^1}+\frac{\bm{\nabla} T^2}{T^2}\right) (\mathbf{x},t)
    \end{bmatrix}  + \int_{\mathcal{D}}  d\mathbf{x'}   \mathbb{L}^{-1}_\Psi (\mathbf{x},\mathbf{x'})\begin{bmatrix}
         \bm{\mJ}'_{12} (\mathbf{x'},t) \\  \bm{J}'_{12} (\mathbf{x'},t) 
    \end{bmatrix}.
    \end{aligned}
\end{equation}
Enforcing the second law  \eqref{eq:2nd_law_VP} requires $\bm{\Xi}\equiv 0 $ (see Sec. \ref{sec:variational_structure}), yielding the FDRs and the total EPR:
\begin{widetext} 
\begin{align}
    & \begin{bmatrix}
        \bm{\mJ}'_{12}(\mathbf{x},t) \\  \bm{J}'_{12} (\mathbf{x},t)
    \end{bmatrix} = \int_{\mathcal{D}} d\mathbf{x'} \,  \mathbb{L}_\Psi (\mathbf{x},\mathbf{x'})  \begin{bmatrix}
        \vspace{2mm}\displaystyle -\left(\f{\bm{\nabla} \mu^2}{T^2}-\f{\bm{\nabla} \mu^1}{T^1}\right)(\mathbf{x'},t) \\\displaystyle -\left(\frac{\bm{\nabla} T^1}{T^1}+\frac{\bm{\nabla} T^2}{T^2}\right) (\mathbf{x'},t)
    \end{bmatrix}  \label{eq:FDR_inter} \\
    & \Rightarrow   \dot{\rm S}_{\rm tot}(t) =  \int_{\o,\mathcal{D}} \f{d\Psi}{\mP_t}   d\mathbf{x} d\mathbf{x'}   \begin{bmatrix}
        \bm{\mathfrak{J}}_\rho (\mathbf{x},t) \\ \bm{\mathfrak{J}}_s  (\mathbf{x},t)
    \end{bmatrix}^T  \mathbb{L}^{-1}_\Psi(\mathbf{x},\mathbf{x'})\begin{bmatrix}
        \bm{\mathfrak{J}}_\rho (\mathbf{x'},t) \\ \bm{\mathfrak{J}}_s (\mathbf{x'},t)
    \end{bmatrix}\geq 0 .\label{eq:total_EPR_inter} 
\end{align}
\end{widetext}

The FDR \eqref{eq:FDR_inter} demonstrates that, owing to cross-correlations, each flux couples simultaneously to temperature and chemical-potential gradients, thereby giving rise to cross-effects. The coupling is encoded in the symmetric positive-definite operator $\mathbb{L}$ [Eq. \eqref{eq:L_inter}], which enforces the Onsager symmetry. Recall that LDB implies microreversibility, from which Onsager symmetry emerges \cite{Onsager_relations,Broeck_1,Broeck_2,Polettini}. Consequently, in the macroscopic limit the formulation reproduces the phenomenological structure of nonequilibrium thermodynamics \cite{Onsager_relations,GBpart2}, mediated by the Onsager operator in spatially extended systems. Namely, the FDR \eqref{eq:FDR_inter} has the same form as its macroscopic counterpart found in \cite{GBpart2} from
\begin{equation}
    I := -\bm{J}_{12}\cdot\left(\frac{\bm{\nabla} T^1}{T^1}+\frac{\bm{\nabla} T^2}{T^2}\right)-\bm{\mJ}_{12}\cdot\left(\f{\bm{\nabla} \mu^2}{T^2}-\f{\bm{\nabla} \mu^1}{T^1}\right)\geq 0,
\end{equation}
where $I$ denotes the macroscopic internal EP. Moreover, this result is the infinite-dimensional extension of that in \cite[Sec. IIIE]{HVP_partI}, where the spatial gradients are replaced by the discrete gradients between spatial cells. For completeness, Appendix~\ref{app:inter} provides an explicit PI computation that benchmarks the LDB condition \eqref{eq:LDB_VP}.

The FDR also accounts for noise-induced drifts, generated by multiplicative noise, which are essential for thermodynamic consistency \cite{Lau_2007, Cates_2022}. For example, choosing $\bm{G} = g^2[T_k\rho_k] \mathbb{I}$ for some nonnegative functional $g[\cdot]$ would provide a multicomponent extension of the Dean-Kawasaki equation \cite{KAWASAKI1993115,Dean_1996} with heat conduction. 

As a remark, the limiting case $\bm{C} = \mathbb{I}$ is degenerate, as the two noise sources coincide, leading to redundancy and an ill-defined model. In particular, we require the covariance kernel $\bm{\mathfrak{C}}$ in Eq. \eqref{cross_correlation} to be positive-definite in order for $\mathbb{L}$ to be positive-definite (it is symmetric by construction), such that its inverse is well-defined and the total EPR \eqref{eq:total_EPR_inter} vanishes only at equilibrium, i.e. when all dissipative currents vanish.

The same equilibrium ansatz \eqref{eq:equilibrium_ansatz} as in the previous example holds in this case,
\begin{equation}\label{eq:equilibrium_ansatz_inter}
\mP_{eq}[\rho_k,s_k] = f(\varepsilon[\rho_k,s_k])e^{\sum_k\int d\mathbf{x} s_k(\mathbf{x})},
\end{equation}
where $f(\varepsilon)$ depends only on the conserved total energy functional $\varepsilon[\rho_k,s_k]$. Enforcing the FDR \eqref{eq:FDR_inter} in the FPE probability density currents yields 
\begin{equation}
     \f{\d }{\d \Psi (\mathbf{x})}\cdot\bm{j}_{eq} (\mathbf{x},t)[\Psi] = 0\Rightarrow \p_t \mP_{eq}[\Psi] = 0.
\end{equation}
Thus, Eq. \eqref{eq:equilibrium_ansatz_inter} solves the FPE steady state under the FDR, ensuring thermodynamic consistency for closed systems. Crucially, the no-flux BCs ensure that boundary contributions vanish. Otherwise, Eq.~\eqref{eq:equilibrium_ansatz_inter} would not satisfy a steady state, as illustrated later in Sec.~\ref{sec:under_open}. This is consistent with the requirement of  isolated systems to converge to thermal equilibrium.

In the special case of constant temperature $T^k=T$, the total energy splits as
\begin{equation}\label{special_case}
\varepsilon[\rho_k,s_k] = F[\rho_k] + T\sum_k\int d\mathbf{x} s_k(\mathbf{x}),
\end{equation}
with $F[\rho_k]$ denoting the Helmholtz free-energy functional satisfying $\mu^k=\dfrac{\d F}{\delta \rho_k}$.
Since $\dot \varepsilon=0$, the energy-resolved PDF reads
\begin{align}
\mP_0(\mathcal{E}) &= \int_\o d\Psi\, f(\varepsilon[\Psi])e^{\sum_k\int d\mathbf{x} s_k(\mathbf{x})}\,\delta(\mathcal{E}-\varepsilon[\Psi]) \nonumber\\
&= f(\mathcal{E}) Z(\mathcal{E}),
\end{align}
such that
\begin{equation}
f(\mathcal{E}) = \frac{\mP_0(\mathcal{E})}{Z(\mathcal{E})}, \quad
Z(\mathcal{E})=\int_\Omega d\Psi\, e^{\sum_k\int d\mathbf{x} s_k(\mathbf{x})}\,\delta(\mathcal{E}-\varepsilon[\Psi]).
\end{equation}
Using the identity for composite Dirac deltas \cite{Arfken,Tarski_1984} yields
\begin{equation}
Z(\mathcal{E}) = \frac{1}{|T|}e^{\mathcal{E}/T}\int_\Omega \mathcal{D}\rho_k\, e^{-F[\rho_k]/T}
= \frac{1}{|T|}e^{\mathcal{E}/T} Z_{G}.
\end{equation}
Hence, the equilibrium PDF, assuming a uniform initial energy distribution $\mP_0 (\mathcal{E})$, reduces to
\begin{align}
\mP_{eq}[\rho_k] \propto \f{e^{\sum_k\int d\mathbf{x} s_k(\mathbf{x})-\mathcal{E}/T}}{Z_{G}}  =  \frac{e^{-F[\rho_k]/T}}{Z_{G}},
\end{align}
in agreement with \cite{Derrida,Leonard_2013}.

We emphasize that the internal energy $\varepsilon[\Psi]$ has not been restricted to a specific form, except when we impose Eq.~\eqref{special_case}. Beyond allowing for irreversible couplings between the species through thermodynamic fluxes, conservative couplings can also arise from a non-separable internal energy. Furthermore, analogous to noise correlations, the internal energy $\varepsilon$ in Eq. \eqref{eq:lagrangian_inter} may in general be nonlocal, as it only affects the definition of the thermodynamic affinities ($T^k,\mu^k$ in this case) and the induced natural BCs, and not the form of the equations of motion or the associated FDRs.

\subsection{\label{sec:under_open} Underdamped dynamics of open system}
As a final example we consider the open-system extension of the underdamped model introduced in Sec. \ref{sec:under_closed}. In addition to the terms in the closed model, the dynamics now include an external drive $\mathbf{f}_+(\mathbf{x},t)[\Psi]$ for $\Psi = (\bm{\phi},\bm{\pi},s)$, which may depend on the system state, as in e.g. the KPZ equation \cite{KPZ}. Moreover, we include heat conduction within the system and an external heat transfer arising from contact with a thermal reservoir at temperature $T_0$, as in e.g. fluctuating hydrodynamics \cite{LLNS,Landau} (not to be confused with the case of an
open thermodynamic system with \textit{prescribed fluxes}, which can also be considered via the present approach). Owing to this driving, the system does not necessarily relax to a steady state and may remain out of equilibrium throughout its evolution. We assume that $\mathcal{D}\subset \mathbb{R}^3$ is a bounded domain, as in Sec. \ref{sec:under_closed}.

The Lagrangian retains the form of Eq. \eqref{eq:closed_lagrangian}, while the variational principle is augmented by the principle of virtual work to incorporate the effect of external manipulation. As the system now includes external thermodynamic fluxes, the full nonholonomic constraints \eqref{eq:nonhol_constraints} must be enforced---including entropy pumping due to the presence of $\mathbf{f}_+$---yielding:
\begin{widetext}
\begin{equation}\label{eq:VP_open}
\begin{aligned}
    &\delta \int_{t_1,\mathcal{D}}^{t_2} dtd\mathbf{x} \bigg(\ell( \bm{\phi}, \bm{u}, \bm{\nabla}\bm{\phi}, s)  + \bm{\pi}\cdot (\dot{\bm{\phi}} - \bm{u}) +(s-\Sigma)\dot \Gamma \bigg)  + \int_{t_1,\mathcal{D}}^{t_2} dtd\mathbf{x}\,\mathbf{f}_+ \cdot\d\bm{\phi}= 0 , \\
     & \int_{\mathcal{D}}d\mathbf{x}\, w\f{\d L}{\d s}\dot \Sigma = \int_{\mathcal{D}}d\mathbf{x}\, w  \bigg[ (\mathbf{f}_-  + \bm{\xi})\cdot \dot{\bm{\phi}} +  (\bm{j}  + \bm{\eta}) \cdot  \bm{\nabla}  \dot\Gamma  + \mI\dot\G\bigg]  -  \int_{\p\mathcal{D}}d\bm{\s}\, w (\dot\G-T_0)(\bm{j}  + \bm{\eta}) \cdot\bm{n}, \\
    &\int_{\mathcal{D}}d\mathbf{x}\, w \f{\d L}{\d s}\d \Sigma=\int_{\mathcal{D}}d\mathbf{x}\, w \bigg[ (\mathbf{f}_-  + \bm{\xi})\cdot \d \bm{\phi}    +  (\bm{j}  + \bm{\eta} ) \cdot \bm{\nabla}  \d \Gamma + \mI\d\G \bigg] -  \int_{\p\mathcal{D}}d\bm{\s}\, w \d\G(\bm{j}  + \bm{\eta}) \cdot\bm{n},
\end{aligned}  
\end{equation}
\end{widetext}
for all $w :\mathcal{D}\rightarrow \mathbb{R}$. Here $T_0:\p \mathcal{D}\rightarrow \mathbb{R}$ is the prescribed temperature field at the boundary. The variational approach can also be adapted to treat prescribed fluxes on the boundary, see Ref. \cite{gawlik24}. These constraints are the continuum analogue to the constraints used for
finite-dimensional open systems in \cite{HVP_partI}.

Recall $\bm{\phi}$ is a vector field and $\bm{\pi}$ its momentum, $\mathbf{f}_-$ denotes a dissipative mechanical force and $\bm{j}$ the entropy flux.  The noises can be expanded as
\begin{subequations}
    \begin{align}
        & \xi^{a}=g^{a}_{\k}(\mathbf{x},t)[\Psi]\circ\zeta^\k_{\rm f} (\mathbf{x},t), \\
        & \eta^{a} =\eta^{a}_{\k}(\mathbf{x},t)[\Psi]\circ\zeta^\k_{\rm s} (\mathbf{x},t),
    \end{align}
\end{subequations}
where $a$ is the spatial $\mathbb{R}^3$ index, $\bm{\zeta}=(\bm{\zeta}_{\rm f}, \bm{\zeta}_{\rm s})$ are Gaussian independent $Q$-Wiener processes [Eq. \eqref{eq:gauss_noise} with $C^{\k\s}_{(f,s)}(\mathbf{x},\mathbf{x}';t) = \d^{\k\s}C_{(f,s)}(\mathbf{x},\mathbf{x}')$], and the covariance of the random forces read:
\begin{equation}
\begin{aligned}
     & g^{a}_{\k}(\mathbf{x},t)[\Psi]g^{b}_{\k}(\mathbf{x}',t)[\Psi]C_f(\mathbf{x},\mathbf{x}')=D^{ab}(\mathbf{x},\mathbf{x}',t)[\Psi],\\
      & \eta^{a}_{\k}(\mathbf{x},t)[\Psi]\eta^{b}_{\k}(\mathbf{x}',t)[\Psi]C_s(\mathbf{x},\mathbf{x}')=K^{ab}(\mathbf{x},\mathbf{x}',t)[\Psi].
\end{aligned}
\end{equation}

The equations of motion yield: 
\begin{subequations}\label{eq:EoM_open}
    \begin{align}
    &\pi^a (\mathbf{x},t) = \p_t\phi^a (\mathbf{x},t),\\
    & T(\mathbf{x},t)[\bm{\phi},s] := \f{\d \varepsilon [\bm{\phi},s] }{\d s (\mathbf{x},t)}=\p_t \G (\mathbf{x},t),\\
    & \p_t \bm{\pi}=-\f{\d\varepsilon}{\d \bm{\phi}}+ \mathbf{f}_+  +\mathbf{f}_- + \bm{\xi}, \\
    &\p_t s +\bm{\nabla}\cdot(\boldsymbol{j}  + \bm{\eta})= -\f{\bm{\pi}}{T}\cdot(\mathbf{f}_- + \bm{\xi}) -\f{\bm{\nabla} T}{T}\cdot(\bm{j}  + \bm{\eta}),\\
    & \p_t \Sigma =  \p_t s +\bm{\nabla}\cdot(\boldsymbol{j}  + \bm{\eta}) - \mI,
\end{align}
\end{subequations}
together with the emerging BC for free flux at the boundary
\begin{equation}
     \int_{\partial\mathcal{D}} d\boldsymbol{\sigma}\,w(\dot\G-T_0)(\bm{j}  + \bm{\eta}) \cdot\bm{n} = 0,\;\forall w \Rightarrow T|_{\partial\mathcal{D}}=T_0 ,
\end{equation}
and the natural BCs given by Eq. \eqref{eq:natural_BC_phi}. Here $\mI = \d_{\bm{\pi}}\cdot\mathbf{f}_+$ following Eq. \eqref{eq:entropy_pump}. See Appendix \ref{app:open} for the derivation.

The first law for this system takes the form
\begin{subequations}\label{energy_balance_III}
\begin{align}
    & \p_t e_{\rm tot} (\mathbf{x},t) = \bm{\pi}\cdot \mathbf{f}_+  -\bm{\nabla}\cdot \left( T(\boldsymbol{j}  + \bm{\eta})\right) + \bm{\nabla}\cdot \left(\frac{\partial e}{\partial \bm{\nabla} \bm{\phi}}\cdot\bm{\pi}\right), \label{eq:1law_open}\\
    & \dot E (t)=\f{d}{dt}\int_{\mathcal{D}} d\mathbf{x} \, e_{\rm tot}(\mathbf{x},t)  \nonumber\\
    & \phantom{\dot E (t)}= \int_{\mathcal{D}}d\mathbf{x}\, \bm{\pi}\cdot \mathbf{f}_+   - \int_{\partial\mathcal{D}} d\boldsymbol{\sigma}\, T_0(\boldsymbol{j}  + \bm{\eta})\cdot\bm{n}\nonumber\\
    & \phantom{\dot E (t)}= P^{ext}_W + P^{ext}_H .
\end{align}
\end{subequations}
Note that the boundary term arising from the third term in Eq. \eqref{eq:1law_open} vanishes due to $\p e / \p \bm{\nabla}\bm{\phi}\cdot\bm{n}|_{\p\mathcal{D}}=0$, as in the closed system example of Sec. \ref{sec:under_closed}. The energy is not conserved due to both the external heat transfer $\bm{j}_Q = T_0(\boldsymbol{j}  + \bm{\eta})$ with the thermal reservoir and the external driving $\mathbf{f}_+$ (analogous to its finite-dimensional counterpart \cite{HVP_partI}). As a consequence, the equilibrium ansatz \eqref{eq:equilibrium_ansatz} no longer satisfies the FPE. 

We define the effective even and odd forces as $F^a_+ =   - \d_{\phi^a (\mathbf{x})} \varepsilon[\bm{\phi},s] + \mathrm{f}^a_+ (\mathbf{x},t)[\Psi]$ and $F'^a_-$ as given in Eq. \eqref{eq:odd_force}, respectively. The FPE \eqref{eq:FPE_VM} is defined by the probability density currents described in \eqref{eq:closed_currents} except for the redefinition of $\bm{F}_+$ and the additional contribution to the entropy current:
\begin{subequations}
    \begin{align}
    & j_{s} (\mathbf{x},t)[\Psi] =  -\f{\bm{\pi} }{T}\cdot \bm{j}_{d}(\mathbf{x},t)[\Psi] -\frac{1}{T}\bm{\nabla} \cdot\left(T \bm{\mathfrak{J}}_s (\mathbf{x},t)[\Psi]\right) ,\\
     & \bm{\mathfrak{J} }_s (\mathbf{x},t)[\Psi]  = \bm{j}'(\mathbf{x},t)[\Psi]\mP_t [\Psi] \nonumber\\
    &\phantom{\bm{\mathfrak{J} }_s(\mathbf{x},t)  } +\int_{\mathcal{D}} d\mathbf{x}'\frac{1}{T(\mathbf{x}')}\bm{\nabla}'\cdot\big(T(\mathbf{x}')\bm{K} (\mathbf{x},\mathbf{x}',t)[\Psi]\big)  \f{\d \mP_t [\Psi]}{\d s (\mathbf{x}')}, \label{eq:J_s_current}
\end{align}
\end{subequations}
where  $\bm{j}'$ accounts for the noise-induced drift on the entropy flux and it is given by, denoting $\bm{\eta}_\k=\eta^a_\k$:
\begin{equation}\label{eq:s_drift}
\begin{aligned}
    &\bm{j}'(\mathbf{x},t)   =\bm{j}(\mathbf{x},t)   \\
    & +  \sum_\k\bm{\eta}_\k(\mathbf{x},t) \int_{\mathcal{D}} d\mathbf{x}'  \f{\d }{\d s (\mathbf{x}')}\frac{\bm{\nabla}'}{T(\mathbf{x}')}\cdot\big(T(\mathbf{x}') \bm{\eta}_\k  (\mathbf{x}',t)C(\mathbf{x},\mathbf{x}')\big).
\end{aligned}
\end{equation}

The averaged first law yields 
\begin{align}\label{avg1law}
    \f{d}{dt}\avg{E}& = -\int_{\o,\mathcal{D}}d\Psi d\mathbf{x} E[\Psi]\f{\d j^a (\mathbf{x},t)[\Psi]}{\d \Psi^a (\mathbf{x})} \nonumber\\
    & = \int_{\o,\mathcal{D}}d\Psi d\mathbf{x} j^a (\mathbf{x},t)[\Psi] \f{\d E[\Psi]}{\d \Psi^a (\mathbf{x})}\nonumber\\
    & = \int_{\o,\mathcal{D}}d\Psi d\mathbf{x} [\bm{\pi}\cdot\mathbf{f}_+ \mP_t+\bm{j}_d\cdot \bm{\pi}+ Tj_s]\nonumber\\
    & =\Bavg{\int_{\mathcal{D}}d\mathbf{x}\,\bm{\pi}\cdot\mathbf{f}_+}-\int_{\o,\mathcal{D}}d\Psi d\mathbf{x}\,\bm{\nabla} \cdot [T\bm{\mathfrak{J}}_s] \nonumber\\
    & =\Bavg{\int_{\mathcal{D}}d\mathbf{x}\,\bm{\pi}\cdot\mathbf{f}_+}-\int_{\o,\p\mathcal{D}}d\Psi d\bm{\s}\,T_0\bm{\mathfrak{J}}_s\cdot\bm{n}.
\end{align}
This is equivalent to taking the double average $\avgg{\cdot}$ of Eq. \eqref{energy_balance_III}. If instead we had Neumann BC, $(\bm{j}+\bm{\eta})\cdot\bm{n} |_{\p\mathcal{D}}=q_0 \Rightarrow T\bm{\mathfrak{J}}_s\cdot\bm{n}|_{\p\mathcal{D}}=q_0\mP_t$ and hence the first law would be of the form:
$$\f{d}{dt}\avg{E} =\Bavg{\int_{\mathcal{D}}d\mathbf{x}\,\bm{\pi}\cdot\mathbf{f}_+}-\Bavg{\int_{\p\mathcal{D}}d\bm{\s}\,q_0}.$$

Applying the double average the medium EPR takes the form
\begin{equation}
    \avgg{\dot\Sigma}=\int_{\o} d\Psi \left(  -\f{\bm{\pi}}{T}\cdot\bm{j}_d -\f{\bm{\nabla} T}{T}\cdot\bm{\mathfrak{J}}_s\right) - \avg{\mI}.
\end{equation}
The system EPR gives (see Appendix \ref{app:open})
\begin{align}
   \dot \mS(t) &= \int_{\o,\mathcal{D}} \f{d\Psi}{\mP_t}  d\mathbf{x} d\mathbf{x'}   \begin{bmatrix}
        \bm{j}_d (\mathbf{x},t) \\ \bm{\mathfrak{J}}_s  (\mathbf{x},t)
    \end{bmatrix}^T  \mathbb{L}^{-1}_\Psi (\mathbf{x},\mathbf{x'})\begin{bmatrix}
        \bm{j}_d (\mathbf{x'},t)  \\ \bm{\mathfrak{J}}_s(\mathbf{x'},t)  
    \end{bmatrix} \nonumber\\
    &- \int_{\o,\mathcal{D}} d\Psi  d\mathbf{x} d\mathbf{x'}   \begin{bmatrix}
       \bm{j}_d (\mathbf{x},t) \\ \bm{\mathfrak{J}}_s  (\mathbf{x},t)
    \end{bmatrix}^T  \mathbb{L}^{-1}_\Psi (\mathbf{x},\mathbf{x'})\begin{bmatrix}
         \bm{F}'_- (\mathbf{x'},t) \\  \bm{j}' (\mathbf{x'},t) 
    \end{bmatrix} \nonumber\\
    & +\Bavg{\int_{\mathcal{D}}d\mathbf{x}\f{\d}{\d \bm{\pi}(\mathbf{x})}\cdot \mathbf{f}_+ (\mathbf{x},t)},
\end{align}
where 
\begin{equation}
    \mathbb{L}^{-1}(\mathbf{x},\mathbf{x}',t)[\Psi]=\begin{bmatrix}
    \bm{D}^{-1}(\mathbf{x},\mathbf{x}',t)[\Psi] & \bm{0}\\ \bm{0} & \bm{K}^{-1}(\mathbf{x},\mathbf{x}',t)[\Psi]
\end{bmatrix}.
\end{equation}
Adding both contributions, where we explicitly use $\Sigma$ to include the whole EP in the medium, yields the total EPR functional:
\begin{equation}
    \begin{aligned}
        & \dot{\rm S}_{\rm tot}(t)  = \int_{\o,\mathcal{D}} \f{d\Psi}{\mP_t} d\mathbf{x}  d\mathbf{x'}   \begin{bmatrix}
        \bm{j}_d (\mathbf{x},t) \\ \bm{\mathfrak{J}}_s  (\mathbf{x},t)
    \end{bmatrix}^T  \mathbb{L}^{-1}_\Psi (\mathbf{x},\mathbf{x'})\begin{bmatrix}
        \bm{j}_d (\mathbf{x'},t)  \\ \bm{\mathfrak{J}}_s(\mathbf{x'},t)  
    \end{bmatrix} \\
    &\phantom{  \dot{\rm S}_{\rm tot}(t) = } - \int_{\o,\mathcal{D}} d\Psi  d\mathbf{x}   \begin{bmatrix}
       \bm{j}_d (\mathbf{x},t) \\ \bm{\mathfrak{J}}_s  (\mathbf{x},t)
    \end{bmatrix}\cdot \bm{\Xi} (\mathbf{x},t) ,\\
    & \bm{\Xi} = \begin{bmatrix}
      \vspace{3mm} \frac{\bm{\pi}}{T}(\mathbf{x},t) \\ \frac{\bm{\nabla} T}{T}(\mathbf{x},t)
    \end{bmatrix}+ \int_{\mathcal{D}} d\mathbf{x'} \mathbb{L}^{-1}_\Psi (\mathbf{x},\mathbf{x'})\begin{bmatrix}
         \bm{F}'_- (\mathbf{x'},t) \\  \bm{j}' (\mathbf{x'},t) 
    \end{bmatrix}.
    \end{aligned}
\end{equation}
Enforcing the second law  \eqref{eq:2nd_law_VP} requires $\bm{\Xi}\equiv 0 $ (see Sec. \ref{sec:variational_structure}), which gives the FDRs and the total EPR:
\begin{widetext}
\begin{align}
    & \begin{bmatrix}
       \bm{F}'_- (\mathbf{x},t) \\ \bm{j}'(\mathbf{x},t)
    \end{bmatrix}=\int_{\mathcal{D}} d\mathbf{x}'\,\mathbb{L}_\Psi  (\mathbf{x},\mathbf{x'}) \begin{bmatrix}
      \vspace{3mm} -\dfrac{\bm{\pi}}{T} (\mathbf{x}',t)\\ -\dfrac{\bm{\nabla} T}{T}(\mathbf{x}',t)
    \end{bmatrix} \label{eq:FDR_open} \\
    &\Rightarrow \dot{\rm S}_{\rm tot}(t) = \int_{\o,\mathcal{D}} \f{d\Psi}{\mP_t} d\mathbf{x}  d\mathbf{x'}   \begin{bmatrix}
        \bm{j}_d (\mathbf{x},t) \\ \bm{\mathfrak{J}}_s (\mathbf{x},t)
    \end{bmatrix}^T  \mathbb{L}^{-1}_\Psi (\mathbf{x},\mathbf{x'})\begin{bmatrix}
        \bm{j}_d   (\mathbf{x}',t)\\ \bm{\mathfrak{J}}_s (\mathbf{x}',t)
    \end{bmatrix}\geq 0,
\end{align}
\end{widetext}
with 
\begin{equation}
    \mathbb{L}(\mathbf{x},\mathbf{x}',t)[\Psi]=\begin{bmatrix}
    \bm{D}(\mathbf{x},\mathbf{x}',t)[\Psi] & \bm{0}\\ \bm{0} & \bm{K}(\mathbf{x},\mathbf{x}',t)[\Psi]
\end{bmatrix}
\end{equation}
and $\bm{F}'_- (\mathbf{x},t)$ given earlier in \eqref{eq:odd_force}.

This FDR encodes the generalized Einstein relation [Eq. \eqref{eq:FDR_closed}] and Fourier’s law for heat conduction \cite{GBpart2,groot} through the operator $\mathbb{L}$, consistent with Onsager symmetry \cite{Onsager_relations}. In the present case, noise cross-correlations are not included---although such couplings could be incorporated---so no cross-effects arise and the off-diagonal terms vanish. It is worth emphasizing that the entropy–pumping contribution $\mI$, entering through the nonholonomic constraints \eqref{eq:VP_open}, plays a crucial role in preserving the quadratic structure of the total EPR , and is unaffected by the FDRs, as expected, since it stems from non-dissipative mechanisms and hence vanishes in the total EPR.

As it was argued in Sec.~\ref{sec:variational_structure}, the FDRs arising from the second law axiom enforce LDB in the extended phase space, as given in Eq.~\eqref{eq:LDB_VP}. In the discrete theory \cite{HVP_partI}, this result was established through an explicit PI computation performed on a case-by-case basis. Since the discrete formulation underlies the continuum theory, the correspondence carries over analogously. To illustrate this general result, we provide an explicit derivation within the continuum formulation for the present example and recall its associated consequences.

To implement the PI formalism for the field dynamics in Eqs.~\eqref{eq:EoM_open}, it is convenient to rewrite the system as a deterministic evolution supplemented by two stochastic current equations (see, e.g., \cite{Cates_2022}):
\begin{equation}
 \bm{\mathfrak{j}}_\pi = \mathbf{f}_+ + \mathbf{f}_- + \bm{\xi}, \quad
 \bm{\mathfrak{j}}_s = \boldsymbol{j} + \bm{\eta},
\end{equation}
where $\bm{\mathfrak{j}}_\pi$ and $\bm{\mathfrak{j}}_s$ are even and odd under time reversal, respectively, as are $\p_t \bm{\pi}$ and $\p_t s$. Alternatively, one may proceed directly from Eqs.~\eqref{eq:EoM_open} using a brute-force construction of the PI, yielding the same result, as shown in \cite{HVP_partI}.

The forward action is then given by
\begin{align}
    & \mathcal{A}[\Psi] = \int_{t_1 , \mathcal{D}}^{t_2}d\t d\mathbf{x}\bigg\{  \d_{\pi (\mathbf{x})}\cdot (\mathbf{f}_+  + \mathbf{f}_-) (\mathbf{x},\t)
     \nonumber\\& + \f{1}{4}\int_{\mathcal{D}} d\mathbf{x'} \begin{bmatrix}
       \bm{\mathfrak{j}}_\pi -\mathbf{f}_+  -\bm{F}'_-\\ \bm{\mathfrak{j}}_s -\boldsymbol{j}'
    \end{bmatrix}^T\mathbb{L}^{-1}_\Psi (\mathbf{x},\mathbf{x'}) \begin{bmatrix}
       \bm{\mathfrak{j}}_\pi -\mathbf{f}_+  - \bm{F}'_-\\ \bm{\mathfrak{j}}_s -\boldsymbol{j}'
    \end{bmatrix}'\bigg\},
\end{align}
with $[\cdot] '$ denoting $\mathbf{x'}$ dependence and $\bm{F}'_-,\boldsymbol{j}' $ the effective drifts given by Eqs.~\eqref{eq:odd_force},\eqref{eq:s_drift} respectively. Here we have omitted the time reversal symmetric contributions of the entropy equation in the first term. The backward action is given by $\hat{\mathcal{A}}[\Psi] = \mathcal{A}[\bm{\phi},-\bm{\pi},s]$ with $(\hat{ \bm{\mathfrak{j}}}_\pi , \hat{ \bm{\mathfrak{j}}}_s )=( \bm{\mathfrak{j}}_\pi , -\bm{\mathfrak{j}}_s )$:
\begin{align}
    &\hat{\mathcal{A}}[\Psi] = \int_{t_1 , \mathcal{D}}^{t_2}d\t d\mathbf{x}\bigg\{  -\d_{\pi (\mathbf{x})}\cdot (\mathbf{f}_+  - \mathbf{f}_- )(\mathbf{x},\t)
     \nonumber\\& + \f{1}{4}\int_{\mathcal{D}} d\mathbf{x'} \begin{bmatrix}
       \bm{\mathfrak{j}}_\pi -\mathbf{f}_+  +\bm{F}'_-\\ -\bm{\mathfrak{j}}_s -\boldsymbol{j}'
    \end{bmatrix}^T\mathbb{L}^{-1}_\Psi (\mathbf{x},\mathbf{x'}) \begin{bmatrix}
       \bm{\mathfrak{j}}_\pi -\mathbf{f}_+  + \bm{F}'_-\\ -\bm{\mathfrak{j}}_s -\boldsymbol{j}'
    \end{bmatrix}'\bigg\}.
\end{align}
The log ratio \eqref{eq:EP} yields, after some algebra:
\begin{equation}
    \begin{aligned}  
    \D \mathfrak{s}_m &= \hat{\mathcal{A}}[\Psi] - \mathcal{A}[\Psi]\\
    & =\int_{t_1 , \mathcal{D}}^{t_2}d\t d\mathbf{x}\bigg\{ -\dfrac{\d }{\d \bm{\pi}(\mathbf{x})}\cdot \mathbf{f}_+ [\Psi] \\
    & + \int_{\mathcal{D}}d\mathbf{x'}\begin{bmatrix}
        (\mathbf{f}_- + \bm{\xi})(\mathbf{x},\t) \\ (\boldsymbol{j}  + \bm{\eta})(\mathbf{x},\t)
    \end{bmatrix}^T\mathbb{L}^{-1}_\Psi (\mathbf{x},\mathbf{x'})\begin{bmatrix}
       \bm{F}'_- (\mathbf{x}',\t) \\ \bm{j}'(\mathbf{x}',\t)
    \end{bmatrix}  \bigg\}.
    \end{aligned}
\end{equation}
Applying the FDRs \eqref{eq:FDR_open} it finally gives: 
\begin{equation}
     \D \mathfrak{s}_m = \int_{t_1 , \mathcal{D}}^{t_2}d\t d\mathbf{x}\bigg\{-\f{\bm{\pi}}{T} \cdot (\mathbf{f}_- + \bm{\xi}) -\f{\bm{\nabla} T}{T}\cdot  (\boldsymbol{j}  + \bm{\eta}) - \mI\bigg\},
\end{equation}
and hence $\D \mathfrak{s}_m = \D\Sigma$ by direct comparison with Eqs. \eqref{eq:EoM_open}, which is the LDB condition in \eqref{eq:LDB_VP}. Recall the variational formulation distinguishes $\D\Sigma$ from $\D s$ by properly accounting for external entropy flows. As a result, it correctly identifies $\D \mathfrak{s}_m = \D\Sigma$, rather than $\D \mathfrak{s}_m = \D s$. Notably, the example of Sec.~\ref{sec:under_closed} constitutes a particular instance of this computation. 

Consequently, the variational formulation satisfies 
\begin{subequations}
\begin{align}
    & \D s_{\rm tot} = \D s_{sys}+\D \Sigma, \\
    & \D s_{sys}=\text{ln}\frac{\mathcal{P}(\Psi_0, t_0)}{\mathcal{P}(\Psi_t,t)} ,\quad \Delta \Sigma  =\text{ln}\frac{P[\Psi|\Psi_0 ]}{P[\hat{\Psi}|\Psi_t ]}.
\end{align}
\end{subequations}
and is therefore consistent with the KL divergence representation of entropy production \eqref{eq:inf_entropy_VP}, i.e. $\langle \Delta s_{\rm tot}\rangle = \langle \Delta\mathfrak{s}_{\rm tot}\rangle $. Accordingly, $\D\Sigma$ satisfies the master FT, valid for arbitrary normalized initial and final distributions $\mP_0,\mP_t$ \cite{Seifert_2012}:
\begin{align}\label{eq:masterFT}
    \avg{e^{-\ln \mP_0/\mP_t - \D \Sigma}} &= \avg{e^{- \D \Sigma} \mP_t /\mP_0 } \nonumber \\
    &= \int_{\o} \mathcal{D}\Psi\,\mP_0 P[\Psi|\Psi_0]e^{- \D \Sigma} \mP_t /\mP_0  \nonumber \\
    & = \int_{\o} \mathcal{D}\Psi\, P[\hat{\Psi}|\Psi_t ]\mP_t = 1,
\end{align}
where the last equality follows by normalization, analogous to the IFT [Eq. \eqref{eq:IFT}]. As anticipated, this result constitutes the functional analogue of that derived in \cite{HVP_partI}, and follows directly from LDB in the extended phase space [Eq.~\eqref{eq:LDB_VP}]. The nontrivial extension lies in the functional character of the probability measures and in the possible spatial nonlocality of the associated operators, such that the theorem now applies to spatially extended systems with nonlinear and/or nonlocal couplings. As an intrinsic feature of the formulation, the master FT also holds for systems with coupled thermal and configurational DoFs, broadening its domain of applicability.

Although we have focused here on the open version of Sec.~\ref{sec:under_closed}, an analogous open formulation of Sec.~\ref{sec:inter} could be considered, where matter and heat are exchanged with the exterior through $\partial\mathcal{D}$ using similar techniques.

\section{\label{sec:discussion} Discussion}
Through the previous examples, we have shown that the variational principle provides a systematic route to the ST of spatially extended systems from first principles. On the one hand, the geometric structure yields well-defined thermodynamic affinities and consistent energetics, i.e., the first law of ST, and incorporates into the dynamical equations the forces induced by different interactions, as well as BCs, in a systematic manner. On the other hand,  the second law leads to the requirement of LDB in an extended phase space, from which generalized FDRs and a well-defined EP satisfying the generalized Clausius relation [Eq.~\eqref{eq:Gen_clausius}] follow. Altogether, the formulation ensures thermodynamic consistency and the associated properties \cite{Broeck_1,Broeck_2,Polettini}, such as thermal equilibrium for isolated systems, Onsager symmetry arising from microreversibility, and FTs with a clear energetic interpretation beyond the log ratio of path weights. Remarkably, the resulting FDRs are independent of the choice of BCs, holding for both open and isolated systems.

In the case of closed systems [Secs.~\ref{sec:under_closed} and \ref{sec:inter}], the FDRs guarantee the existence of a steady state of the associated FPE consistent with thermodynamic equilibrium. Since the PDF in the thermodynamic phase space $\o$ is expressed in terms of the entropy $s$, any steady state of the FPE corresponds to an equilibrium state. We demonstrated that this steady state realizes the Einstein's equilibrium ansatz [Eq.~\eqref{eq:equilibrium_ansatz}] under the assumption of Markovian dynamics. In appropriate limits, such as weak coupling, this ansatz reduces to the MB distribution. Moreover, a property specific to closed/isolated systems is that entropy is a state function, so that the EPR reduces to a total time derivative satisfying Eq.~\eqref{eq:real_time_rate}. 

Notably, for systems with conserved scalar quantities their respective densities satisfy a (stochastic) continuity equation of the form $\p_t \varphi + \bm{\nabla}\cdot\bm{j}=0$, such as the energy density for closed systems [Eqs.~\eqref{eq:1law_closed}, \eqref{eq:1law_inter}] or mass densities in multicomponent systems [Eq.~\eqref{eq:EoM_interc}].

The FDRs encode the Onsager symmetry through symmetric positive-definite matrices $\mathbb{L}$ that consistently incorporate cross-correlations. In this way, the variational approach incorporates fluctuations compatible with the structure of macroscopic nonequilibrium thermodynamics, as a direct consequence of LDB \cite{Broeck_1,Broeck_2,Polettini,PhysRevX.15.021050}. In the stochastic setting, cross-correlations between thermodynamic fluxes are shown to arise from noise cross-correlations in order to maintain thermodynamic consistency \cite{Cates_2022,HVP_partI}. Notably, the FDRs also determine the parity of irreversible processes under time reversal, in agreement with results from the PI formulation of standard ST. The key difference is that the parity need not be specified \textit{a priori} before entropy computations, further illustrating the internal consistency of this framework.

Section~\ref{sec:under_open} demonstrated how the variational formulation naturally incorporates external interactions. Nonholonomic constraints represent a broad class of irreversible processes, including couplings to equilibrium reservoirs, while external work enters through the principle of virtual work. For open systems, the resulting FDRs continue to enforce LDB, a well-defined entropy, and Onsager symmetry. However, due to the absence of energy conservation, the equilibrium ansatz is no longer satisfied, as expected for systems under continuous driving. Importantly, in the limit where external interactions vanish, relaxation to thermal equilibrium is recovered. A further contribution of the variational principle is the unambiguous identification of entropy fluxes, distinguishing intrinsic thermodynamic entropy production from entropy exchanges with the environment (entropy flow), including information-related contributions such as entropy pumping.

A central outcome of this framework is that the EPRs allow for a reformulation of standard ST within the complete thermodynamic phase space. The present formulation therefore extends key results of ST, including trajectory-level descriptions of both configurational and thermal variables, as well as FTs, explicitly demonstrated here through the master FT, to the full thermodynamic phase space. Thermodynamic consistency, enforced through LDB, ensures that these information-theoretic properties remain systematically connected to the underlying physical description.

The extension to the full thermodynamic phase space also reveals general FDRs that capture additional couplings between configurational and thermal variables. For instance, in Secs.~\ref{sec:under_closed}–\ref{sec:under_open} these correspond to system–bath couplings, while bath–bath couplings arise in Sec.~\ref{sec:inter}. In particular, noise-induced drifts appearing in the FDRs guarantee thermodynamic consistency for nonlinear couplings and multiplicative noise.

Although this work focused on the extension to spatially extended systems, taken together with Ref.~\cite{HVP_partI} the variational formulation naturally accommodates both finite- and infinite-dimensional DoFs simultaneously. Hybrid models of this type have been applied, for instance, to the study of a probe coupled to a fluctuating medium~\cite{Venturelli_2024}.

Finally, given the equivalence of FDRs derived via the second law or LDB, both are natural candidates for the axiomatization of the variational principle. Taking the second law as the fundamental principle highlights universal thermodynamic constraints and consistency across macroscopic and mesoscopic scales, while providing a direct connection with classical thermodynamics and projection-operator methods~\cite{zwanzig,Grabert1982}. Conversely, taking LDB as fundamental directly encodes time reversal symmetry breaking at the trajectory level, providing a clear physical interpretation of entropy production~\cite{Maes2003} and a natural connection to fluctuation theorems and information theory. From a practical point of view, one applies the FPE while the other relies on the path integral formulation. Notably, while microreversibility offers a compelling physical rationale for viewing LDB as fundamental, standard stochastic formulations must introduce \textit{ad hoc} closures to formulate LDB; in contrast, the present variational structure systematically generates the governing evolution equation for $\dot\Sigma$, thereby providing the closure required for the LDB relation.

\section{\label{sec:conclusion} Conclusion}
Stochastic field theories are often constructed through phenomenological arguments. As a result, models of complex systems frequently lack a systematic assessment of thermodynamic consistency, relying instead on measures of irreversibility such as informatic entropy production \cite{PhysRevLett.117.038103,Fodor_2022,ActiveFields}. A more rigorous route is to derive models by coarse-graining microscopic dynamics through methods which guarantee LDB by construction \cite{zwanzig,Grabert1982,PhysRevX.15.021050}. However, this bottom-up approach remains challenging for many applications and, in practice, is often replaced by top-down modeling guided by symmetry considerations. The latter, while powerful, does not by itself provide systematic means of enforcing LDB or, equivalently, deriving FDRs from first principles.

The variational principle introduced here offers a unifying top-down framework for stochastic field theories that systematically enforces thermodynamic consistency. In contrast to coarse-graining, it can accommodate complex models with relative simplicity, though the formulation retains free parameters to be specified by the physical context. Crucially, the framework enables an efficient treatment of Lie group symmetries and the associated reduction processes \cite{reduction_nematic,GAYBALMAZ2009}, which we aim to exploit in applications to complex fluids, and it provides a foundation for structure-preserving numerical schemes \cite{Gawlik2020,Gawlik_2021,gawlik24,GAWLIK2025114336}. 

Contrary to traditional phenomenological approaches, this formulation subsumes Lagrangian (variational) mechanics as a particular case, extending its scope and the theories built upon it. In doing so, ST and the modern tools of theoretical physics are placed on the same footing, unified by an underlying geometric structure.

In a subsequent manuscript, the application of the method to stochastic hydrodynamics will be illustrated, offering a generalization of the derivations here presented.

\begin{acknowledgments}
H. V.P. acknowledges the Provost Graduate Award scholarship from Nanyang Technological University, Singapore.  F. G.-B. is partially supported by a startup grant from Nanyang Technological University and by the Ministry of Education, Singapore, under Academic Research Fund (AcRF) Tier 1 Grant RG99/24. H.Y. is partially supported by JSPS Grant-in-Aid for Scientific Research (22K03443),
JST CREST (JPMJCR24Q5), and Waseda University (SR 2025C-095).

H. V.P. derived the results and prepared the original draft. F. G.-B. and L. Y.C. supervised the work. F. G.-B., H. Y., and L. Y.C. contributed to the review and editing of the manuscript. All authors participated in discussion.
\end{acknowledgments}

\appendix
\renewcommand{\thetable}{\thesection}
\setcounter{table}{0}

\section{\label{app:FPE_BC} Functional Fokker-Planck boundary conditions}
Suitable BCs for the (unbounded) functional phase space are natural BCs \cite{Risken1996,gardiner1985handbook}:
\begin{equation}\label{eq:natural_BC}
    \lim_{\|\phi\|_{L^2} \to \infty} \mathcal{P}_t[\bm{\phi}] = 0,
\end{equation}
where $\|\cdot\|_{L^2}$ denotes the norm induced by the inner product  \eqref{eq:inner_prod} in $\Omega$. The BCs for the probability current density $\bm{j} (\mathbf{x},t)[\bm{\phi}]$ in the spatial domain $\mathcal{D}$ can be defined, for instance, as follows \cite{Risken1996,gardiner1985handbook}:
\begin{itemize}
    \item Natural BC (for unbounded domains): 
    \begin{equation}
        \lim_{\|\mathbf{x}\| \to \infty} \bm{j} (\mathbf{x},t)[\bm{\phi}] = \bm{0}.
    \end{equation} 
    \item Reflecting BC (for bounded domains): 
    \begin{equation}
        \bm{j} (\mathbf{x},t)[\bm{\phi}] \cdot \bm{n} \big|_{\p \mathcal{D}} = 0,
    \end{equation}
    where $\bm{n}$ is normal to $\p\mathcal{D}$. 
    \item Absorbing BC (for bounded domains): 
    \begin{equation}
        \bm{j} (\mathbf{x},t)[\bm{\phi}] \big|_{\p \mathcal{D}} = \bm{0}.
    \end{equation}
    \item Periodic BC: $\bm{j} (\mathbf{x},t)[\bm{\phi}]$ is periodic across $\partial\mathcal{D}$, ensuring that the net contribution from boundary terms cancels between opposite faces of the domain.
\end{itemize}

\section{\label{app:discrete_theory} Discrete formulation and continuum limit of SPDEs}

The continuum equations in Sec.~\ref{sec:representations} are obtained as the continuum limit of their discrete counterparts. The latter follow from a Langevin equation defined on a spatial lattice, where the theory of finite-dimensional stochastic processes applies. Taking the continuum limit of this lattice formulation then yields the equations used in the main text. Throughout, we use $\l,\mu,\t,\a$ to denote spatial cells, and the corresponding coarse-grained field in cell $\l$ as $\phi^a_\l$, which evolves according to \cite{ojalvo}
\begin{equation}\label{B1}
    d\phi^a_\l  = F^a_\l (\{\phi\})\,dt + g^{a\k}_{\l \tau} (\{\phi\}) \circ dW^\k_{\tau}(t).
\end{equation}
Note that we have given $g^{a\k}_{\l \tau} (\{\phi\})$ a multiple-point coupling because, if $g^{a}_{\k}(\mathbf{x})[\bm{\phi},\bm{\nabla}]$ involves an operator such as $\bm{\nabla}$, its discrete representation $\nabla_{\tau \alpha}$ naturally couples multiple lattice points \cite{ojalvo,Toral}. Thus, for the sake of generality, we consider a multiple-point coupling in this Appendix. 

The covariance kernel and covariance operator are given, respectively, by
\begin{align}
   & \avg{dW^\k_{\t} (t)dW^\s_{\a} (t') }=2C^{\k\s}_{\t\a}\d(t-t')dt,\\
    &D^{ab}_{\l\mu} (\{\phi\},t) =g^{a\k}_{\l\t} (\{\phi\},t)g^{b\s}_{\mu\a}(\{\phi\},t)C^{\k\s}_{\t\a} .
\end{align}
Applying standard rules from finite-dimensional stochastic processes \cite{ojalvo, gardiner1985handbook} the associated Fokker-Planck equation and Onsager-Machlup action take the form:
\begin{widetext}
\begin{align} 
        &\p_t \mP_t (\{\phi\}) = -\f{\p}{\p \phi^a_\l}\bigg[\left(F^a_\l (\{\phi\},t) - g^{a\k}_{\l\t} (\{\phi\},t)\f{\p }{\p \phi^b_\mu}g^{b\s}_{\mu\a}(\{\phi\},t)C^{\k\s}_{\t\a} \right)\mP_t (\{\phi\})  - D^{ab}_{\l\mu} (\{\phi\},t) \f{\p \mP_t (\{\phi\}) }{\p \phi^b_\mu}\bigg], \label{eq:FPE_lattice}\\
        & \mathcal{A}(\{\phi\}) = \int dt\bigg[ \f{1}{4}\D V_\l \D V_\mu\left(\dot \phi^a_\l - F'^a_\l (\{\phi\})\right)[D^{ab}_{\l\mu}]^{-1}(\{\phi\})\left(\dot \phi^b_\mu - F'^b_\mu (\{\phi\})\right) \nonumber\\
        &\phantom{\mathcal{A}(\{\phi\}) = \int dt\bigg[ } +\f{1}{2}\f{\p F^a_\l (\{\phi\})}{\p \phi^a_\l}  +\f{1}{4}\left(\f{\p g^{a\k}_{\l\t}}{\p \phi^b_\mu}\f{\p g^{b\s}_{\mu\a}}{\p \phi^a_\l}-\f{\p g^{a\k}_{\l\t}}{\p \phi^a_\l}\f{\p g^{b\s}_{\mu\a}}{\p \phi^b_\mu}\right)C^{\k\s}_{\t\a} \bigg]. \label{eq:OM_lattice}
\end{align}
\end{widetext}
Thus, Eqs.~\eqref{eq:FPE} and \eqref{eq:PI_multid} are obtained by applying the rules summarized in Table~\ref{tab:discr_cont_limit}, and should be understood as shorthand notation for the lattice expressions in Eqs.~\eqref{eq:FPE_lattice} and \eqref{eq:OM_lattice}. 

In the continuum limit, the identity element diverges as \begin{equation}
    \d(0) = \lim_{\D V \rightarrow 0} \f{\d_{\l\l}}{\D V} \rightarrow \infty .
\end{equation} 
This term can emerge, for instance, in the noise-induced drifts, the entropy-pumping or the OM action, due to the property \cite{Cates_2022}
\begin{equation}
    \f{\d \phi(y)}{\d \phi(x)} = \d(x-y).
\end{equation}
For example, for a velocity-dependent control force due to an external agent implementing feedback cooling $(\hat{\g}'=-\g')$ \cite{Munakata_2012,Kim&Qian}, the entropy-pumping contribution yields 
\begin{equation}
  \mathbf{f}_+(x,t)[\bm{\pi}]=-\g'\bm{\pi}(x,t)\Rightarrow  \f{\d \mathbf{f}_+(x,t)[\bm{\pi}]}{\d \bm{\pi}(x)} = -3\g'\d(0),
\end{equation}
yielding the ill-definition of the continuum interpretation.

\begin{table}[h]
\centering
{\caption{Correspondence between discrete and continuum formulations used in the derivation of the continuum FPE and path integral \cite{ojalvo,WuWang,bellac_quantum_1992,peskinQFT,parisi_statistical_1988}. }}
\label{tab:discr_cont_limit}
\renewcommand{\arraystretch}{1.5}
{\begin{tabular}{c c c}
\hline
\textbf{Discrete} & $\Longleftrightarrow$ & \textbf{Continuum} \\
\hline
$\phi^a_\lambda$ & $\Longleftrightarrow$ & $\phi^a(\mathbf{x})$ \\
$\mP(\{\phi\})$ & $\Longleftrightarrow$ & $\mP[\bm{\phi}]$ \\
$F^a_\lambda(\{\phi\})$ & $\Longleftrightarrow$ & $F^a(\mathbf{x})[\bm{\phi}]$ \\
$\D V\displaystyle\sum_\lambda$ & $\Longleftrightarrow$ &  $\displaystyle\int d\mathbf{x}$ \\
\vspace{2mm}$\displaystyle \f{1}{\D V}\delta_{\lambda\mu}$ & $\Longleftrightarrow$ & $\delta(\mathbf{x} - \mathbf{x}')$ \\
\vspace{2mm}$\displaystyle\f{1}{\D V}\frac{\partial}{\partial \phi^a_\lambda}$ & $\Longleftrightarrow$ & $\displaystyle \frac{\delta}{\delta \phi^a(\mathbf{x})}$ \\
\vspace{2mm}$\displaystyle\prod_{a\l}d\phi^a_\lambda$ & $\Longleftrightarrow$ & $\displaystyle\int d\bm{\phi}$\\
\hline
\end{tabular}}
\end{table}

\section{\label{app:closed} Closed system derivations}
Taking variations of the variational principle \eqref{eq:VP_closed} and taking into account
the fixed endpoints condition gives
\begin{align}
    &\int_{t_1,\mathcal{D}}^{t_2}dtd\mathbf{x}\bigg( \f{\d L}{\d \bm{\phi}}\d \bm{\phi}+\f{\d L}{\d \bm{u}}\d \bm{u} +\f{\d L}{\d s}\d s +\d \bm{\pi}\cdot(\dot{\bm{\phi}} - \bm{u}) \nonumber \\
   & - \dot{\bm{\pi}}\cdot\d \bm{\phi}- \bm{\pi}\cdot\d \bm{u} -  \d\Gamma(\dot s - \dot \Sigma) + \dot \Gamma(\d s - \d \Sigma)\bigg) \nonumber \\
    &+\int_{t_1,\partial \mathcal{D}}^{t_2}dt d\boldsymbol{\sigma}\frac{\partial \ell}{\partial\bm{\nabla}\bm{\phi}}\cdot\boldsymbol{n}\delta \bm{\phi}=0.
\end{align}
Enforcing the variational constraints yields
\begin{subequations}
    \begin{align}
        & \d \bm{\pi}: \bm{u}=\dot{\bm{\phi}}, \\
        & \d \bm{u}: \bm{\pi}= \dfrac{\d L}{\d \bm{u}} = \bm{u}, \\
        & \d\G: \dot s = \dot \Sigma,\\
        & \d s: \dot\G = -\dfrac{\d L}{\d s} = T,\\
        & \d \bm{\phi}: \dot \pi^a = -\f{\d \varepsilon}{\d \phi^a }+ {\rm f}^a_-  + g^{a}_{\k} \circ\zeta^\k ,
    \end{align}
\end{subequations}
which together with the kinematic constraints results in the system \eqref{eq:EoM_closed}. The variations $\delta\phi$ at the boundary yield the BC in \eqref{eq:natural_BC_phi}.

The Shannon entropy change is computed as
\begin{equation}
\begin{aligned}
     \dot \mS  &= - \int_\o d\Psi\, \ln \mP_t\, \f{\p \mP_t}{\p t} = \int_{\o,\mathcal{D}} d\Psi d\mathbf{x}\, \ln \mP_t\, \dfrac{\d j^a(\mathbf{x},t)[\Psi]}{\d \Psi^a (\mathbf{x})}\\
&= -\int_{\o,\mathcal{D}}d\Psi d\mathbf{x}\bigg[\f{j_{\bm{\phi}}^a (\mathbf{x},t)}{ \mP_t}\f{\d \mP_t [\Psi]}{\d \phi^a (\mathbf{x})}  + \f{j_{r}^a (\mathbf{x},t)}{ \mP_t}\f{\d \mP_t [\Psi]}{\d \pi^a (\mathbf{x})} \\
& \qquad + \f{j_{d}^a (\mathbf{x},t)}{ \mP_t}\left(\f{\d \mP_t [\Psi]}{\d \pi^a (\mathbf{x})} - \f{\pi_a}{T}\f{\d \mP_t [\Psi] }{\d s (\mathbf{x})}\right)\bigg]\\
&= \int_{\o,\mathcal{D}}d\Psi d\mathbf{x}d\mathbf{x}'\bigg[\f{j_{d}^{a} (\mathbf{x},t)}{\mP_t} M_{ab}(\mathbf{x},\mathbf{x}',t)j_{d}^{b} (\mathbf{x}',t) \\
&\qquad - j_{d}^{a} (\mathbf{x},t) M_{ab}(\mathbf{x},\mathbf{x}',t)F'^b_- (\mathbf{x}',t)\bigg],
\end{aligned}
\end{equation}
where the terms $\propto j^a_{\bm{\phi}},j^a_r$ vanish via functional integration by parts and we have substituted the inverse of relation Eq. \eqref{eq:j_d_closed} in the last step.

\section{\label{app:inter} Multicomponent systems derivations}
Taking variations of the variational principle \eqref{eq:VP_inter} and taking into account fixed endpoints conditions gives
\begin{align}
    &\int_{t_1,\mathcal{D}}^{t_2}dtd\mathbf{x}\sum_k\left( \f{\d L}{\d \rho_k}\d \rho_k +\f{\d L}{\d s_k}\d s_k+\dot W^k\d \rho_k - \dot \rho_k\d W^k \nonumber \right.\\
    &-  \d\Gamma^k(\dot s_k - \dot \Sigma_k) + \dot \Gamma^k(\d s_k - \d \Sigma_k)\bigg) \nonumber\\
    &+\int_{t_1,\partial \mathcal{D}}^{t_2}dt d\boldsymbol{\sigma}\sum_k\left(\frac{\partial \ell}{\partial\bm{\nabla} \rho_k }\cdot\boldsymbol{n}\delta \rho_k+\frac{\partial \ell}{\partial\bm{\nabla} s_k }\cdot\boldsymbol{n}\delta s_k\right)=0.
\end{align}
Enforcing the variational constraints yields
\begin{subequations}
    \begin{align}
        & \d \rho_k: -\f{\d L}{\d \rho_k}=\dot W^k =\mu^k,\\
        &  \d s_k:-\f{\d L}{\d s_k} =\dot \Gamma ^k= T^k,\\
        & \d \Gamma^k: \dot\Sigma_k = \dot s_k + \bm{\nabla}\cdot (\bm{J}_k+\bm{\eta}_k),\\
        & \d W^k: \dot \rho_k =- \bm{\nabla}\cdot(\bm{\mJ}_k+\bm{\xi}_k),
    \end{align}
\end{subequations}
which together with the kinematic constraints results in the system \eqref{eq:EoM_interc}. 
The variations $\delta\rho_k$, $\delta s_k$ at the boundary yield the natural BCs. The Neumann no-flux BCs follow directly from integration by parts. Indeed, from
\begin{equation}
    \begin{aligned}
       &\int_{\mathcal{D}}d\mathbf{x}\frac{\d L}{\d s_k}\d \Sigma_k \\ & =\int_{\mathcal{D}}d\mathbf{x}\bigg[(\bm{J}_{k}+\bm{\eta}_{k})\cdot\bm{\nabla}\d\Gamma^k +(\bm{\mJ}_{k}+\bm{\xi}_{k})\cdot\bm{\nabla}\d W^k\bigg] \\
       & = \int_{\p\mathcal{D}}d\bm{\s}\bigg[(\bm{J}_{k}+\bm{\eta}_{k})\cdot\bm{n}\d\Gamma^k +(\bm{\mJ}_{k}+\bm{\xi}_{k})\cdot\bm{n}\d W^k\bigg]\\
       &\phantom{=} -\int_{\mathcal{D}}d\mathbf{x}\bigg[\bm{\nabla}\cdot (\bm{J}_{k}+\bm{\eta}_{k})\d\Gamma^k + \bm{\nabla}\cdot(\bm{\mJ}_{k}+\bm{\xi}_{k})\d W^k\bigg],
    \end{aligned}
\end{equation}
the variations $\d\Gamma^k,\d W^k$ at the boundary lead to  $(\bm{J}_k+\bm{\eta}_k) \cdot\bm{n}|_{\p\mathcal{D}} = (\bm{\mJ}_k+\bm{\xi}_k)\cdot\bm{n}|_{\p\mathcal{D}} = 0$.

The dissipative mass and entropy probability density currents are given by [Eq. \eqref{eq:FPE}]
\begin{widetext}
   \begin{align}
    & \mathfrak{J}^a_\rho(\mathbf{x},t)[\Psi] = (\mJ^a_{12} - \nu^a_\rho)(\mathbf{x},t)[\Psi]\mP_t[\Psi] \nonumber\\
    & \phantom{\bm{\mathfrak{J}}_\rho(\mathbf{x},t)[\Psi] =} + \int_{\mathcal{D}} d\mathbf{x}'\bigg[\p'_b G^{ab}(\mathbf{x},\mathbf{x}')\left(\f{\d\mP_t [\Psi]}{\d \rho_2 (\mathbf{x}')}-\f{\d\mP_t [\Psi]}{\d \rho_1 (\mathbf{x}')}\right) +  G^{ab}(\mathbf{x},\mathbf{x}')\left(\f{\p'_b \mu^2 (\mathbf{x'})}{T^2 (\mathbf{x'})}\f{\d\mP_t [\Psi]}{\d s_2 (\mathbf{x}')}-\f{\p'_b \mu^1 (\mathbf{x'})}{T^1 (\mathbf{x'})}\f{\d\mP_t [\Psi]}{\d s_1 (\mathbf{x}')}\right) \nonumber\\
    & \phantom{\bm{\mathfrak{J}}_\rho(\mathbf{x},t)[\Psi] =} + g^a_\k(\mathbf{x})\left(\frac{\p'_b(T^1 (\mathbf{x'}) C^{\k\s}(\mathbf{x},\mathbf{x}') \eta^b_\s (\mathbf{x}'))}{T^1 (\mathbf{x'})}\f{\d\mP_t [\Psi]}{\d s_1 (\mathbf{x}')}+\frac{\p'_b (T^2 (\mathbf{x'}) C^{\k\s}(\mathbf{x},\mathbf{x}') \eta^b_\s(\mathbf{x}'))}{T^2 (\mathbf{x'})}\f{\d\mP_t [\Psi]}{\d s_2 (\mathbf{x}')}\right)\bigg],\\
    &\mathfrak{J}^a_s (\mathbf{x},t)[\Psi] = (J^a_{12} - \nu^a_s)(\mathbf{x},t)[\Psi]\mP_t[\Psi] \nonumber\\
    & \phantom{\bm{\mathfrak{J}}_s } + \int_{\mathcal{D}} d\mathbf{x}'\bigg[  \left(\frac{\p_b' (T^1 (\mathbf{x'}) K^{ab} (\mathbf{x},\mathbf{x}'))}{T^1 (\mathbf{x'})}\f{\d\mP_t [\Psi]}{\d s_1 (\mathbf{x}')}+\frac{\p'_b (T^2 (\mathbf{x'}) K^{ab} (\mathbf{x},\mathbf{x}'))}{T^2 (\mathbf{x'})}\f{\d\mP_t [\Psi]}{\d s_2 (\mathbf{x}')}\right) \nonumber\\
    &\phantom{\bm{\mathfrak{J}}_s } +   \eta^a_\k(\mathbf{x}) \bigg\{\p'_b (C^{\k\s}(\mathbf{x},\mathbf{x}')g^b_\s (\mathbf{x}'))\left(\f{\d\mP_t [\Psi]}{\d \rho_2 (\mathbf{x}')}-\f{\d\mP_t [\Psi]}{\d \rho_1 (\mathbf{x}')}\right)  +C^{\k\s}(\mathbf{x},\mathbf{x}') g^b_\s (\mathbf{x}') \left(\f{\p'_b \mu^2 (\mathbf{x'})}{T^2 (\mathbf{x'})}\f{\d\mP_t [\Psi]}{\d s_2 (\mathbf{x}')}-\f{\p'_b \mu^1 (\mathbf{x'})}{T^1 (\mathbf{x'})}\f{\d\mP_t [\Psi]}{\d s_1 (\mathbf{x}')}\right) \bigg\}\bigg],
\end{align} 
with
\begin{align}
    &\begin{bmatrix}
        \vspace{3mm}\nu_\rho^{a}(\mathbf{x},t) \\ \nu_s^{a} (\mathbf{x},t)
    \end{bmatrix} =\int_{\mathcal{D}}d\mathbf{x'} \begin{bmatrix}
        \vspace{3mm}g^a_\k (\mathbf{x},t) & 0 \\ 0 & \eta^a_\k (\mathbf{x},t) 
    \end{bmatrix}\begin{bmatrix}
       \vspace{1mm} \d_{\rho_1 (\mathbf{x}')} \\ \vspace{1mm}\d_{\rho_2 (\mathbf{x}')} \\ \vspace{1mm}\d_{s_1 (\mathbf{x}')} \\ \d_{s_2 (\mathbf{x}')}
    \end{bmatrix} \cdot \begin{bmatrix}
         \vspace{1mm} \bm{\chi}_1 (\mathbf{x'},t) \\ \bm{\chi}_2 (\mathbf{x'},t)
    \end{bmatrix} \nonumber\\
    &\begin{cases}
        \vspace{2mm}\bm{\chi}_1 =\bigg[\p_b (\d^{\k\s}g^b_\s) , - \p_b (\d^{\k\s}g^b_\s ),   \left(\frac{\p_b \mu^1}{T^1}\d^{\k\s}g^b_\s-\frac{\p_b (T^1 C^{\k\s}\eta^b_\s)}{T^1}\right),\left( -\frac{\p_b \mu^2}{T^2}\d^{\k\s}g^b_\s-\frac{\p_b (T^2 C^{\k\s}\eta^b_\s)}{T^2}\right) \bigg],\\
        \bm{\chi}_2 = \bigg[\p_b (C^{\k\s}g^b_\s), - \p_b (C^{\k\s}g^b_\s) ,\left(\frac{\p_b \mu^1}{T^1}C^{\k\s}g^b_\s-\frac{\p_b (T^1 \d^{\k\s}\eta^b_\s)}{T^1}\right), \left( -\frac{\p_b  \mu^2}{T^2}C^{\k\s}g^b_\s- \frac{\p_b (T^2 \d^{\k\s}\eta^b_\s)}{T^2}\right) \bigg].
    \end{cases}
\end{align}

Using the no-flux BCs $\bm{J}_{12}\cdot\bm{n}|_{\p\mathcal{D}}=\bm{\eta}_{12}\cdot\bm{n}|_{\p\mathcal{D}}=0$ (and analogously for the mass fluxes), the currents $  \bm{\mathfrak{J}}_\rho,\bm{\mathfrak{J}}_s $ can be simplified by integration by parts. Defining
\begin{equation}
\begin{bmatrix}
    \vspace{2mm} \bm{A}(\mathbf{x}) \\ \bm{B}(\mathbf{x})
\end{bmatrix}=\begin{bmatrix}
    \vspace{2mm}\sum_k (-1)^k\left(\f{\bm{\nabla} \mu^k}{T^k}\f{\d\mP_t}{\d s_k (\mathbf{x})}  -  \bm{\nabla}\f{\d\mP_t}{\d \rho_k (\mathbf{x})}\right)\\
    -\sum_k T^k(\mathbf{x}) \bm{\nabla}\big( \frac{1}{T^k}\f{\d\mP_t}{\d s_k (\mathbf{x})}\big)
\end{bmatrix},
\end{equation}
the currents can be rewritten as
\begin{equation}\label{eq:inv_currents_inter}
    \begin{bmatrix}
        \bm{\mathfrak{J}}_\rho - \bm{\mJ}'_{12}\mP_t \\ \bm{\mathfrak{J}}_s -\bm{J}'_{12}\mP_t 
    \end{bmatrix}(\mathbf{x},t)[\Psi]=\int_{\mathcal{D}} d\mathbf{x'}\mathbb{L}_\Psi(\mathbf{x},\mathbf{x}')\begin{bmatrix}
        \bm{A}(\mathbf{x'}) \\ \bm{B}(\mathbf{x}')
    \end{bmatrix}.
\end{equation}
Recall the no-flux BCs implied $J^a_{12} n_a|_{\p\mathcal{D}}=\eta^a_{12} n_a|_{\p\mathcal{D}}=0 \implies \eta^a _\kappa n_a |_{\p\mathcal{D}}=0$ for each $\kappa$. This in turn implies both $K^{ab}n_a|_{\p\mathcal{D}}=0$ and $K^{ab}n_b|_{\p\mathcal{D}}=0$, c.f. Eq.~\eqref{J_noise}, from which $\mathfrak{J}^a_s n_a |_{\p\mathcal{D}}=0$ follows by direct computation (and analogously for $\bm{\mathfrak{J}}_\rho$). Here $\mathbb{L}_\Psi(\mathbf{x},\mathbf{x}')$ is given by \eqref{eq:L_inter}.

The Shannon entropy change is computed as
\begin{equation}\label{D8}
\begin{aligned}
     \dot \mS  &= - \int_\o d\Psi\, \ln \mP_t\, \f{\p \mP_t}{\p t} = \int_{\o,\mathcal{D}} d\Psi d\mathbf{x}\, \ln \mP_t\, \dfrac{\d j^a(\mathbf{x},t)[\Psi]}{\d \Psi^a (\mathbf{x})}= -\int_{\o,\mathcal{D}}d\Psi d\mathbf{x}\bigg[\f{j_{\rho_k} (\mathbf{x},t)}{ \mP_t }\f{\d \mP_t [\Psi]}{\d \rho_k (\mathbf{x})}  + \f{j_{s_k} (\mathbf{x},t)}{ \mP_t}\f{\d \mP_t [\Psi]}{\d s_k (\mathbf{x})} \bigg]\\
&= \int_{\o,\mathcal{D}}\f{d\Psi}{\mP_t} d\mathbf{x}\bigg[ \bm{\nabla}\cdot \bm{\mathfrak{J}}_\rho (\mathbf{x},t)\sum_k (-1)^k\f{\d \mP_t [\Psi]}{\d \rho_k (\mathbf{x})} +\bm{\mathfrak{J}}_\rho (\mathbf{x},t)\sum_k (-1)^k \f{\bm{\nabla} \mu^k}{T^k}\f{\d\mP_t [\Psi]}{\d s_k (\mathbf{x})} +\sum_k\f{\bm{\nabla}\cdot(T^k\bm{\mathfrak{J}}_s)}{T^k}\f{\d \mP_t}{\d s_k (\mathbf{x})}\bigg]\\
   &= \int_{\o,\mathcal{D}} \f{d\Psi}{\mP_t}  d\mathbf{x} d\mathbf{x'}   \begin{bmatrix}
        \bm{\mathfrak{J}}_\rho (\mathbf{x},t) \\ \bm{\mathfrak{J}}_s  (\mathbf{x},t)
    \end{bmatrix}^T \cdot \begin{bmatrix}
     \bm{A}(\mathbf{x}) \\ \bm{B}(\mathbf{x}) 
\end{bmatrix},
\end{aligned}
\end{equation}
where we applied ordinary and functional integration by parts. Applying the inverse of relation \eqref{eq:inv_currents_inter}, we then recover the result stated in the main text. 

For completeness, the LDB condition \eqref{eq:LDB_VP} is validated through an explicit PI computation. The field dynamics in Eqs.~\eqref{eq:EoM_interc} can be recast as a deterministic system supplemented by two stochastic current equations (see e.g. \cite{Cates_2022,HVP_partI}):
\begin{equation}
 \bm{\mathfrak{j}}_\rho = \bm{\mJ}_{12}+\bm{\xi}_{12}, \quad
 \bm{\mathfrak{j}}_s = \bm{J}_{12}+\bm{\eta}_{12},
\end{equation}
where $\bm{\mathfrak{j}}_\rho$ and $\bm{\mathfrak{j}}_s$ are odd under time reversal, as are $\p_t \rho_k$ and $\p_t s_k$. 

The forward and backward actions are then given by
\begin{align}
    & \mathcal{A}[\Psi] = \f{1}{4} \int_{t_1 , \mathcal{D}}^{t_2}d\t d\mathbf{x} d\mathbf{x'} \begin{bmatrix}
       \bm{\mathfrak{j}}_\rho -\bm{\mJ}'_{12}\\ \bm{\mathfrak{j}}_s -\bm{J}'_{12}
    \end{bmatrix}^T\mathbb{L}^{-1}_\Psi (\mathbf{x},\mathbf{x'}) \begin{bmatrix}
       \bm{\mathfrak{j}}_\rho -\bm{\mJ}'_{12}\\ \bm{\mathfrak{j}}_s -\bm{J}'_{12}
    \end{bmatrix}',\\
    & \hat{\mathcal{A}}[\Psi] = \f{1}{4} \int_{t_1 , \mathcal{D}}^{t_2}d\t d\mathbf{x} d\mathbf{x'} \begin{bmatrix}
       -\bm{\mathfrak{j}}_\rho -\bm{\mJ}'_{12}\\ -\bm{\mathfrak{j}}_s -\bm{J}'_{12}
    \end{bmatrix}^T\mathbb{L}^{-1}_\Psi (\mathbf{x},\mathbf{x'}) \begin{bmatrix}
       -\bm{\mathfrak{j}}_\rho -\bm{\mJ}'_{12}\\ -\bm{\mathfrak{j}}_s -\bm{J}'_{12}
    \end{bmatrix}',
\end{align}
with $[\cdot] '$ denoting $\mathbf{x'}$ dependence and where the Jacobian contribution (divergence) has been omitted due to being symmetric under time reversal. The log ratio \eqref{eq:EP} yields, after some algebra:
\begin{equation}
    \begin{aligned}  
    \D \mathfrak{s}_m &= \hat{\mathcal{A}}[\Psi] - \mathcal{A}[\Psi]\\
    & =\int_{t_1 , \mathcal{D}}^{t_2}d\t d\mathbf{x} d\mathbf{x'}\begin{bmatrix}
        (\bm{\mJ}_{12}+\bm{\xi}_{12})(\mathbf{x},\t) \\ (\bm{J}_{12}+\bm{\eta}_{12})(\mathbf{x},\t)
    \end{bmatrix}^T\mathbb{L}^{-1}_\Psi (\mathbf{x},\mathbf{x'})\begin{bmatrix}
       \bm{\mJ}'_{12} (\mathbf{x}',\t) \\ \bm{J}'_{12}(\mathbf{x}',\t)
    \end{bmatrix} .
    \end{aligned}
\end{equation}
Applying the FDRs \eqref{eq:FDR_inter} it finally gives: 
\begin{equation}
     \D \mathfrak{s}_m = \int_{t_1 , \mathcal{D}}^{t_2}d\t d\mathbf{x}\bigg\{-\left(\frac{\bm{\nabla} T^1}{T^1}+\frac{\bm{\nabla} T^2}{T^2}\right)\cdot(\bm{J}_{12}+\bm{\eta}_{12})  + \left(\f{\bm{\nabla} \mu^1}{T^1}-\f{\bm{\nabla} \mu^2}{T^2}\right)\cdot(\bm{\mJ}_{12}+\bm{\xi}_{12})\bigg\},
\end{equation}
and hence $\D \mathfrak{s}_m = \D\Sigma =\sum_k  \D\Sigma_k$ by direct comparison with Eqs. \eqref{eq:EoM_interc}, which is the LDB condition in \eqref{eq:LDB_VP}.
\end{widetext}

\section{\label{app:open} Open system derivations}
Taking variations of the variational principle \eqref{eq:VP_open} and taking into account
the fixed endpoints condition gives
\begin{align}
    &\int_{t_1,\mathcal{D}}^{t_2}dtd\mathbf{x}\bigg( \f{\d L}{\d \bm{\phi}}\d \bm{\phi}+\f{\d L}{\d \bm{u}}\d \bm{u} +\f{\d L}{\d s}\d s +\d \bm{\pi}\cdot(\dot{\bm{\phi}} - \bm{u}) - \dot{\bm{\pi}}\cdot\d \bm{\phi}\nonumber \\
    &- \bm{\pi}\cdot\d \bm{u} -  \d\Gamma(\dot s - \dot \Sigma) + \dot \Gamma(\d s - \d \Sigma) + \mathbf{f}_+ \cdot\d\bm{\phi}\bigg) \nonumber\\
    &+\int_{t_1,\partial \mathcal{D}}^{t_2}dt d\boldsymbol{\sigma}\frac{\partial \ell}{\partial\bm{\nabla}\bm{\phi}}\cdot\boldsymbol{n}\delta \bm{\phi}=0.
\end{align}
\\ \\
Enforcing the variational constraints yields
\begin{subequations}
    \begin{align}
        & \d \bm{\pi}: \bm{u}=\dot{\bm{\phi}}, \\
        & \d \bm{u}: \bm{\pi}= \dfrac{\d L}{\d \bm{u}} = \bm{u}, \\
        & \d\G: \dot \Sigma =  \dot s +\bm{\nabla}\cdot(\boldsymbol{j}  + \bm{\eta}) - \mI,\\
        & \d s: \dot\G = -\dfrac{\d L}{\d s} = T,\\
        & \d \bm{\phi}: \dot \pi^a = -\f{\d \varepsilon}{\d \phi^a }+{\rm f}^a_+ +{\rm f}^a_-  + g^{a}_{\k} \circ\zeta^\k ,
    \end{align}
\end{subequations}
which together with the kinematic constraints results in the system \eqref{eq:EoM_open} together with the natural BCs.  

\begin{widetext}
The Shannon entropy change is computed as
\begin{equation}\label{dot_S}
\begin{aligned}
     \dot \mS  &= - \int_\o d\Psi\, \ln \mP_t\, \f{\p \mP_t}{\p t} = \int_{\o,\mathcal{D}} d\Psi d\mathbf{x}\, \ln \mP_t\,\dfrac{\d j^a(\mathbf{x},t)[\Psi]}{\d \Psi^a (\mathbf{x})}\\
&= -\int_{\o,\mathcal{D}}d\Psi d\mathbf{x}\bigg[\f{j_{\bm{\phi}}^a (\mathbf{x},t)}{ \mP_t}\f{\d \mP_t [\Psi]}{\d \phi^a (\mathbf{x})}  + \f{j_{r}^a (\mathbf{x},t)}{ \mP_t}\f{\d \mP_t [\Psi]}{\d \pi^a (\mathbf{x})} + \f{j_{d}^a (\mathbf{x},t)}{ \mP_t}\left(\f{\d \mP_t [\Psi]}{\d \pi^a (\mathbf{x})} - \f{\pi_a}{T}\f{\d \mP_t [\Psi] }{\d s (\mathbf{x})}\right)-\frac{1}{T}\bm{\nabla}\cdot(T\bm{\mathfrak{J}}_s )\f{\d \mP_t [\Psi]}{\d s (\mathbf{x})} \bigg].
\end{aligned}
\end{equation}
The first two contributions vanish via functional integration by parts except for the term
\begin{equation}\label{eq:E4}
    \Bavg{\int_{\mathcal{D}}d\mathbf{x}\f{\d }{\d \bm{\pi}(\mathbf{x})}\cdot \mathbf{f}_+ [\Psi]}=\Bavg{\int_{\mathcal{D}}d\mathbf{x}\,\mI}.
\end{equation}
To proceed, we use the inverse relation [Eq. \eqref{eq:inverse_def}] applied to Eq. \eqref{eq:j_d_closed}, following the derivation in Appendix \ref{app:closed}:
\begin{equation}\label{eq:E5}
    \begin{aligned}
    & -\int_{\o,\mathcal{D}}d\Psi d\mathbf{x} \f{j_{d}^a (\mathbf{x},t)}{ \mP_t}\left(\f{\d \mP_t [\Psi]}{\d \pi^a (\mathbf{x})} - \f{\pi_a}{T}\f{\d \mP_t [\Psi] }{\d s (\mathbf{x})}\right)=\int_{\o,\mathcal{D}}d\Psi d\mathbf{x}d\mathbf{x}'\f{j_{d}^{a} (\mathbf{x},t)}{\mP_t} [D^{-1}]_{ab}(\mathbf{x},\mathbf{x}',t)\bigg[j_{d}^{b} (\mathbf{x}',t)- F'^b_- (\mathbf{x}',t)\mP_t\bigg].
    \end{aligned}
\end{equation}

For the last term, we illustrate how the FDRs are independent of the BCs (both spatial and of FPE). Applying ordinary integration by parts:
\begin{equation}\label{eq:E6}
    \begin{aligned}
    &\int_{\mathcal{D}} d\mathbf{x}\frac{1}{T}\bm{\nabla}\cdot [T\bm{\mathfrak{J}}_s(\mathbf{x},t) ]\f{\d \mP_t [\Psi]}{\d s (\mathbf{x})}  = \int_{\p\mathcal{D}} d\bm{\s}\,\bm{\mathfrak{J}}_s (\mathbf{x},t)\cdot\bm{n}\f{\d \mP_t [\Psi]}{\d s (\mathbf{x})} - \int_{\mathcal{D}} d\mathbf{x}\, T\bm{\mathfrak{J}}_s (\mathbf{x},t)\cdot\bm{\nabla}\left(\frac{1}{T}\f{\d \mP_t [\Psi]}{\d s (\mathbf{x})} \right).
    \end{aligned}
\end{equation}
Applying ordinary integration by parts in the definition of $\bm{\mathfrak{J}}_s$ [Eq. \eqref{eq:J_s_current}] and multiplying on both sides by the inverse of $\bm{K}$ as defined in Eq. \eqref{eq:inverse_def}:
\begin{equation}\label{eq:E7}
    \begin{aligned}
        &\bm{\mathfrak{J} }_s (\mathbf{x},t)[\Psi]  - \bm{j}'(\mathbf{x},t)[\Psi]\mP_t [\Psi] = \int_{\p\mathcal{D}} d\bm{\s}'\bm{K} (\mathbf{x},\mathbf{x}',t)[\Psi]\cdot\bm{n}  \f{\d \mP_t [\Psi]}{\d s (\mathbf{x}')} - \int_{\mathcal{D}} d\mathbf{x}'T(\mathbf{x}')\bm{K} (\mathbf{x},\mathbf{x}',t)[\Psi]\cdot\bm{\nabla}'  \bigg(\frac{1}{T(\mathbf{x}')}\f{\d \mP_t [\Psi]}{\d s (\mathbf{x}')}\bigg), \\
        &\times \int_{\mathcal{D}}d\mathbf{x}\,\bm{K}^{-1}(\mathbf{x}'',\mathbf{x}) \implies\\
        & \int_{\mathcal{D}}d\mathbf{x}'\,\bm{K}^{-1}(\mathbf{x},\mathbf{x}')\big[\bm{\mathfrak{J} }_s (\mathbf{x}',t)  - \bm{j}'(\mathbf{x}',t)\mP_t \big] = \int_{\p\mathcal{D}} d\bm{\s}' \bm{n}  \f{\d \mP_t [\Psi]}{\d s (\mathbf{x}')}\d(\mathbf{x}-\mathbf{x}') - T(\mathbf{x})\cdot\bm{\nabla}  \bigg(\frac{1}{T(\mathbf{x})}\f{\d \mP_t [\Psi]}{\d s (\mathbf{x})}\bigg),
    \end{aligned}
\end{equation}
where we have applied $\int_{\mathcal{D}}d\mathbf{x}\,\bm{K}^{-1}(\mathbf{x}'',\mathbf{x}) \bm{K}(\mathbf{x},\mathbf{x}')=\mathbb{I}\d(\mathbf{x}''-\mathbf{x}')$ and relabeled the free variable $\mathbf{x}''\rightarrow \mathbf{x}$. 

Substituting Eq. \eqref{eq:E7} in the last term of Eq. \eqref{eq:E6} yields:
\begin{equation}
    \begin{aligned}
    &\int_{\mathcal{D}} d\mathbf{x}\frac{1}{T}\bm{\nabla}\cdot [T\bm{\mathfrak{J}}_s(\mathbf{x},t) ]\f{\d \mP_t [\Psi]}{\d s (\mathbf{x})}  = \int_{\p\mathcal{D}} d\bm{\s}\,\bm{\mathfrak{J}}_s (\mathbf{x},t)\cdot\bm{n}\f{\d \mP_t [\Psi]}{\d s (\mathbf{x})} +  \int_{\mathcal{D}}d\mathbf{x}d\mathbf{x'}\,\bm{\mathfrak{J}}_s (\mathbf{x},t)\bm{K}^{-1}(\mathbf{x},\mathbf{x}')\big[\bm{\mathfrak{J} }_s (\mathbf{x}',t)  - \bm{j}'(\mathbf{x}',t)\mP_t \big] \\
    &\phantom{\int_{\mathcal{D}} d\mathbf{x}\frac{1}{T}\bm{\nabla}\cdot [T\bm{\mathfrak{J}}_s(\mathbf{x},t) ]\f{\d \mP_t [\Psi]}{\d s (\mathbf{x})}  =} - \int_{\p\mathcal{D},\mathcal{D}} d\bm{\s}'d\mathbf{x}\, \bm{\mathfrak{J}}_s (\mathbf{x},t)\cdot \bm{n}  \f{\d \mP_t [\Psi]}{\d s (\mathbf{x}')}\d(\mathbf{x}-\mathbf{x}')\\
    &\phantom{\int_{\mathcal{D}} d\mathbf{x}\frac{1}{T}\bm{\nabla}\cdot [T\bm{\mathfrak{J}}_s(\mathbf{x},t) ]\f{\d \mP_t [\Psi]}{\d s (\mathbf{x})}  }  = \int_{\p\mathcal{D}} d\bm{\s}\,\bm{\mathfrak{J}}_s (\mathbf{x},t)\cdot\bm{n}\f{\d \mP_t [\Psi]}{\d s (\mathbf{x})} +  \int_{\mathcal{D}}d\mathbf{x}d\mathbf{x'}\,\bm{\mathfrak{J}}_s (\mathbf{x},t)\bm{K}^{-1}(\mathbf{x},\mathbf{x}')\big[\bm{\mathfrak{J} }_s (\mathbf{x}',t)  - \bm{j}'(\mathbf{x}',t)\mP_t \big] \\
    &\phantom{\int_{\mathcal{D}} d\mathbf{x}\frac{1}{T}\bm{\nabla}\cdot [T\bm{\mathfrak{J}}_s(\mathbf{x},t) ]\f{\d \mP_t [\Psi]}{\d s (\mathbf{x})}  =} - \int_{\p\mathcal{D}} d\bm{\s}'\, \bm{\mathfrak{J}}_s (\mathbf{x}',t)\cdot \bm{n}  \f{\d \mP_t [\Psi]}{\d s (\mathbf{x}')}.
    \end{aligned}
\end{equation}
Since the surface integrals cancel each other (note $\mathbf{x}$ is a dummy variable), we finally arrive at:
\begin{equation}\label{eq:E9}
    \int_{\o,\mathcal{D}}\f{d\Psi}{\mP_t} d\mathbf{x}\frac{1}{T}\bm{\nabla}\cdot [T\bm{\mathfrak{J}}_s(\mathbf{x},t) ]\f{\d \mP_t [\Psi]}{\d s (\mathbf{x})}  =   \int_{\o,\mathcal{D}}\f{d\Psi}{\mP_t} d\mathbf{x}d\mathbf{x}'\, \bm{\mathfrak{J}}_s (\mathbf{x},t)\bm{K}^{-1}(\mathbf{x},\mathbf{x}',t)\bigg[\bm{\mathfrak{J} }_s (\mathbf{x}',t)  - \bm{j}'(\mathbf{x}',t)\mP_t \bigg].
\end{equation}
Collecting all contributions \eqref{eq:E4},\eqref{eq:E5},\eqref{eq:E9} in \eqref{dot_S} yield the result stated in the main text. Remarkably, this result is independent of any BCs, and a similar procedure is applicable for any conserved dynamics of a scalar field of the form $\p_t \varphi=-\bm{\nabla}\cdot(\bm{j}+\bm{\xi})$, including, for instance the example of Sec. \ref{sec:inter} for any choice of BCs.
\end{widetext}


\bibliography{bibliography}

\end{document}